\documentclass{aa}  

\usepackage{graphicx}
\usepackage{txfonts}
\usepackage{xcolor}
\usepackage{placeins}

\begin{document}

   \title{Reconnaissance ultracool spectra in the \textit{Euclid} Deep Fields}

   \author{
  J.-Y.\ Zhang 
  \inst{1,2}   
     \and  N. Lodieu \inst{1,2}
     \and  E.\ L.\ Mart\'in \inst{1,2}
        }

   \institute{Instituto de Astrof\'isica de Canarias (IAC), Calle V\'ia L\'actea s/n, 38200 La Laguna, Tenerife, Spain \\
       \email{jzhang@iac.es}
       \and
       Departamento de Astrof\'isica, Universidad de La Laguna (ULL), Avenida Astrofísico Francisco Sánchez s/n, 38206 La Laguna, Tenerife, Spain
       }

   \date{\today{}}

 
  \abstract
   {The \textit{Euclid} spacecraft has been launched and will carry out a deep survey benefiting the discovery and characterisation of ultracool dwarfs (UCDs), especially in the \textit{Euclid} Deep Fields (EDFs), which the telescope will scan repeatedly throughout its mission. The photometric and spectroscopic standards 
   in the EDFs are important benchmarks, crucial for the classification and characterisation of new UCD discoveries and for the calibration of the mission itself.}
   {We aim to provide a list of photometric UCD candidates and collect near-infrared reconnaissance spectra for M, L, and T-type UCDs in the EDFs as future $Euclid$ UCD references.}
   {In EDF North, we cross-matched public optical and infrared surveys with certain photometric criteria to select UCDs. In EDF Fornax and EDF South, we used photometrically classified samples from the literature. We also include UCDs identified by \textit{Gaia} DR2.
   We selected seven UCD targets with different spectral types from the lists and obtained low-resolution 0.9--2.5 $\mu$m spectra of them  using GTC/EMIR and the VLT/X-shooter.  We also selected a young, bright L dwarf near EDF Fornax to test the coherence of these two facilities. We included one extra T dwarf in EDF North with its published $J$-band spectrum.}
   {We retrieved a list of 81 (49, 231) M, eight (29, 115) L, and one (0, 2) T dwarf candidates in EDF North, Fornax, and South, respectively. They are provided to guide future UCD discoveries and characterisations by $Euclid$. 
   In total, we collected near-infrared spectra for nine UCDs,  including two M types, three L types, and four T types in or close to the three EDFs.
  The Euclidised spectra show consistency in their spectral classification, which demonstrates that slitless \textit{Euclid} spectroscopy will recover the spectral types with high fidelity for UCDs, both in the EDFs and in the wide survey. We also demonstrate that \textit{Euclid} will be able to distinguish different age groups of UCDs. }
   {
   }

               
\maketitle
   
%

\section{Introduction}

Ultracool dwarfs (UCDs), including the lowest-mass stars, brown dwarfs (BDs), and planetary mass objects (PMOs), are the dimmest, coldest objects among our Galactic population. As the name suggests, these objects 
can have effective temperatures from 2700\,K \citep{kirkpatrick1995first_solar_neighbour_UCD}  down to around 250\,K, the lowest known so far \citep{luhman2014W0855}. The established spectral classes for UCDs are the late-M type \citep[M7 and later, ][]{kirkpatrick1997ultracoolM}, the 
L type \citep{1997Martin,1998Martin,martin1999Lclassification,kirkpatrick1999L,2000Kirkpatrick}, the T type \citep{burgasser2002dT_spec_classification,geballe2002dT_classification, burgasser2006unifiedT}, and recently the Y type \citep{delorme2008TY_J0059, cushing2011dY_WISE, kirkpatrick2011hundredBD_WISE} with decreasing temperature.  

Nowadays, we believe that the census of UCD is still incomplete beyond 20 pc. The coldest population with spectral types later than T8 has not been explored thoroughly even within 20 pc \citep{kirkpatrick2021LTY20pc}. An appropriate method of discovering a UCD population is a deep, wide survey at the near-infrared (NIR) wavelength \citep{cushing2014bookUCD_LTY}. This is because UCDs are faint and have their emission peaks in the NIR, and they are scattered all over the sky. 

Space telescopes such as the Hubble Space Telescope (HST) and the James Webb Space Telescope (JWST), or space surveys like the Gaia mission \citep{gaia2016} and the Wide-field Infrared Survey Explorer \citep[WISE;][]{wright2010WISE}, have the great advantage of being free of atmospheric influences compared with ground-based facilities. HST has discovered many UCDs in its deep fields \citep{pirzkal2005HubbleDeepField_star,pirzkal2009PEARS,stanway2008GOODS_dM,ryan2005HST_LTdistribution,Ryan2011HST_UCD_highLat,holwerda2014dwarf_distribution_WFC3,vanVledder2016BDdM_BoRG,aganze2022HST_parallel_UCDdensity,aganze2022HST_parallel_UCDage_scaleheights}. JWST has also discovered many UCDs in several fields of its Early Release Observations \citep{wang_poya2023dT_NIRCamEarly,nonino2023GLASS_JWST,holwerda2023JWST_CEERS_dwarf} and galactic surveys \citep{hainline2023BD_JADES_CEERS}, including a discovery of the farthest BD to date at 4.8 kpc \citep{langeroodi2023farthestBD}. JWST also demonstrated its power in characterising UCDs \citep{beiler2023dY_SED,burgasser2023JWST_kpcBD,holwerda2023JWST_CEERS_dwarf,manjavacas2024JWST_youngBD}. Gaia not only discovers UCDs in the whole sky \citep{reyle2018GaiaDR2UCD,sarro2023gaiaDR3_UCD} but also performs low-resolution spectroscopy on UCDs \citep{cooper2024gaia_ucd_spectra}; its precise astrometry reveals UCD companions of the other stars \citep{stevenson2023BD_desert_binary_Gaia}. WISE has also contributed to UCD discoveries \citep{kirkpatrick2011hundredBD_WISE}, especially the coldest population \citep{cushing2011dY_WISE,cushing2014_WISE_3coolBD}, thanks to its specially designed passbands \citep{wright2010WISE} and whole-sky coverage. Next-generation space telescopes like the Nancy Grace Roman Space Telescope will offer a great opportunity for UCD sciences as well \citep{holwerda2023MLTY_Roman}.

$Euclid$\footnote{\url{https://www.cosmos.esa.int/web/euclid}} \citep{laureijs2011euclid}, a 1.2-m space telescope launched on July 1 2023, is a European Space Agency (ESA) mission aiming to conduct a deep survey covering $\approx$ 14500 deg$^2$ of the sky in the coming six years.  \textit{Euclid} is 
equipped with optical and infrared capabilities, with the Visible Instrument (VIS) and the Near Infrared Spectrometer and Photometer (NISP) on board. Compared with HST or JWST, a huge advantage of $Euclid$ is its wide field of view (FoV). The VIS has an FoV of 0.787 deg $\times$ 0.709 deg  in the $I_E$ band; the NISP covers the $Y_E$, $J_E$, and $H_E$ bands and has an FoV of 0.763 deg $\times$ 0.722 deg. The \textit{Euclid} Wide Surveys (EWSs) will have a 5-$\sigma$ limit of 26.2 mag, 24.3 mag, 24.5 mag, and 24.4 mag in the $I_E$, $Y_E$, $J_E$, and $H_E$ bands, respectively \citep{euclid2022i.EWS}. Compared with Gaia, $Euclid$ is much more sensitive to UCDs. $Euclid$ has a much better spatial resolution than WISE. The NISP can also perform low-resolution slitless spectroscopy ($R\approx380$ for point sources) from 1.25--1.85 $\mu$m by the red grism and 0.92--1.30 $\mu$m by the blue grism, but only in the EDFs \citep{euclid2023NISPperformance}. The spectroscopy enables the spectral classification of UCDs. \textit{Euclid} will be a powerful tool with which to hunt and characterise undiscovered UCDs \citep{solano2021virtualUCDs,martin2021ch4nh3}.

\begin{figure*}
    \centering
    \includegraphics[width=\textwidth]{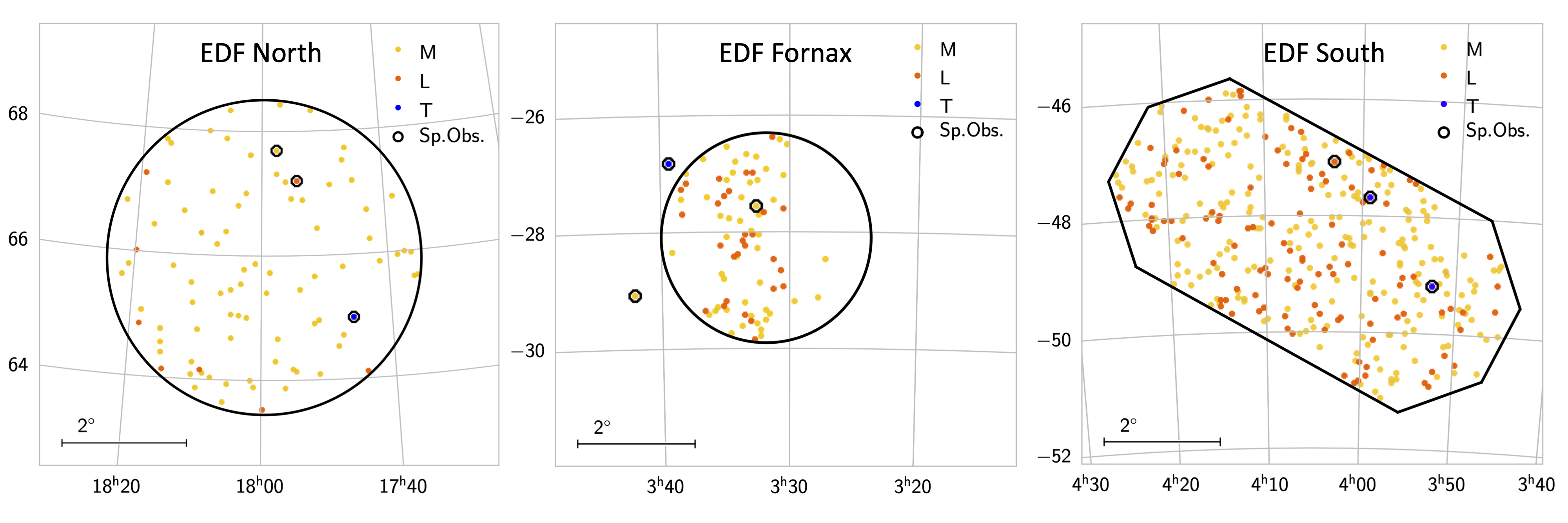}
    \caption{UCD candidates in the three EDFs (black borders). The circled objects are the ones for which we will get spectra.}
    \label{3EDFs}
\end{figure*}

Three EDFs,\footnote{\url{https://www.cosmos.esa.int/web/euclid/euclid-survey}} North, Fornax, and South, are the patches of the sky that \textit{Euclid} will repeatedly scan during its lifetime. They consist of a total of 53 deg$^2$ of sky. After the co-adding, the depth in these regions will be about two magnitudes deeper than the EWSs. For point sources, it will reach an unprecedented 5-$\sigma$ limit about 26.3, 26.5, 26.4 mag (AB system) in the $Y_E$, $J_E$, and $H_E$ bands, respectively \citep{euclid2022i.EWS,euclid2022Spitzer_EDF,euclid2023NISPperformance}. \textit{Euclid} will unveil UCDs far beyond the Galactic disc. 
In addition, monthly visits makes precise astrometry possible, which is key to measure parallaxes and proper motions of UCDs. The parallax is indispensable in determining the lumininosity of UCDs; the proper motion together with the radial velocity from a coarse spectroscopic measurement can classify the objects into different kinematic groups. 
The Large Synoptic Survey Telescope (LSST) of the Vera C. Rubin Observatory at Cerro Pachón, Chile will complement \textit{Euclid}'s coverage, especially the two southern EDFs at optical wavelengths in depth. \citep{rhodes2017lsst_euclid}.

$Euclid$ has $Y_E$, $J_E$, $H_E$-band filters and photometric systems that are different from the NIR systems of ground-based surveys such as the 2MASS, Mauna Kea Observatories (MKO), or the VHS \citep{euclid2022nisp_photometry}.  The \textit{Euclid} optical filter is much broader than the $i$ or $z$ filters of the PS1 and the DES as well. In addition, the ground-based surveys suffer a lot from telluric attenuation. A correction must be applied to the photometric measurements, if we are to use the ground-based data, before they can serve towards the calibration of $Euclid$. \citet{euclid2022nisp_photometry} presented linear transformations between the \textit{Euclid} NISP photometric system and the other ground-based photometric system for M, L, and T-type UCDs. It fitted the simulated photometry of M, L, and T dwarfs from \citet{euclid2019quazar}, who used observational templates to calculate corresponding colours. However, the transformations have great residuals. Therefore, having a list of UCDs characterised in the three EDFs prior to the start of the \textit{Euclid} surveys is important. They serve as photometric and spectroscopic standard anchors before the very first nominal survey observations of \textit{Euclid} and they will help \textit{Euclid} to calibrate itself every time it goes back and points at the EDFs.

In this work, we selected UCD candidates as photometric and spectroscopic standards in the three EDFs: North, Fornax, and South. These photometric standards and reconnaissance spectral templates are ready for the first data release of $Euclid$.   Section~\ref{selection} explains the UCD selection.
We spectroscopically characterised their representatives using the 10.4-m Gran Telescopio Canarias (GTC) on the Spanish island of La Palma (for EDFs North and Fornax) and the European Southern Observatory (ESO) 8.2-m Very Large Telescope (VLT) on Cerro Paranal, Chile (for EDF South). Section~\ref{observation} describes the observation details. Section~\ref{methods} explains the spectral classification method. Section~\ref{results} shows the results of spectral analysis for each object, and discusses the capability of \textit{Euclid} in terms of UCD characterisation.

\section{UCD samples in the EDFs}
\label{selection}
We used the Tool for OPerations on Catalogues And Tables \citep[TOPCAT;][]{taylor2005topcat} to cross-match existing catalogues and select UCD candidates in specific areas. We used different criteria and techniques to make the selection in the three different regions.

\subsection{Euclid Deep Field North}


\begin{table*}
      \caption[]{Targets selected for spectroscopic observation. }
         \label{targets}
         \centering
         \begin{tabular}{ccccccc}
            \hline\hline
            \noalign{\smallskip}
            Name   & Location & $Y$& $J$&  $H$  & SpT. & Note \\
            
            \hline
            
        WISEA J175730.71+674138.4 & EDF North  & 15.2 & 13.7 & 13.2  & $^{(1)}$M7.0
            \\
        WISEA J175410.34+671212.0 & EDF North & 18.0 &  16.1 & 15.5  & L & Potential binary
            \\
        $^{(2)}$WISE J174556.65+645933.8 & EDF North & - &      - &  - & T7.0 & $J$
 band spectrum by (2)
            \\


        WISEA J033234.35$-$273333.8  & EDF Fornax & 18.9 & 17.2 & 16.7  & M9.0
            \\
        $^{(3)}$WISEA J033921.70$-$264906.5 & EDF Fornax$^*$ & 20.9  & 18.4 & 18.7 & T6.0
            \\
            
        WISEA J040254.85-470440.9  & EDF South & 20.9 & 18.9 & - & L5.0
            \\
        WISEA J035231.80-491059.4  & EDF South & 20.3 & 17.8 & 18.2 & T7.0
            \\
        WISEA J035909.75-474056.8  & EDF South & 20.6 & 18.1 & 18.3 & T8.0
            \\
            
        $^{(3)}$WISEA J034209.37$-$290431.9 & EDF Fornax$^*$ & 18.1 & 15.9 & 15.1 & L4.0, $^{(6)}$L0$\beta$   &  GTC \& VLT common target
            \\
            \hline
         \end{tabular}

                   \vspace{1ex}

    {\raggedright \textbf{Note
:} In the rest of the paper, we subsequently use WISE\textit{hhmm} or W\textit{hhmm}, or 2MASS\textit{hhmm},
as abbreviations. Tables in the appendices provide precise photometry with errors and references.\par}

{\raggedright \textbf{*:} WISE0339 and WISE0342 are outside but close to
the EDF Fornax border, illustrated in the middle panel of Figure~\ref{3EDFs}. \par}
    \tablebib{(1). \citet{reyle2018GaiaDR2UCD}; (2). \citet{mace2013WISE_dT}; (3). \citet{rosell2019DES_BD}; (4). \citet{burgasser2015dML_RV}, photometry from the 2MASS; (5). \citet{hsu2021dT_RV}, photometry from the 2MASS; (6). \citet{gagne2015BANYAN_substellar_movinggroup}.} 
   \end{table*}

EDF North originally was a 10 deg$^2$ circled area centered at the coordinate 17h58m55.9s $+$66$^\circ$01'03.7" with a radius of 1.78 deg.  In April 2022, a request to extend EDF North's area to 20 deg$^2$ (radius extended to 2.52 deg) 
was endorsed by ESA based on the need for a larger area of spectroscopic calibration. 

We collected information on the $i$ and $z$ bands from the Panoramic Survey Telescope and Rapid Response System release 1 \citep[Pan-STARRS, PS1;][]{chambers2016ps1}, $J$, $H$, and $K$ bands from the Two Micron All Sky Survey \citep[2MASS; ][]{cutri2003_2MASSpoint, skrutskie2006_2MASS}, and $W1,2,3,4$ bands from the AllWISE programme \citep{cutri2014allwise} of WISE. We cross-matched all the point sources in these three catalogues with a radius of 3 arcsec inside EDF North.

To select the UCDs, we applied photometric criteria from \citet{rosell2019DES_BD}, but we generalised and slightly loosened the criteria according to the colours presented in \citet{schmidt2015MLcolor, skrzypeck2016phometric_classification_LTcolor}.
In practice, for late-M UCDs, we required 
\begin{itemize}
    \item $i-z > 0.65 $ mag;
    \item $2.68 < i-J < 4.0 $ mag;
    \item $1.59 < z-J < 2.55$ mag;
    \item $z-W1 < 4.5$ mag;
    \item $J-H > 0.47$ mag;
    \item $J-W2 > 1.36$ mag;
    \item $H-K > 0.17$ mag;
    \item $ 0.13 < W1-W2 < 0.25 $ mag; and
    \item $SNR_{W4} \leq 3$.
\end{itemize}
The colour criterion, $W1-W2 > 0.13$ mag, already separates most of the red giants from the dwarfs \citep{li2016redgiantsIRphot}. To further ensure that we would not include any possible red giants, which could possess big and cooler envelopes, a non-detection criterion in the $W4$ was set by limiting the signal-to-noise ratio (SNR) of this band. We found 90 late-M UCDs. We note that in our search we did not classify the objects into subclasses.

\citet{reyle2018GaiaDR2UCD} selected UCDs from \textit{Gaia} DR2 based on the locus of spectroscopically confirmed UCDs in the Hertzsprung–Russel (HR) diagram. There are in total 12 UCDs, all late-M type with tentative spectral classification (eight M7 dwarfs, three M7.5 dwarfs, and one M7 subdwarf) in EDF North. Our search recovered ten of these 12 M dwarfs in the catalogue of \citet{reyle2018GaiaDR2UCD}. The other two have a slightly bluer $J-W2$ colour. We decided to include these two objects. Therefore, we have 81 late-M UCDs in EDF North (Table~\ref{EDFNcatalogM}).

For L-type UCDs, we required redder $i-z$, $i-J$, $z-J$, and $z-W1$ colours \citep{rosell2019DES_BD}:
\begin{itemize}
    \item $i-z > 1.35 $ mag;
    \item $i-J \geq 4.0 $ mag;
    \item $z-J \geq 2.55$ mag;
    \item $z-W1 \geq 4.5$ mag;
    \item $J-H > 0.47$ mag;
    \item $J-W2 > 1.36$ mag;
    \item $H-K > 0.17$ mag;
    \item $W1-W2 \geq 0.25 $ mag;
    \item $SNR_{W4} \leq 3$.
\end{itemize}
We found eight L-type UCDs (Table~\ref{EDFNcatalogL}).

For T dwarfs, because methane absorption starts to play a role in the NIR, the $J-H$ and $H-K$ colours turn bluer when the type gets later \citep{leggett2003NIRcolorLT}. Hence, we separated the criteria of the early-type and late-type T dwarfs. For early-T-type UCDs, we required 
\begin{itemize}
    \item $z-J > 1.59$ mag;
    \item $1 < W1-W2 < 2$ mag;
    \item $SNR_{W4} \leq 3$,
\end{itemize}
and for late-T-type UCDs, we required 
\begin{itemize}
    \item $J-H < - 0.3$ mag;
    \item $H-K < 0.2$ mag;
    \item $W1-W2 > 1.5$ mag; and
    \item $SNR_{W4} \leq 3$.
\end{itemize}

Unfortunately our search found no T dwarf in EDF North. However, we noticed that there is a T7 dwarf WISE\,J174556.65$+$645933.8 in EDF North revealed by \citet{mace2013WISE_dT} in WISE. This object is too faint to be detected by the PS1 or the 2MASS\@. The $J$-band spectrum was already obtained by \citet{mace2013WISE_dT} using the Near-Infrared Spectrometer \citep[NIRSPEC;][]{mclean1998nirspec,mclean2000nirspec} of the 10-m Keck telescope on Mauna Kea, and was kindly provided to us by Dr.\ Gregory Mace.

We selected the M dwarf WISE1757 and the L dwarf WISE1754 for spectroscopic characterisation (Table~\ref{targets}). We note that WISE1754 could be an L-dwarf binary system consisting of \textit{Gaia} DR3 1633622185772907392 (hereafter WISE1754 A) and \textit{Gaia} DR3 1633622181475606528 (hereafter WISE1754 B), which were not resolved in \textit{WISE} and the 2MASS. WISE1754 B is bluer than WISE1754 A.  The left panel in Figure~\ref{3EDFs} shows the UCD candidates in EDF North.

\begin{table*}
      \centering
      \caption{Summary of GTC/EMIR and VLT/X-shooter observations.}
         \label{Obs}
         \begin{tabular}{ccccccc}
            \hline\hline
            \noalign{\smallskip}
            Name  &  MJD & Seeing &  Configuration & Grism/Arm/Filter & Exposures    & Std. \ star 
            \\
            \hline
    WISE0332 & 59890.09 & 0.7" & EMIR Long Slit& $YJ$ & 360s$\times$8 & $\tau$ For 
           \\       
    WISE0332 & 59890.13 & 0.6" & EMIR Long Slit& $HK$ & 240s$\times$12 & $\tau$ For 
           \\          
    WISE0339 & 59931.95 & 0.8" & EMIR Long Slit& $YJ$ & 360s$\times$8 &$\tau$ For 
           \\
    WISE0339 & 59932.93 & 0.7" & EMIR Long Slit& $YJ$ & 360s$\times$8 &$\tau$ For 
           \\
    WISE0339 & 59931.99 & 0.7" & EMIR Long Slit& $HK$ & 240s$\times$12 &$\tau$ For 
           \\
    WISE0339 & 59932.96 & 0.8" & EMIR Long Slit& $HK$ & 240s$\times$12 &$\tau$ For 
           \\     
    WISE0342 & 59862.11 & 0.8"  & EMIR Long Slit& $YJ$ & 200s$\times$8 & HD 20423 
           \\
    WISE0342 & 59905.12 & 1.0" & EMIR Long Slit& $HK$ & 240s$\times$8 & HD 20423 
           \\
   WISE0342 & 60247.70 & 2.3" & X-shooter& UVB, VIS, NIR & (276s, 292s, 300s)$\times$8 & HD 16226 
           \\
    WISE0352 & 60246.71 & 1.4" & X-shooter& UVB, VIS, NIR & (276s, 292s, 300s)$\times$12 & HD 207158 
           \\
    WISE0359 & 60246.76 & 1.6" & X-shooter& UVB, VIS, NIR & (276s, 292s, 300s)$\times$12 & HD 207158 
           \\
    WISE0402 & 60247.78  & 3.3" & X-shooter& UVB, VIS, NIR & (276s, 292s, 300s)$\times$12  & HD 16226 
           \\
    WISE1754\,AB & 59831.91 & 0.7" & EMIR Imaging & $J$ & 5s$\times$14 & -- 
           \\
    WISE1754\,AB & 59834.98 & 0.9" & EMIR Long Slit& $YJ$ & 320s$\times$8 &55 Dra 
           \\
    WISE1754\,AB & 59835.03 & 0.8" & EMIR Long Slit& $HK$ & 280s$\times$8 &55 Dra 
           \\
    WISE1757  & 59831.93 & 0.7" & EMIR Long Slit& $YJ$ & 120s$\times$8 &55 Dra 
           \\
    WISE1757 &  59831.95 & 0.7" & EMIR Long Slit& $HK$ & 60s$\times$16 &55 Dra 
           \\
            \hline
         \end{tabular}
          \vspace{1ex}

     {\raggedright 
         \textbf{Note:} Candidates are ordered by right ascension. MJD indicates the Modified Julian Date at the middle of each observation. \par}
   \end{table*}


\subsection{Euclid Deep Field Fornax}

\textit{Euclid} Deep Field Fornax is a 10 deg$^2$ circled area centered on the coordinate 03h31m43.6s $-$28$^\circ$05'18.6" with a radius of 1.78 deg. \citet{rosell2019DES_BD} did a UCD census in an area about 2400 deg$^2$, using the $i$, $z$, and $Y$ bands of the Dark Energy Survey \citep[DES;][]{des_collaboration2016} Year 3 release, matched to the $J$, $H$, and $K_s$ bands of the Vista Hemisphere Survey \citep[VHS;][]{mcmahon2013vhs} DR3 and the $W1, W2$ bands of WISE\@. We directly selected the UCD samples from this catalogue in the EDF Fornax field.

This catalogue returned 45 late-M dwarfs. We decided to include four extra late-M dwarfs from the \textit{Gaia} UCD catalogue \citep{reyle2018GaiaDR2UCD}. 
At the end, we have 49 late-M dwarfs (Table~\ref{EDFFcatalogM}). It also returned 29 L dwarfs (Table~\ref{EDFFcatalogL}) and no T dwarf in EDF Fornax. All the L dwarfs are photometrically classified as early-L dwarfs (all of them have a spectral class $\leq$ L2.0). 


Despite no T dwarf being found in EDF Fornax, there is a T6 dwarf WISEA J033921.70$-$264906.5 (hereafter WISE0339) outside but close to the field (0.34 deg or 20 arcmin away from the northeast border). It is worth including it, considering the large and rectangular FoV of $Euclid$.

We selected an M dwarf, WISE0332, and the T dwarf WISE0339 for spectroscopic characterisation (Table~\ref{targets}). The middle panel in Figure~\ref{3EDFs} shows the UCD candidates in and near EDF Fornax. The uneven UCD distribution is because the $H$-band VHS only covers about half of EDF Fornax.

\subsection{Euclid Deep Field South}

\textit{Euclid} Deep Field South is a 23 deg$^2$ octagon stadium-shaped area\footnote{The envelope of EDF South is 63.25$^\circ$ $-$45.67$^\circ$,
        65.35$^\circ$ $-$46.10$^\circ$,
        66.40$^\circ$ $-$47.25$^\circ$,
        65.99$^\circ$ $-$48.72$^\circ$,
        59.25$^\circ$ $-$51.19$^\circ$,
        56.95$^\circ$ $-$50.82$^\circ$,
        55.90$^\circ$ $-$49.40$^\circ$,
        56.80$^\circ$ $-$47.99$^\circ$.} centered on the coordinate 04h04m57.84s $-$48$^\circ$25'22.8". We directly used the catalogues of \citet{rosell2019DES_BD} and \citet{reyle2018GaiaDR2UCD} as well.

We retrieved 213 late-M dwarfs (Table~\ref{EDFScatalogM}). Only one late-M dwarf is in the 19 M dwarfs from the \textit{Gaia} UCD catalogue of \citet{reyle2018GaiaDR2UCD}. We end up with 231 late-M dwarfs in EDF South. We retrieved 115 L dwarfs (Table~\ref{EDFScatalogL}) and two T dwarfs (Table~\ref{EDFScatalogT}) in EDF South. Ten out of 115 L dwarfs are photometrically classified as mid-L dwarfs (L4.0-L6.0) and the rest are all early-L dwarfs. The two T dwarfs are late-T dwarfs: one T7.0 and the other T8.0. Figure~\ref{3EDFs} plots all the UCD candidates in EDF South.




\subsection{Object for coherence check}
It is essential to have a shared UCD target to evaluate the coherence between the two facilities. We thus chose a relatively bright L dwarf, WISEA\,J034209.37$-$290431.9 \citep{gagne2015BANYAN_substellar_movinggroup}, close to the EDF Fornax border. This object will still be covered by the EWSs and its position is shown in the middle panel of Figure~\ref{3EDFs}.



\section{Observations and data reduction}
\label{observation}
   
   We proposed a long-slit spectroscopic observation on a total of five UCD candidates in EDFs North and Fornax with their spectroscopic standards 
   using the Espectrógrafo Multiobjeto Infra-Rojo (EMIR) on the GTC, programme GTC97-22B (PI J.-Y.\ Zhang). The Espectrógrafo Multiobjeto Infra-Rojo is a NIR wide-field imager and a multi-object spectrograph with a HAWAII-2 detector. The pixel size is 0.195"/pix.  We used a 0.6 arcsec slit with $YJ$ and $HK$ grisms, covering 0.85 -- 1.35 $\mu$m and 1.45 -- 2.42 $\mu$m, respectively, yielding a low spectral resolution of $R\,\approx\,987$. 
   We used a standard ABBA observing block sequence. Meanwhile, we asked for a dedicated imaging block of WISE1754\,AB in the $J$ band with a 7-point dithering pattern before its spectroscopic observation to confirm the companionship. During the spectroscopic observations, we put the slit directly on both objects to obtain their spectra simultaneously.  
   We requested a seeing better than 0.9 arcsec, a clear sky, and no restriction on the moon phase for the GTC observations.

We also proposed a long-slit spectroscopic observation on a total of three UCD candidates in EDF South with spectroscopic standards
using the X-shooter \citep{vernet2011xshooter} on the VLT, ESO programme 112.261S (PI J.-Y. Zhang). The X-shooter consists of three spectroscopic arms: UVB, VIS, and NIR. Each of the arms is an independent cross-dispersed echelle spectrograph. The UVB arm has an E2V CCD44-82 detector covering 0.30 -- 0.56 $\mu$m with a pixel scale of 0.161"/pix; the VIS arm has a MIT/LL CCID 20 detector covering 0.56 -- 1.02 $\mu$m with a pixel scale of 0.158"/pix; and the NIR arm has a Hawaii 2RG detector covering 1.02 -- 2.48 $\mu$m with a pixel scale of 0.258"/pix. We asked for a standard ABBA observing block sequence and simultaneous observations in the three arms. For the UVB arm we used a 1.3 arcsec slit ($R \approx 4100$). We used a 1.2 arcsec slit for both the VIS arm ($R \approx 6500$) and the NIR arm ($R \approx 4300$). We requested a seeing better than 1.5 arcsec, a sky condition of thin cirrus or better, and no restriction on the moon phase for the VLT observations. A summary of all the observations is provided in Table~\ref{Obs}.


The 2D EMIR spectral frames were preliminarily reduced by the EMIR default pipeline PyEMIR.\footnote{\url{https://pyemir.readthedocs.io/en/stable/#}} The pipeline performed flat field correction, geometric distortion rectification, wavelength calibration (first using the empirical wavelength profile, and then using OH airglow lines in the images to refine the calibration), and ABBA stacking. The outputs are rectified and wavelength-calibrated 2D spectra. We also used the same pipeline to reduce the $J$-band image of WISE1754\,AB. The pipeline performed flat field correction, sky subtraction, and stacking under the knowledge of the dithering pattern position.  We then extracted 1D spectra in IRAF \citep{tody1986iraf}.  

The $YJ$ and $HK$ spectra of GTC/EMIR were not observed simultaneously. We convolved and normalised the spectra with the $J$ and $H$ filter profiles of full throughput of the 2MASS and the VHS.\footnote{For the 2MASS:\url{http://svo2.cab.inta-csic.es/theory/fps3/index.php?mode=browse&gname=2MASS&asttype=}; For the VHS: \url{http://svo2.cab.inta-csic.es/theory/fps3/index.php?mode=browse&gname=Paranal&gname2=VISTA&asttype=}} We calculated the $J-H$ colour according to the corresponding zero points of each filter. The difference between the observed $J-H$ colour and the colour from the spectra is the coefficient that we should apply to adjust the $YJ$ and $HK$ spectra.

The 2D X-shooter data were reduced by the X-shooter pipeline inside the EsoReflex environment \citep{freudling2013esoreflex}. The pipeline produces flux-calibrated 1D spectra for each arm of X-shooter. Then the 1D spectra were corrected for the telluric contribution by Molecfit \citep{smette2015molecfit,kausch2015molecfit_xshooter} in EsoReflex. Because of the low SNR in UVB and VIS arm spectra, we only used NIR arm spectra starting from 1.02 $\mu$m in the future analysis.


\section{Euclidisation and spectral classification}
\label{methods}

To `Euclidise' the spectra, we used a Gaussian profile with a full width half maximum (FHWM) of 40 \AA{} to convolve with our spectra ($R\approx380$ at 1.5 $\mu$m). We only consider the wavelength range of 0.92-1.86 $\mu$m for our spectra. The results mimic the spectra as if they were taken by the \textit{Euclid} NISP with both blue and red grisms. 

We collected the spectral templates from the L and T dwarf data archive\footnote{\url{http://svocats.cab.inta-csic.es/chiu06/index.php}} of \citet{chiu2006_71LT_sdss,golimowski2004LM_phot_LT,knapp2004nir_spectra_LT} and the NIRSPEC Brown Dwarf Spectroscopic Survey\footnote{\url{https://www.astro.ucla.edu/~mclean/BDSSarchive/}.} \citep[BDSS;][]{mclean2003BDSS} to spectroscopically classify L and T dwarfs among our targets. We collected late-M dwarf templates and also early-L dwarf ones from the NASA Infrared Telescope Facility (IRTF) Spectral Library\footnote{\url{http://irtfweb.ifa.hawaii.edu/~spex/IRTF_Spectral_Library/}} from \citet{cushing2005IR_MLT, rayner2009IRTF_coolstars}.   

We normalised both the templates and our spectra. We used linearly distributed sampling points with a step of 0.1 \AA{} in four wavelength ranges: 1.02--1.12 $\mu$m, 1.15--1.34 $\mu$m, 1.47--1.79 $\mu$m, and 1.97--2.30 $\mu$m to interpolate our spectra. Ultracool dwarf spectra in the EDFs will be covered by both red and blue grisms; hence, we interpolated the Euclidised spectra without the $K$ band, which is not covered by $Euclid$. In particular, to Euclidise the $J$-band spectrum of WISE1745, we used the second wavelength range only. We emphasise that only the red grism will be used in the EWSs\@. We then interpolated the Euclidised spectra in two ranges: 1.25--1.34 $\mu$m and 1.47--1.79 $\mu$m. We applied the least squares method and found the best fit for the original spectrum and also the Euclidised spectrum of each object. The error was determined by the difference between the spectral subclass of the optimal solution and that of the two nearest points in the template grid.





\section{Results}
\label{results}

\subsection{WISE1754 AB}
\begin{figure}[htbp]
    \centering
    \includegraphics[width=0.49\textwidth]{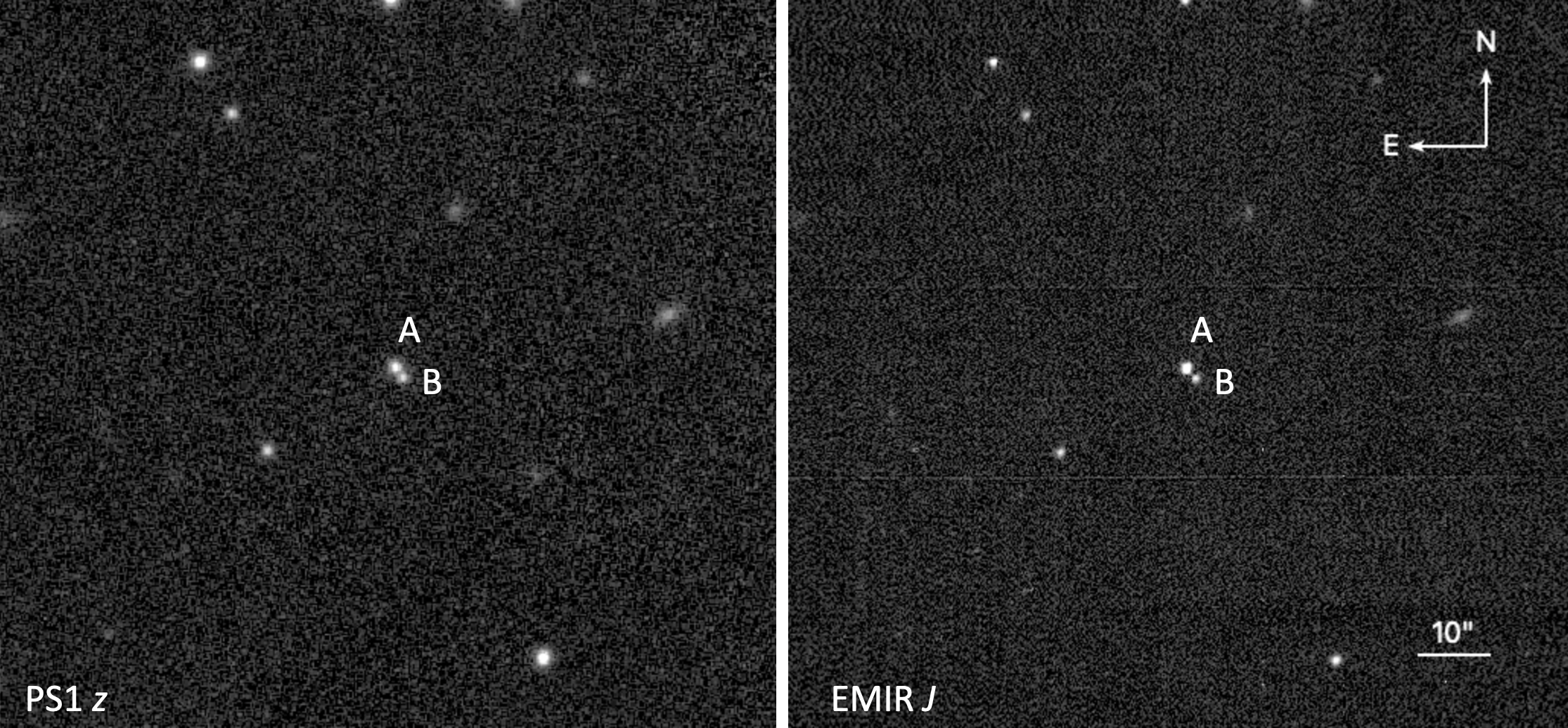}
    \caption{PS1 $z$-band image (left) and GTC/EMIR $J$-band image (right) of WISE1754\,AB\@. The A component has a proper motion towards south-east direction, while the B component does not. }
    \label{W1754_image}
\end{figure}

With the $J$-band EMIR image and the images from the PS1 (Figure~\ref{W1754_image}), we excluded the possibility that WISE1754\,AB is a co-moving binary system. WISE1754\,A has a proper motion in a southeasterly direction, while WISE1754\,B does not.  The spectroscopy also proves that WISE1754\,A is an L1 dwarf, while WISE1754\,B is not a UCD at all. 

\subsection{Common target WISE0342}

\begin{figure}[htbp]
    \centering
    \includegraphics[width=0.49\textwidth]{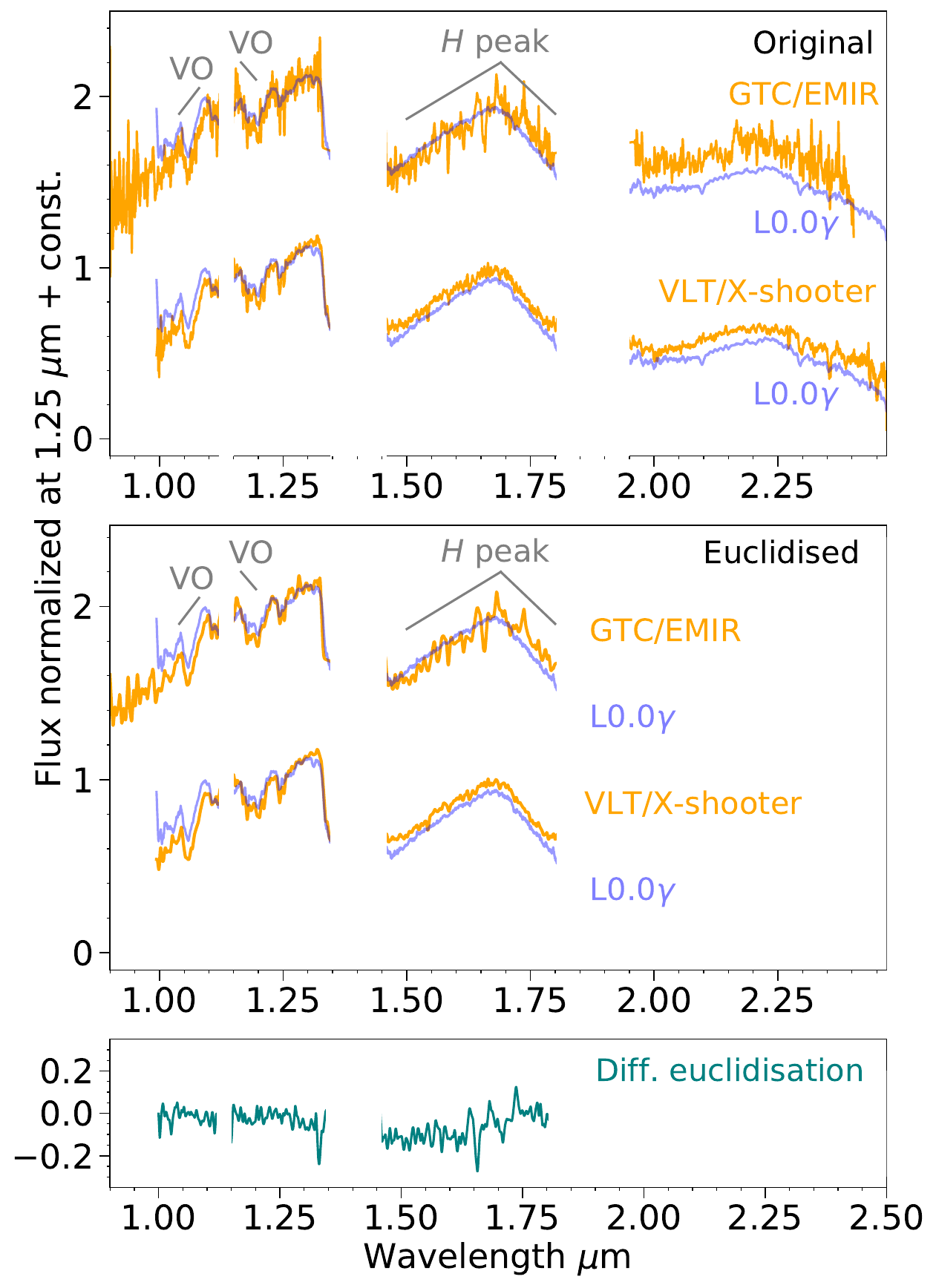}
    \caption{Normalised and Euclidised spectra of W0342 from both GTC/EMIR and the VLT/X-shooter. Strong telluric bands are masked. Features indicating the youth of the object are denoted. Both original spectra were classified as L0.0$\gamma$ (upper panel). The spectral similarity underscores that GTC/EMIR and the VLT/X-shooter are coherent in characterising UCDs for $Euclid$. The two Euclidised spectra were all classified as L0.0$\gamma$ (middle panel). The difference between the two Euclidised spectra is visualised in the lower panel.}
    \label{W0342}
\end{figure}

The GTC/EMIR spectrum and the VLT/X-shooter spectrum exhibit strong agreement with each other, as is indicated in the upper panel of Figure~\ref{W0342}. The Euclidised spectra also demonstrate a high degree of similarity, as is illustrated in the middle panel of Figure~\ref{W0342}. The differences between two Euclidised spectra are within 0.1 unit of normalised flux, with some spikes of noise around 0.2 (the lower panels of Figure~\ref{W0342}). The coherence between the two instruments is good enough to characterise UCDs for \textit{Euclid} in different EDFs.

In comparison with the field dwarf templates, both GTC/EMIR and VLT/X-shooter spectra have two similar local least squares minima at L4.5 and L7.0. The Euclidised GTC/EMIR spectrum has been classified as L4.5, and the Euclidised VLT/X-shooter spectrum has two similar local minima of the least squares at L5.5 and L7.0. The confusion of spectral type classification may be due to the youth and, hence, lower surface gravity of the object. 
For lower-gravity objects, in the $J$ band, the FeH feature at 0.99 $\mu$m, and the potassium doublets are weaker than those of higher-gravity objects, while the VO absorption features at 1.06 and 1.18 $\mu$m are much more prominent.  The $H$-band spectrum has a peaked, triangular shape for lower-gravity objects, due to stronger water absorption \citep{zapatero2000PMO_sOri,cushing2000_HK_rhoOph,lucas2001Ori_young,gorlova2003NIRgravity_BD,mcgovern2004lowgravity_BD,lodieu2008NIR_upSCo,scholz2012young_substellar_NGC1333,manjavacas2014atm_young_ML,muzic2015young_substellar_lupus3,lodieu2018dL_upperSco}.

We decided to compare the W0342 spectra against young L$\gamma$-type UCD templates in the Upper Scorpius association (5--10 Myr) \citep{lodieu2018dL_upperSco}.
Our classification suggests a L0.0$\gamma$ subclass, which agrees with the result of L0$\beta$ from \citet{gagne2015BANYAN_substellar_movinggroup}, but not with the photometric classification of L4.0 from \citet{rosell2019DES_BD}. We demonstrate that \textit{Euclid} slitless spectroscopy will be able to detect spectral indicators sensitive to the surface gravity or age, at least for late-M to early-L-type UCDs.


\subsection{Spectral classes}
 
\begin{figure*}
\centering
    \includegraphics[width=0.97\textwidth]{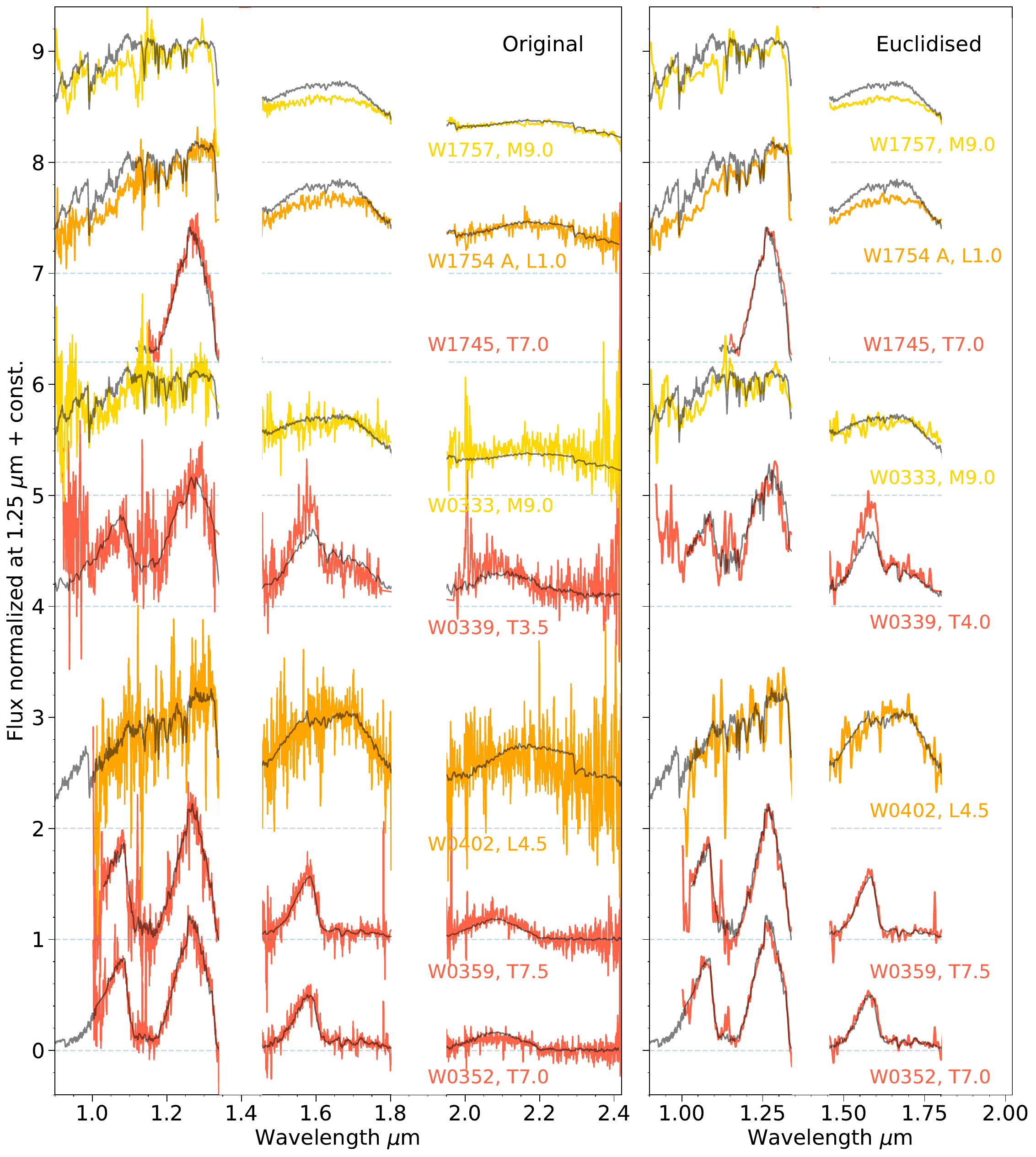}

   \caption{Normalised UCD spectra in the three EDFs. From the top to the bottom are UCDs from EDFs North, Fornax, and South, respectively. The VLT/X-shooter and GTC/EMIR spectra were smoothed with a 31 and 5-pixel-size window, respectively. All the spectra were classified into spectral subclasses, and the best-fit templates are plotted in grey (left panel). We Euclidised the spectra and the classifications remain consistent within 0.5 subclasses (right panel).}
    \label{EDFspectra}
\end{figure*}

Figure~\ref{EDFspectra} depicts the original spectra and Euclidised spectra of the eight reconnaissance UCDs, except the common target WISE0342 (Figure~\ref{W0342}). The best-fit spectral templates are also overplotted. Saturated telluric wavelength regions were masked manually. The results and comparisons of spectral classification are summarised in Table~\ref{result_table}. The accuracy of the spectral type depends on the density of the template grid.

The spectral classifications of the reconnaissance UCDs are in line with their photometric spectral classifications from our search and from the selection of \citet{rosell2019DES_BD} and \citet{reyle2018GaiaDR2UCD}, except for the young object WISE0342\@. The Euclidised spectra maintains the spectral classification results. We conclude that \textit{Euclid} NISP slitless spectroscopy using two grisms in the EDFs can bring out an excellent spectral classification for UCDs, with a precision of 0.5 subclass without any major problems. With only the red grism in the EWSs, the spectral classification for mid-L to T dwarfs will be as precise as that in the EDFs; while for those field late-M to early-L dwarfs the precision is of two subclasses because those hotter UCDs have less characteristic features in the $H$ band. For young early-L dwarfs like W0342, their peculiar $H$ band improves the spectral classification precision of the red grism.

\begingroup
\setlength{\tabcolsep}{4pt} 

      \begin{table}[htbp]
      \centering
      \caption{Summary of the spectral classification.}
         \label{result_table}
         \begin{tabular}{ccccc}
            \hline\hline
            \noalign{\smallskip}
            Name  &  Phot.  & Spec.&  Euc. & Euc. red
            \\
            \hline \\
    WISE0332 & M9.0 & M9.0$\pm$1.0  & M9.0$\pm$1.0 & M7.0$\pm$1.0 
           \\            
    WISE0339 & T6.0 & T3.5$\pm$0.5  & T4.0$\pm$0.5 & T3.0$\pm$0.5 
           \\
    WISE0342 & L4.0 & L0.0$\pm$1.0$\gamma$  & L0.0$\pm$1.0$\gamma$ & L0.0$\pm$1.0$\gamma$
            \\
    WISE0352 & T7.0 & T7.0$\pm$0.5  & T7.0$\pm$0.5 & T7.0$\pm$0.5
           \\
    WISE0359 & T8.0 & T7.5$\pm$0.5  & T7.5$\pm$0.5  & T7.5$\pm$0.5
           \\
    WISE0402 & L5.0  & L4.5$\pm$0.5  & L4.5$\pm$0.5 & L4.5$\pm$0.5
           \\
    WISE1745& - & T7.0$\pm$0.5  & T7.0$\pm$0.5  & - 
           \\
    WISE1754 A& L & L1.0$\pm$0.5  & L1.0$\pm$0.5 & M9.0$\pm$1.0   
           \\
    WISE1757 & M7.0 & M9.0$\pm$1.0  & M9.0$\pm$1.0 & M7.0$\pm$1.0 
           \\
            \hline
         \end{tabular}
          \vspace{1ex}

     {\raggedright 
         \textbf{Note:} Candidates are ordered by right ascension. Euc. means that the results are derived from the whole Euclidised spectra (for UCDs in the EDFs); and Euc red means only the part covered by the red grism was used (for UCDs in the EWSs).\par}
   \end{table}
\endgroup

\subsection{Sample purity}
We selected eight UCD candidates from the whole sample pool for the spectroscopy and all of them were spectroscopically confirmed as UCDs. Assuming an overall purity of our selections in the three EDFs is $p$, the probability $P(X=k,n)$ with $k$ true UCD in $n$ candidates follows a binomial distribution,
\begin{equation}
P(X=k, n)=\frac{n!}{k!(n-k)!}p^k(1-p)^{n-k}.
\end{equation}
Hence, we estimated a 90\% confidence interval (with only a lower tail) of the purity by $P(X=8, 8)=p^8\geq1-90\%$, yielding $75\%\leq p\leq100\%$.  The lower limit is extremely conservative since it is completely based on our spectroscopic confirmation.


\section{Conclusions}
   
$Euclid$, a space-borne telescope dedicated to cosmology, will also be powerful for UCD science.  It has advantages compared with other current space telescopes, especially the combination of its big FoV and its depth in the NIR range. It will start nominal surveys soon and will visit the three EDFs repeatedly.  In this work, we provide a list of UCDs in the three EDFs for $Euclid$ to characterise them and use them as references to identify new UCDs. 
We found 81 (49, 231) M, eight (29, 115) L, and one (0, 2) T dwarfs in EDFs North, Fornax, and South, respectively.  The total number of these potential UCD standards is enough to provide a fairly good connection between $Euclid$ and ground-based data, despite the possibility of photometric or spectroscopic variability of any single object. 

From the candidate list in or next to all the EDFs, we selected in total eight UCD candidates of different spectral classes, for which we collected spectra with GTC/EMIR and the VLT/X-shooter, confirming their UCD nature. We used a common object, W0342, to demonstrate that these two telescopic instrument combinations provide a coherent UCD characterisation for \textit{Euclid}.  All of the eight selected candidates were spectroscopically confirmed as UCDs, suggesting a high UCD purity of the selection that we made. Together with a $J$-band spectrum of a T dwarf in EDF North from the literature, these nine UCD spectra will serve as reconnaissance spectra for UCD discoveries of $Euclid$. We estimated the spectral classification ability of \textit{Euclid} by analysing Euclidised spectra and comparing them to the original spectra. We concluded that \textit{Euclid} NISP spectroscopy will be able to deliver a precise spectral class for UCDs in the EDFs, and to determine their age groups.

At the end of the \textit{Euclid} mission, with about 40--50 visits in each EDF, the co-added spectra will have an SNR at least six times better than the spectra from a single epoch. This may even permit a radial velocity measurement for many UCDs. Their parallaxes and proper motions will be also measured through multi-epoch astrometry. Multi-epoch and multi-band photometry will enable one to monitor the monthly photometric variability of UCDs, which could be caused by magnetic activity, inhomogeneous cloud distribution, rotation, or complex atmosphere dynamics. Combining the determination of spectral classes, distances, kinematics, long-term photometric variabilities, and the ages of thousands of new UCD discoveries, the \textit{Euclid} deep surveys will bring us to a new era in UCD science, with unprecedented statistics for the understanding of the lowest-mass population in the Milky Way.

\begin{acknowledgements}
 We thank the referee for the useful and insightful report.
NL acknowledges support from the Agencia Estatal de Investigaci\'on del Ministerio de Ciencia e Innovaci\'on (AEI-MCINN) under grant PID2019-109522GB-C53\@. 
Funding for this research was provided by the European Union (ERC, SUBSTELLAR, project number 101054354).   
Based on observations made with the Gran Telescopio Canarias (GTC), installed at the Spanish Observatorio del Roque de los Muchachos of the Instituto de Astrofísica de Canarias, on the island of La Palma. 
Based on observations collected at the European Southern Observatory under ESO programme 112.261S. 
This research has made use of the Spanish Virtual Observatory (https://svo.cab.inta-csic.es) project funded by MCIN/AEI/10.13039/501100011033/ through grant PID2020-112949GB-I00. 
This work has used the Pan-STARRS1 Surveys (PS1) and the PS1 public science archive that have been made possible through contributions by the Institute for Astronomy, the University of Hawaii, the Pan-STARRS Project Office, the Max-Planck Society and its participating institutes, the Max Planck Institute for Astronomy, Heidelberg and the Max Planck Institute for Extraterrestrial Physics, Garching, The Johns Hopkins University, Durham University, the University of Edinburgh, the Queen's University Belfast, the Harvard-Smithsonian Center for Astrophysics, the Las Cumbres Observatory Global Telescope Network Incorporated, the National Central University of Taiwan, the Space Telescope Science Institute, the National Aeronautics and Space Administration under Grant No. NNX08AR22G issued through the Planetary Science Division of the NASA Science Mission Directorate, the National Science Foundation Grant No. AST-1238877, the University of Maryland, Eotvos Lorand University (ELTE), the Los Alamos National Laboratory, and the Gordon and Betty Moore Foundation. This publication makes use of data products from the Wide-field Infrared Survey Explorer, which is a joint project of the University of California, Los Angeles, and the Jet Propulsion Laboratory/California Institute of Technology, funded by the National Aeronautics and Space Administration. This project used public archival data from the Dark Energy Survey (DES). Funding for the DES Projects has been provided by the U.S. Department of Energy, the U.S. National Science Foundation, the Ministry of Science and Education of Spain, the Science and Technology Facilities Council of the United Kingdom, the Higher Education Funding Council for England, the National Center for Supercomputing Applications at the University of Illinois at Urbana–Champaign, the Kavli Institute of Cosmological Physics at the University of Chicago, the Center for Cosmology and Astro-Particle Physics at the Ohio State University, the Mitchell Institute for Fundamental Physics and Astronomy at Texas A\&M University, Financiadora de Estudos e Projetos, Fundação Carlos Chagas Filho de Amparo à Pesquisa do Estado do Rio de Janeiro, Conselho Nacional de Desenvolvimento Científico e Tecnológico and the Ministério da Ciência, Tecnologia e Inovação, the Deutsche Forschungsgemeinschaft and the Collaborating Institutions in the Dark Energy Survey.The Collaborating Institutions are Argonne National Laboratory, the University of California at Santa Cruz, the University of Cambridge, Centro de Investigaciones Enérgeticas, Medioambientales y Tecnológicas–Madrid, the University of Chicago, University College London, the DES-Brazil Consortium, the University of Edinburgh, the Eidgenössische Technische Hochschule (ETH) Zürich, Fermi National Accelerator Laboratory, the University of Illinois at Urbana-Champaign, the Institut de Ciències de l’Espai (IEEC/CSIC), the Institut de Física d’Altes Energies, Lawrence Berkeley National Laboratory, the Ludwig-Maximilians Universität München and the associated Excellence Cluster Universe, the University of Michigan, the National Optical Astronomy Observatory, the University of Nottingham, The Ohio State University, the OzDES Membership Consortium, the University of Pennsylvania, the University of Portsmouth, SLAC National Accelerator Laboratory, Stanford University, the University of Sussex, and Texas A\&M University. 
This work made use of Astropy:\footnote{http://www.astropy.org} a community-developed core Python package and an ecosystem of tools and resources for astronomy \citep{astropy2013, astropy2018, astropy2022}.
\end{acknowledgements}

%
%

%
\bibliographystyle{aa} 
\bibliography{bibliography.bib} 

\begin{thebibliography}{92}
\expandafter\ifx\csname natexlab\endcsname\relax\def\natexlab#1{#1}\fi

\bibitem[{{Aganze} {et~al.}(2022{\natexlab{a}}){Aganze}, {Burgasser}, {Malkan},
  {Theissen}, {Tejada Arevalo}, {Hsu}, {Bardalez Gagliuffi}, {Ryan}, \&
  {Holwerda}}]{aganze2022HST_parallel_UCDdensity}
{Aganze}, C., {Burgasser}, A.~J., {Malkan}, M., {et~al.} 2022{\natexlab{a}},
  \apj, 924, 114

\bibitem[{{Aganze} {et~al.}(2022{\natexlab{b}}){Aganze}, {Burgasser}, {Malkan},
  {Theissen}, {Tejada Arevalo}, {Hsu}, {Bardalez Gagliuffi}, {Ryan}, \&
  {Holwerda}}]{aganze2022HST_parallel_UCDage_scaleheights}
{Aganze}, C., {Burgasser}, A.~J., {Malkan}, M., {et~al.} 2022{\natexlab{b}},
  \apj, 934, 73

\bibitem[{{Astropy Collaboration} {et~al.}(2022){Astropy Collaboration},
  {Price-Whelan}, {Lim}, {Earl}, {Starkman}, {Bradley}, {Shupe}, {Patil},
  {Corrales}, {Brasseur}, {N{\"o}the}, {Donath}, {Tollerud}, {Morris},
  {Ginsburg}, {Vaher}, {Weaver}, {Tocknell}, {Jamieson}, {van Kerkwijk},
  {Robitaille}, {Merry}, {Bachetti}, {G{\"u}nther}, {Aldcroft},
  {Alvarado-Montes}, {Archibald}, {B{\'o}di}, {Bapat}, {Barentsen},
  {Baz{\'a}n}, {Biswas}, {Boquien}, {Burke}, {Cara}, {Cara}, {Conroy},
  {Conseil}, {Craig}, {Cross}, {Cruz}, {D'Eugenio}, {Dencheva}, {Devillepoix},
  {Dietrich}, {Eigenbrot}, {Erben}, {Ferreira}, {Foreman-Mackey}, {Fox},
  {Freij}, {Garg}, {Geda}, {Glattly}, {Gondhalekar}, {Gordon}, {Grant},
  {Greenfield}, {Groener}, {Guest}, {Gurovich}, {Handberg}, {Hart},
  {Hatfield-Dodds}, {Homeier}, {Hosseinzadeh}, {Jenness}, {Jones}, {Joseph},
  {Kalmbach}, {Karamehmetoglu}, {Ka{\l}uszy{\'n}ski}, {Kelley}, {Kern},
  {Kerzendorf}, {Koch}, {Kulumani}, {Lee}, {Ly}, {Ma}, {MacBride}, {Maljaars},
  {Muna}, {Murphy}, {Norman}, {O'Steen}, {Oman}, {Pacifici}, {Pascual},
  {Pascual-Granado}, {Patil}, {Perren}, {Pickering}, {Rastogi}, {Roulston},
  {Ryan}, {Rykoff}, {Sabater}, {Sakurikar}, {Salgado}, {Sanghi}, {Saunders},
  {Savchenko}, {Schwardt}, {Seifert-Eckert}, {Shih}, {Jain}, {Shukla}, {Sick},
  {Simpson}, {Singanamalla}, {Singer}, {Singhal}, {Sinha}, {Sip{\H{o}}cz},
  {Spitler}, {Stansby}, {Streicher}, {{\v{S}}umak}, {Swinbank}, {Taranu},
  {Tewary}, {Tremblay}, {de Val-Borro}, {Van Kooten}, {Vasovi{\'c}}, {Verma},
  {de Miranda Cardoso}, {Williams}, {Wilson}, {Winkel}, {Wood-Vasey}, {Xue},
  {Yoachim}, {Zhang}, {Zonca}, \& {Astropy Project Contributors}}]{astropy2022}
{Astropy Collaboration}, {Price-Whelan}, A.~M., {Lim}, P.~L., {et~al.} 2022,
  \apj, 935, 167

\bibitem[{{Astropy Collaboration} {et~al.}(2018){Astropy Collaboration},
  {Price-Whelan}, {Sip{\H{o}}cz}, {G{\"u}nther}, {Lim}, {Crawford}, {Conseil},
  {Shupe}, {Craig}, {Dencheva}, {Ginsburg}, {VanderPlas}, {Bradley},
  {P{\'e}rez-Su{\'a}rez}, {de Val-Borro}, {Aldcroft}, {Cruz}, {Robitaille},
  {Tollerud}, {Ardelean}, {Babej}, {Bach}, {Bachetti}, {Bakanov}, {Bamford},
  {Barentsen}, {Barmby}, {Baumbach}, {Berry}, {Biscani}, {Boquien}, {Bostroem},
  {Bouma}, {Brammer}, {Bray}, {Breytenbach}, {Buddelmeijer}, {Burke},
  {Calderone}, {Cano Rodr{\'\i}guez}, {Cara}, {Cardoso}, {Cheedella}, {Copin},
  {Corrales}, {Crichton}, {D'Avella}, {Deil}, {Depagne}, {Dietrich}, {Donath},
  {Droettboom}, {Earl}, {Erben}, {Fabbro}, {Ferreira}, {Finethy}, {Fox},
  {Garrison}, {Gibbons}, {Goldstein}, {Gommers}, {Greco}, {Greenfield},
  {Groener}, {Grollier}, {Hagen}, {Hirst}, {Homeier}, {Horton}, {Hosseinzadeh},
  {Hu}, {Hunkeler}, {Ivezi{\'c}}, {Jain}, {Jenness}, {Kanarek}, {Kendrew},
  {Kern}, {Kerzendorf}, {Khvalko}, {King}, {Kirkby}, {Kulkarni}, {Kumar},
  {Lee}, {Lenz}, {Littlefair}, {Ma}, {Macleod}, {Mastropietro}, {McCully},
  {Montagnac}, {Morris}, {Mueller}, {Mumford}, {Muna}, {Murphy}, {Nelson},
  {Nguyen}, {Ninan}, {N{\"o}the}, {Ogaz}, {Oh}, {Parejko}, {Parley}, {Pascual},
  {Patil}, {Patil}, {Plunkett}, {Prochaska}, {Rastogi}, {Reddy Janga},
  {Sabater}, {Sakurikar}, {Seifert}, {Sherbert}, {Sherwood-Taylor}, {Shih},
  {Sick}, {Silbiger}, {Singanamalla}, {Singer}, {Sladen}, {Sooley},
  {Sornarajah}, {Streicher}, {Teuben}, {Thomas}, {Tremblay}, {Turner},
  {Terr{\'o}n}, {van Kerkwijk}, {de la Vega}, {Watkins}, {Weaver}, {Whitmore},
  {Woillez}, {Zabalza}, \& {Astropy Contributors}}]{astropy2018}
{Astropy Collaboration}, {Price-Whelan}, A.~M., {Sip{\H{o}}cz}, B.~M., {et~al.}
  2018, \aj, 156, 123

\bibitem[{{Astropy Collaboration} {et~al.}(2013){Astropy Collaboration},
  {Robitaille}, {Tollerud}, {Greenfield}, {Droettboom}, {Bray}, {Aldcroft},
  {Davis}, {Ginsburg}, {Price-Whelan}, {Kerzendorf}, {Conley}, {Crighton},
  {Barbary}, {Muna}, {Ferguson}, {Grollier}, {Parikh}, {Nair}, {Unther},
  {Deil}, {Woillez}, {Conseil}, {Kramer}, {Turner}, {Singer}, {Fox}, {Weaver},
  {Zabalza}, {Edwards}, {Azalee Bostroem}, {Burke}, {Casey}, {Crawford},
  {Dencheva}, {Ely}, {Jenness}, {Labrie}, {Lim}, {Pierfederici}, {Pontzen},
  {Ptak}, {Refsdal}, {Servillat}, \& {Streicher}}]{astropy2013}
{Astropy Collaboration}, {Robitaille}, T.~P., {Tollerud}, E.~J., {et~al.} 2013,
  \aap, 558, A33

\bibitem[{{Beiler} {et~al.}(2023){Beiler}, {Cushing}, {Kirkpatrick},
  {Schneider}, {Mukherjee}, \& {Marley}}]{beiler2023dY_SED}
{Beiler}, S.~A., {Cushing}, M.~C., {Kirkpatrick}, J.~D., {et~al.} 2023, \apjl,
  951, L48

\bibitem[{{Burgasser} {et~al.}(2024){Burgasser}, {Bezanson}, {Labbe},
  {Brammer}, {Cutler}, {Furtak}, {Greene}, {Gerasimov}, {Leja}, {Pan}, {Price},
  {Wang}, {Weaver}, {Whitaker}, {Fujimoto}, {Kokorev}, {Dayal}, {Nanayakkara},
  {Williams}, {Marchesini}, {Zitrin}, \& {van
  Dokkum}}]{burgasser2023JWST_kpcBD}
{Burgasser}, A.~J., {Bezanson}, R., {Labbe}, I., {et~al.} 2024, \apj, 962, 177

\bibitem[{{Burgasser} {et~al.}(2006){Burgasser}, {Geballe}, {Leggett},
  {Kirkpatrick}, \& {Golimowski}}]{burgasser2006unifiedT}
{Burgasser}, A.~J., {Geballe}, T.~R., {Leggett}, S.~K., {Kirkpatrick}, J.~D.,
  \& {Golimowski}, D.~A. 2006, \apj, 637, 1067

\bibitem[{{Burgasser} {et~al.}(2002){Burgasser}, {Kirkpatrick}, {Brown},
  {Reid}, {Burrows}, {Liebert}, {Matthews}, {Gizis}, {Dahn}, {Monet}, {Cutri},
  \& {Skrutskie}}]{burgasser2002dT_spec_classification}
{Burgasser}, A.~J., {Kirkpatrick}, J.~D., {Brown}, M.~E., {et~al.} 2002, \apj,
  564, 421

\bibitem[{{Burgasser} {et~al.}(2015){Burgasser}, {Logsdon}, {Gagn{\'e}},
  {Bochanski}, {Faherty}, {West}, {Mamajek}, {Schmidt}, \&
  {Cruz}}]{burgasser2015dML_RV}
{Burgasser}, A.~J., {Logsdon}, S.~E., {Gagn{\'e}}, J., {et~al.} 2015, \apjs,
  220, 18

\bibitem[{{Carnero Rosell} {et~al.}(2019){Carnero Rosell}, {Santiago}, {dal
  Ponte}, {Burningham}, {da Costa}, {James}, {Marshall}, {McMahon}, {Bechtol},
  {De Paris}, {Li}, {Pieres}, {Garc{\'\i}a-Bellido}, {Abbott}, {Annis},
  {Avila}, {Bernstein}, {Brooks}, {Burke}, {Carrasco Kind}, {Carretero}, {De
  Vicente}, {Drlica-Wagner}, {Fosalba}, {Frieman}, {Gaztanaga}, {Gruendl},
  {Gschwend}, {Gutierrez}, {Hollowood}, {Maia}, {Menanteau}, {Miquel},
  {Plazas}, {Roodman}, {Sanchez}, {Scarpine}, {Schindler}, {Serrano},
  {Sevilla-Noarbe}, {Smith}, {Sobreira}, {Soares-Santos}, {Suchyta}, {Swanson},
  {Tarle}, {Vikram}, {Walker}, \& {DES Collaboration}}]{rosell2019DES_BD}
{Carnero Rosell}, A., {Santiago}, B., {dal Ponte}, M., {et~al.} 2019, \mnras,
  489, 5301

\bibitem[{{Chambers} {et~al.}(2016){Chambers}, {Magnier}, {Metcalfe},
  {Flewelling}, {Huber}, {Waters}, {Denneau}, {Draper}, {Farrow}, {Finkbeiner},
  {Holmberg}, {Koppenhoefer}, {Price}, {Rest}, {Saglia}, {Schlafly}, {Smartt},
  {Sweeney}, {Wainscoat}, {Burgett}, {Chastel}, {Grav}, {Heasley}, {Hodapp},
  {Jedicke}, {Kaiser}, {Kudritzki}, {Luppino}, {Lupton}, {Monet}, {Morgan},
  {Onaka}, {Shiao}, {Stubbs}, {Tonry}, {White}, {Ba{\~n}ados}, {Bell},
  {Bender}, {Bernard}, {Boegner}, {Boffi}, {Botticella}, {Calamida},
  {Casertano}, {Chen}, {Chen}, {Cole}, {Deacon}, {Frenk}, {Fitzsimmons},
  {Gezari}, {Gibbs}, {Goessl}, {Goggia}, {Gourgue}, {Goldman}, {Grant},
  {Grebel}, {Hambly}, {Hasinger}, {Heavens}, {Heckman}, {Henderson}, {Henning},
  {Holman}, {Hopp}, {Ip}, {Isani}, {Jackson}, {Keyes}, {Koekemoer}, {Kotak},
  {Le}, {Liska}, {Long}, {Lucey}, {Liu}, {Martin}, {Masci}, {McLean}, {Mindel},
  {Misra}, {Morganson}, {Murphy}, {Obaika}, {Narayan}, {Nieto-Santisteban},
  {Norberg}, {Peacock}, {Pier}, {Postman}, {Primak}, {Rae}, {Rai}, {Riess},
  {Riffeser}, {Rix}, {R{\"o}ser}, {Russel}, {Rutz}, {Schilbach}, {Schultz},
  {Scolnic}, {Strolger}, {Szalay}, {Seitz}, {Small}, {Smith}, {Soderblom},
  {Taylor}, {Thomson}, {Taylor}, {Thakar}, {Thiel}, {Thilker}, {Unger},
  {Urata}, {Valenti}, {Wagner}, {Walder}, {Walter}, {Watters}, {Werner},
  {Wood-Vasey}, \& {Wyse}}]{chambers2016ps1}
{Chambers}, K.~C., {Magnier}, E.~A., {Metcalfe}, N., {et~al.} 2016, arXiv
  e-prints, arXiv:1612.05560

\bibitem[{{Chiu} {et~al.}(2006){Chiu}, {Fan}, {Leggett}, {Golimowski}, {Zheng},
  {Geballe}, {Schneider}, \& {Brinkmann}}]{chiu2006_71LT_sdss}
{Chiu}, K., {Fan}, X., {Leggett}, S.~K., {et~al.} 2006, \aj, 131, 2722

\bibitem[{{Cooper} {et~al.}(2024){Cooper}, {Smart}, {Jones}, \&
  {Sarro}}]{cooper2024gaia_ucd_spectra}
{Cooper}, W.~J., {Smart}, R.~L., {Jones}, H.~R.~A., \& {Sarro}, L.~M. 2024,
  \mnras, 527, 1521

\bibitem[{{Cushing}(2014)}]{cushing2014bookUCD_LTY}
{Cushing}, M.~C. 2014, in Astrophysics and Space Science Library, Vol. 401, 50
  Years of Brown Dwarfs, ed. V.~{Joergens}, 113

\bibitem[{{Cushing} {et~al.}(2011){Cushing}, {Kirkpatrick}, {Gelino},
  {Griffith}, {Skrutskie}, {Mainzer}, {Marsh}, {Beichman}, {Burgasser},
  {Prato}, {Simcoe}, {Marley}, {Saumon}, {Freedman}, {Eisenhardt}, \&
  {Wright}}]{cushing2011dY_WISE}
{Cushing}, M.~C., {Kirkpatrick}, J.~D., {Gelino}, C.~R., {et~al.} 2011, \apj,
  743, 50

\bibitem[{{Cushing} {et~al.}(2014){Cushing}, {Kirkpatrick}, {Gelino}, {Mace},
  {Skrutskie}, \& {Gould}}]{cushing2014_WISE_3coolBD}
{Cushing}, M.~C., {Kirkpatrick}, J.~D., {Gelino}, C.~R., {et~al.} 2014, \aj,
  147, 113

\bibitem[{{Cushing} {et~al.}(2005){Cushing}, {Rayner}, \&
  {Vacca}}]{cushing2005IR_MLT}
{Cushing}, M.~C., {Rayner}, J.~T., \& {Vacca}, W.~D. 2005, \apj, 623, 1115

\bibitem[{{Cushing} {et~al.}(2000){Cushing}, {Tokunaga}, \&
  {Kobayashi}}]{cushing2000_HK_rhoOph}
{Cushing}, M.~C., {Tokunaga}, A.~T., \& {Kobayashi}, N. 2000, \aj, 119, 3019

\bibitem[{{Cutri} {et~al.}(2003){Cutri}, {Skrutskie}, {van Dyk}, {Beichman},
  {Carpenter}, {Chester}, {Cambresy}, {Evans}, {Fowler}, {Gizis}, {Howard},
  {Huchra}, {Jarrett}, {Kopan}, {Kirkpatrick}, {Light}, {Marsh}, {McCallon},
  {Schneider}, {Stiening}, {Sykes}, {Weinberg}, {Wheaton}, {Wheelock}, \&
  {Zacarias}}]{cutri2003_2MASSpoint}
{Cutri}, R.~M., {Skrutskie}, M.~F., {van Dyk}, S., {et~al.} 2003, {2MASS All
  Sky Catalog of point sources.}

\bibitem[{{Cutri} {et~al.}(2021){Cutri}, {Wright}, {Conrow}, {Fowler},
  {Eisenhardt}, {Grillmair}, {Kirkpatrick}, {Masci}, {McCallon}, {Wheelock},
  {Fajardo-Acosta}, {Yan}, {Benford}, {Harbut}, {Jarrett}, {Lake}, {Leisawitz},
  {Ressler}, {Stanford}, {Tsai}, {Liu}, {Helou}, {Mainzer}, {Gettngs},
  {Gonzalez}, {Hoffman}, {Marsh}, {Padgett}, {Skrutskie}, {Beck}, {Papin}, \&
  {Wittman}}]{cutri2014allwise}
{Cutri}, R.~M., {Wright}, E.~L., {Conrow}, T., {et~al.} 2021, VizieR Online
  Data Catalog, II/328

\bibitem[{{Dark Energy Survey Collaboration} {et~al.}(2016){Dark Energy Survey
  Collaboration}, {Abbott}, {Abdalla}, {Aleksi{\'c}}, {Allam}, {Amara},
  {Bacon}, {Balbinot}, {Banerji}, {Bechtol}, {Benoit-L{\'e}vy}, {Bernstein},
  {Bertin}, {Blazek}, {Bonnett}, {Bridle}, {Brooks}, {Brunner}, {Buckley-Geer},
  {Burke}, {Caminha}, {Capozzi}, {Carlsen}, {Carnero-Rosell}, {Carollo},
  {Carrasco-Kind}, {Carretero}, {Castander}, {Clerkin}, {Collett}, {Conselice},
  {Crocce}, {Cunha}, {D'Andrea}, {da Costa}, {Davis}, {Desai}, {Diehl},
  {Dietrich}, {Dodelson}, {Doel}, {Drlica-Wagner}, {Estrada}, {Etherington},
  {Evrard}, {Fabbri}, {Finley}, {Flaugher}, {Foley}, {Fosalba}, {Frieman},
  {Garc{\'\i}a-Bellido}, {Gaztanaga}, {Gerdes}, {Giannantonio}, {Goldstein},
  {Gruen}, {Gruendl}, {Guarnieri}, {Gutierrez}, {Hartley}, {Honscheid}, {Jain},
  {James}, {Jeltema}, {Jouvel}, {Kessler}, {King}, {Kirk}, {Kron}, {Kuehn},
  {Kuropatkin}, {Lahav}, {Li}, {Lima}, {Lin}, {Maia}, {Makler}, {Manera},
  {Maraston}, {Marshall}, {Martini}, {McMahon}, {Melchior}, {Merson}, {Miller},
  {Miquel}, {Mohr}, {Morice-Atkinson}, {Naidoo}, {Neilsen}, {Nichol}, {Nord},
  {Ogando}, {Ostrovski}, {Palmese}, {Papadopoulos}, {Peiris}, {Peoples},
  {Percival}, {Plazas}, {Reed}, {Refregier}, {Romer}, {Roodman}, {Ross},
  {Rozo}, {Rykoff}, {Sadeh}, {Sako}, {S{\'a}nchez}, {Sanchez}, {Santiago},
  {Scarpine}, {Schubnell}, {Sevilla-Noarbe}, {Sheldon}, {Smith}, {Smith},
  {Soares-Santos}, {Sobreira}, {Soumagnac}, {Suchyta}, {Sullivan}, {Swanson},
  {Tarle}, {Thaler}, {Thomas}, {Thomas}, {Tucker}, {Vieira}, {Vikram},
  {Walker}, {Wechsler}, {Weller}, {Wester}, {Whiteway}, {Wilcox}, {Yanny},
  {Zhang}, \& {Zuntz}}]{des_collaboration2016}
{Dark Energy Survey Collaboration}, {Abbott}, T., {Abdalla}, F.~B., {et~al.}
  2016, \mnras, 460, 1270

\bibitem[{{Delorme} {et~al.}(2008){Delorme}, {Delfosse}, {Albert}, {Artigau},
  {Forveille}, {Reyl{\'e}}, {Allard}, {Homeier}, {Robin}, {Willott}, {Liu}, \&
  {Dupuy}}]{delorme2008TY_J0059}
{Delorme}, P., {Delfosse}, X., {Albert}, L., {et~al.} 2008, \aap, 482, 961

\bibitem[{{Euclid Collaboration} {et~al.}(2019){Euclid Collaboration},
  {Barnett}, {Warren}, {Mortlock}, {Cuby}, {Conselice}, {Hewett}, {Willott},
  {Auricchio}, {Balaguera-Antol{\'\i}nez}, {Baldi}, {Bardelli}, {Bellagamba},
  {Bender}, {Biviano}, {Bonino}, {Bozzo}, {Branchini}, {Brescia}, {Brinchmann},
  {Burigana}, {Camera}, {Capobianco}, {Carbone}, {Carretero}, {Carvalho},
  {Castander}, {Castellano}, {Cavuoti}, {Cimatti}, {Cl{\'e}dassou}, {Congedo},
  {Conversi}, {Copin}, {Corcione}, {Coupon}, {Courtois}, {Cropper}, {Da Silva},
  {Duncan}, {Dusini}, {Ealet}, {Farrens}, {Fosalba}, {Fotopoulou},
  {Fourmanoit}, {Frailis}, {Fumana}, {Galeotta}, {Garilli}, {Gillard},
  {Gillis}, {Graci{\'a}-Carpio}, {Grupp}, {Hoekstra}, {Hormuth}, {Israel},
  {Jahnke}, {Kermiche}, {Kilbinger}, {Kirkpatrick}, {Kitching}, {Kohley},
  {Kubik}, {Kunz}, {Kurki-Suonio}, {Laureijs}, {Ligori}, {Lilje}, {Lloro},
  {Maiorano}, {Mansutti}, {Marggraf}, {Martinet}, {Marulli}, {Massey}, {Mauri},
  {Medinaceli}, {Mei}, {Mellier}, {Metcalf}, {Metge}, {Meylan}, {Moresco},
  {Moscardini}, {Munari}, {Neissner}, {Niemi}, {Nutma}, {Padilla}, {Paltani},
  {Pasian}, {Paykari}, {Percival}, {Pettorino}, {Polenta}, {Poncet},
  {Pozzetti}, {Raison}, {Renzi}, {Rhodes}, {Rix}, {Romelli}, {Roncarelli},
  {Rossetti}, {Saglia}, {Sapone}, {Scaramella}, {Schneider}, {Scottez},
  {Secroun}, {Serrano}, {Sirri}, {Stanco}, {Sureau}, {Tallada-Cresp{\'\i}},
  {Tavagnacco}, {Taylor}, {Tenti}, {Tereno}, {Toledo-Moreo}, {Torradeflot},
  {Valenziano}, {Vassallo}, {Wang}, {Zacchei}, {Zamorani}, {Zoubian}, \&
  {Zucca}}]{euclid2019quazar}
{Euclid Collaboration}, {Barnett}, R., {Warren}, S.~J., {et~al.} 2019, \aap,
  631, A85

\bibitem[{{Euclid Collaboration} {et~al.}(2023){Euclid Collaboration},
  {Gabarra}, {Mancini}, {Rodriguez Mu{\~n}oz}, {Rodighiero}, {Sirignano},
  {Scodeggio}, {Talia}, {Dusini}, {Gillard}, {Granett}, {Maiorano}, {Moresco},
  {Paganin}, {Palazzi}, {Pozzetti}, {Renzi}, {Rossetti}, {Vergani}, {Allevato},
  {Bisigello}, {Castignani}, {De Caro}, {Fumana}, {Ganga}, {Garilli},
  {Hirschmann}, {La Franca}, {Laigle}, {Passalacqua}, {Schirmer}, {Stanco},
  {Troja}, {Yung}, {Zamorani}, {Zoubian}, {Anselmi}, {Oppizzi}, {Verza},
  {Aghanim}, {Amara}, {Auricchio}, {Baldi}, {Bender}, {Bodendorf}, {Bonino},
  {Branchini}, {Brescia}, {Brinchmann}, {Camera}, {Capobianco}, {Carbone},
  {Carretero}, {Castander}, {Castellano}, {Cavuoti}, {Cledassou}, {Congedo},
  {Conselice}, {Conversi}, {Copin}, {Corcione}, {Costille}, {Courbin}, {Da
  Silva}, {Degaudenzi}, {Dinis}, {Dubath}, {Dupac}, {Ealet}, {Farrens},
  {Ferriol}, {Frailis}, {Franceschi}, {Franzetti}, {Galeotta}, {Gillis},
  {Giocoli}, {Grazian}, {Grupp}, {Guzzo}, {Holmes}, {Hornstrup}, {Hudelot},
  {Jahnke}, {K{\"u}mmel}, {Kermiche}, {Kiessling}, {Kilbinger}, {Kitching},
  {Kohley}, {Kubik}, {Kunz}, {Kurki-Suonio}, {Ligori}, {Lilje}, {Lloro},
  {Mansutti}, {Marggraf}, {Markovic}, {Marulli}, {Massey}, {Maurogordato},
  {Mei}, {Meneghetti}, {Meylan}, {Moscardini}, {Munari}, {Nichol}, {Niemi},
  {Nightingale}, {Padilla}, {Paltani}, {Pasian}, {Pedersen}, {Percival},
  {Pettorino}, {Polenta}, {Poncet}, {Raison}, {Rhodes}, {Riccio}, {Romelli},
  {Roncarelli}, {Saglia}, {Sapone}, {Schneider}, {Secroun}, {Seidel},
  {Serrano}, {Sirri}, {Surace}, {Tallada-Cresp{\'\i}}, {Tavagnacco}, {Taylor},
  {Tereno}, {Toledo-Moreo}, {Torradeflot}, {Trifoglio}, {Tutusaus},
  {Valentijn}, {Valenziano}, {Vassallo}, {Wang}, {Weller}, {Zacchei},
  {Andreon}, {Aussel}, {Bardelli}, {Bolzonella}, {Boucaud}, {Bozzo},
  {Colodro-Conde}, {Di Ferdinando}, {Farina}, {Graci{\'a}-Carpio},
  {Keih{\"a}nen}, {Lindholm}, {Maino}, {Mauri}, {Mellier}, {Neissner},
  {Scottez}, {Tenti}, {Zucca}, {Akrami}, {Baccigalupi}, {Ballardini},
  {Bernardeau}, {Biviano}, {Borlaff}, {Borsato}, {Burigana}, {Cabanac},
  {Cappi}, {Carvalho}, {Casas}, {Castro}, {Chambers}, {Cooray}, {Coupon},
  {Courtois}, {Davini}, {de la Torre}, {De Lucia}, {Desprez}, {Dole},
  {Escartin}, {Escoffier}, {Ferrero}, {Finelli}, {Fotopoulou},
  {Garcia-Bellido}, {George}, {Giacomini}, {Gozaliasl}, {Hildebrandt}, {Hook},
  {Ilbert}, {Jimenez Mu{\~n}oz}, {Kajava}, {Kansal}, {Kirkpatrick}, {Legrand},
  {Loureiro}, {Macias-Perez}, {Magliocchetti}, {Mainetti}, {Maoli}, {Marcin},
  {Martinelli}, {Martinet}, {Martins}, {Matthew}, {Maurin}, {Metcalf},
  {Morgante}, {Nadathur}, {Nucita}, {Patrizii}, {Popa}, {Porciani}, {Potter},
  {P{\"o}ntinen}, {S{\'a}nchez}, {Sakr}, {Schneider}, {Sefusatti}, {Sereno},
  {Shulevski}, {Spurio Mancini}, {Stadel}, {Steinwagner}, {Teyssier},
  {Valiviita}, {Veropalumbo}, {Viel}, \&
  {Zinchenko}}]{euclid2023NISPperformance}
{Euclid Collaboration}, {Gabarra}, L., {Mancini}, C., {et~al.} 2023, \aap, 676,
  A34

\bibitem[{{Euclid Collaboration} {et~al.}(2022{\natexlab{a}}){Euclid
  Collaboration}, {Moneti}, {McCracken}, {Shuntov}, {Kauffmann}, {Capak},
  {Davidzon}, {Ilbert}, {Scarlata}, {Toft}, {Weaver}, {Chary}, {Cuby},
  {Faisst}, {Masters}, {McPartland}, {Mobasher}, {Sanders}, {Scaramella},
  {Stern}, {Szapudi}, {Teplitz}, {Zalesky}, {Amara}, {Auricchio}, {Bodendorf},
  {Bonino}, {Branchini}, {Brau-Nogue}, {Brescia}, {Brinchmann}, {Capobianco},
  {Carbone}, {Carretero}, {Castander}, {Castellano}, {Cavuoti}, {Cimatti},
  {Cledassou}, {Congedo}, {Conselice}, {Conversi}, {Copin}, {Corcione},
  {Costille}, {Cropper}, {Da Silva}, {Degaudenzi}, {Douspis}, {Dubath},
  {Duncan}, {Dupac}, {Dusini}, {Farrens}, {Ferriol}, {Fosalba}, {Frailis},
  {Franceschi}, {Fumana}, {Garilli}, {Gillis}, {Giocoli}, {Granett}, {Grazian},
  {Grupp}, {Haugan}, {Hoekstra}, {Holmes}, {Hormuth}, {Hudelot}, {Jahnke},
  {Kermiche}, {Kiessling}, {Kilbinger}, {Kitching}, {Kohley}, {K{\"u}mmel},
  {Kunz}, {Kurki-Suonio}, {Ligori}, {Lilje}, {Lloro}, {Maiorano}, {Mansutti},
  {Marggraf}, {Markovic}, {Marulli}, {Massey}, {Maurogordato}, {Meneghetti},
  {Merlin}, {Meylan}, {Moresco}, {Moscardini}, {Munari}, {Niemi}, {Padilla},
  {Paltani}, {Pasian}, {Pedersen}, {Pires}, {Poncet}, {Popa}, {Pozzetti},
  {Raison}, {Rebolo}, {Rhodes}, {Rix}, {Roncarelli}, {Rossetti}, {Saglia},
  {Schneider}, {Secroun}, {Seidel}, {Serrano}, {Sirignano}, {Sirri}, {Stanco},
  {Tallada-Cresp{\'\i}}, {Taylor}, {Tereno}, {Toledo-Moreo}, {Torradeflot},
  {Wang}, {Welikala}, {Weller}, {Zamorani}, {Zoubian}, {Andreon}, {Bardelli},
  {Camera}, {Graci{\'a}-Carpio}, {Medinaceli}, {Mei}, {Polenta}, {Romelli},
  {Sureau}, {Tenti}, {Vassallo}, {Zacchei}, {Zucca}, {Baccigalupi},
  {Balaguera-Antol{\'\i}nez}, {Bernardeau}, {Biviano}, {Bolzonella}, {Bozzo},
  {Burigana}, {Cabanac}, {Cappi}, {Carvalho}, {Casas}, {Castignani},
  {Colodro-Conde}, {Coupon}, {Courtois}, {Di Ferdinando}, {Farina}, {Finelli},
  {Flose-Reimberg}, {Fotopoulou}, {Galeotta}, {Ganga}, {Garcia-Bellido},
  {Gaztanaga}, {Gozaliasl}, {Hook}, {Joachimi}, {Kansal}, {Keihanen},
  {Kirkpatrick}, {Lindholm}, {Mainetti}, {Maino}, {Maoli}, {Martinelli},
  {Martinet}, {Maturi}, {Metcalf}, {Morgante}, {Morisset}, {Nucita},
  {Patrizii}, {Potter}, {Renzi}, {Riccio}, {S{\'a}nchez}, {Sapone}, {Schirmer},
  {Schultheis}, {Scottez}, {Sefusatti}, {Teyssier}, {Tubio}, {Tutusaus},
  {Valiviita}, {Viel}, \& {Hildebrandt}}]{euclid2022Spitzer_EDF}
{Euclid Collaboration}, {Moneti}, A., {McCracken}, H.~J., {et~al.}
  2022{\natexlab{a}}, \aap, 658, A126

\bibitem[{{Euclid Collaboration} {et~al.}(2022{\natexlab{b}}){Euclid
  Collaboration}, {Scaramella}, {Amiaux}, {Mellier}, {Burigana}, {Carvalho},
  {Cuillandre}, {Da Silva}, {Derosa}, {Dinis}, {Maiorano}, {Maris}, {Tereno},
  {Laureijs}, {Boenke}, {Buenadicha}, {Dupac}, {Gaspar Venancio},
  {G{\'o}mez-{\'A}lvarez}, {Hoar}, {Lorenzo Alvarez}, {Racca},
  {Saavedra-Criado}, {Schwartz}, {Vavrek}, {Schirmer}, {Aussel}, {Azzollini},
  {Cardone}, {Cropper}, {Ealet}, {Garilli}, {Gillard}, {Granett}, {Guzzo},
  {Hoekstra}, {Jahnke}, {Kitching}, {Maciaszek}, {Meneghetti}, {Miller},
  {Nakajima}, {Niemi}, {Pasian}, {Percival}, {Pottinger}, {Sauvage},
  {Scodeggio}, {Wachter}, {Zacchei}, {Aghanim}, {Amara}, {Auphan}, {Auricchio},
  {Awan}, {Balestra}, {Bender}, {Bodendorf}, {Bonino}, {Branchini},
  {Brau-Nogue}, {Brescia}, {Candini}, {Capobianco}, {Carbone}, {Carlberg},
  {Carretero}, {Casas}, {Castander}, {Castellano}, {Cavuoti}, {Cimatti},
  {Cledassou}, {Congedo}, {Conselice}, {Conversi}, {Copin}, {Corcione},
  {Costille}, {Courbin}, {Degaudenzi}, {Douspis}, {Dubath}, {Duncan}, {Dusini},
  {Farrens}, {Ferriol}, {Fosalba}, {Fourmanoit}, {Frailis}, {Franceschi},
  {Franzetti}, {Fumana}, {Gillis}, {Giocoli}, {Grazian}, {Grupp}, {Haugan},
  {Holmes}, {Hormuth}, {Hudelot}, {Kermiche}, {Kiessling}, {Kilbinger},
  {Kohley}, {Kubik}, {K{\"u}mmel}, {Kunz}, {Kurki-Suonio}, {Lahav}, {Ligori},
  {Lilje}, {Lloro}, {Mansutti}, {Marggraf}, {Markovic}, {Marulli}, {Massey},
  {Maurogordato}, {Melchior}, {Merlin}, {Meylan}, {Mohr}, {Moresco}, {Morin},
  {Moscardini}, {Munari}, {Nichol}, {Padilla}, {Paltani}, {Peacock},
  {Pedersen}, {Pettorino}, {Pires}, {Poncet}, {Popa}, {Pozzetti}, {Raison},
  {Rebolo}, {Rhodes}, {Rix}, {Roncarelli}, {Rossetti}, {Saglia}, {Schneider},
  {Schrabback}, {Secroun}, {Seidel}, {Serrano}, {Sirignano}, {Sirri},
  {Skottfelt}, {Stanco}, {Starck}, {Tallada-Cresp{\'\i}}, {Tavagnacco},
  {Taylor}, {Teplitz}, {Toledo-Moreo}, {Torradeflot}, {Trifoglio}, {Valentijn},
  {Valenziano}, {Verdoes Kleijn}, {Wang}, {Welikala}, {Weller}, {Wetzstein},
  {Zamorani}, {Zoubian}, {Andreon}, {Baldi}, {Bardelli}, {Boucaud}, {Camera},
  {Di Ferdinando}, {Fabbian}, {Farinelli}, {Galeotta}, {Graci{\'a}-Carpio},
  {Maino}, {Medinaceli}, {Mei}, {Neissner}, {Polenta}, {Renzi}, {Romelli},
  {Rosset}, {Sureau}, {Tenti}, {Vassallo}, {Zucca}, {Baccigalupi},
  {Balaguera-Antol{\'\i}nez}, {Battaglia}, {Biviano}, {Borgani}, {Bozzo},
  {Cabanac}, {Cappi}, {Casas}, {Castignani}, {Colodro-Conde}, {Coupon},
  {Courtois}, {Cuby}, {de la Torre}, {Desai}, {Dole}, {Fabricius}, {Farina},
  {Ferreira}, {Finelli}, {Flose-Reimberg}, {Fotopoulou}, {Ganga}, {Gozaliasl},
  {Hook}, {Keihanen}, {Kirkpatrick}, {Liebing}, {Lindholm}, {Mainetti},
  {Martinelli}, {Martinet}, {Maturi}, {McCracken}, {Metcalf}, {Morgante},
  {Nightingale}, {Nucita}, {Patrizii}, {Potter}, {Riccio}, {S{\'a}nchez},
  {Sapone}, {Schewtschenko}, {Schultheis}, {Scottez}, {Teyssier}, {Tutusaus},
  {Valiviita}, {Viel}, {Vriend}, \& {Whittaker}}]{euclid2022i.EWS}
{Euclid Collaboration}, {Scaramella}, R., {Amiaux}, J., {et~al.}
  2022{\natexlab{b}}, \aap, 662, A112

\bibitem[{{Euclid Collaboration} {et~al.}(2022{\natexlab{c}}){Euclid
  Collaboration}, {Schirmer}, {Jahnke}, {Seidel}, {Aussel}, {Bodendorf},
  {Grupp}, {Hormuth}, {Wachter}, {Appleton}, {Barbier}, {Brinchmann},
  {Carrasco}, {Castander}, {Coupon}, {De Paolis}, {Franco}, {Ganga}, {Hudelot},
  {Jullo}, {Lan{\c{c}}on}, {Nucita}, {Paltani}, {Smadja}, {Strafella},
  {Venancio}, {Weiler}, {Amara}, {Auphan}, {Auricchio}, {Balestra}, {Bender},
  {Bonino}, {Branchini}, {Brescia}, {Capobianco}, {Carbone}, {Carretero},
  {Casas}, {Castellano}, {Cavuoti}, {Cimatti}, {Cledassou}, {Congedo},
  {Conselice}, {Conversi}, {Copin}, {Corcione}, {Costille}, {Courbin}, {Da
  Silva}, {Degaudenzi}, {Douspis}, {Dubath}, {Dupac}, {Dusini}, {Ealet},
  {Farrens}, {Ferriol}, {Fosalba}, {Frailis}, {Franceschi}, {Franzetti},
  {Fumana}, {Garilli}, {Gillard}, {Gillis}, {Giocoli}, {Grazian}, {Guzzo},
  {Haugan}, {Hoekstra}, {Holmes}, {Hornstrup}, {K{\"u}mmel}, {Kermiche},
  {Kiessling}, {Kilbinger}, {Kitching}, {Kohley}, {Kunz}, {Kurki-Suonio},
  {Laureijs}, {Ligori}, {Lilje}, {Lloro}, {Maciaszek}, {Maiorano}, {Mansutti},
  {Marggraf}, {Markovic}, {Marulli}, {Massey}, {Maurogordato}, {Mellier},
  {Meneghetti}, {Merlin}, {Meylan}, {Moresco}, {Moscardini}, {Munari},
  {Nakajima}, {Nichol}, {Niemi}, {Padilla}, {Pasian}, {Pedersen}, {Percival},
  {Pettorino}, {Pires}, {Poncet}, {Popa}, {Pozzetti}, {Prieto}, {Raison},
  {Rhodes}, {Rix}, {Roncarelli}, {Rossetti}, {Saglia}, {Sartoris},
  {Scaramella}, {Schneider}, {Secroun}, {Serrano}, {Sirignano}, {Sirri},
  {Stanco}, {Tallada-Cresp{\'\i}}, {Taylor}, {Teplitz}, {Tereno},
  {Toledo-Moreo}, {Torradeflot}, {Trifoglio}, {Valentijn}, {Valenziano},
  {Wang}, {Weller}, {Zamorani}, {Zoubian}, {Andreon}, {Bardelli}, {Boucaud},
  {Camera}, {Farinelli}, {Graci{\'a}-Carpio}, {Maino}, {Medinaceli}, {Mei},
  {Morisset}, {Polenta}, {Renzi}, {Romelli}, {Tenti}, {Vassallo}, {Zacchei},
  {Zucca}, {Baccigalupi}, {Balaguera-Antol{\'\i}nez}, {Biviano}, {Blanchard},
  {Borgani}, {Bozzo}, {Burigana}, {Cabanac}, {Cappi}, {Carvalho}, {Casas},
  {Castignani}, {Colodro-Conde}, {Cooray}, {Courtois}, {Crocce}, {Cuby},
  {Davini}, {de la Torre}, {Di Ferdinando}, {Escartin}, {Farina}, {Ferreira},
  {Finelli}, {Fotopoulou}, {Galeotta}, {Garcia-Bellido}, {Gaztanaga}, {George},
  {Gozaliasl}, {Hook}, {Ili{\'c}}, {Kansal}, {Kashlinsky}, {Keihanen},
  {Kirkpatrick}, {Lindholm}, {Mainetti}, {Maoli}, {Martinelli}, {Martinet},
  {Maturi}, {Mauri}, {McCracken}, {Metcalf}, {Monaco}, {Morgante},
  {Nightingale}, {Patrizii}, {Peel}, {Popa}, {Porciani}, {Potter}, {Reimberg},
  {Riccio}, {S{\'a}nchez}, {Sapone}, {Scottez}, {Sefusatti}, {Teyssier},
  {Tutusaus}, {Valieri}, {Valiviita}, {Viel}, \&
  {Hildebrandt}}]{euclid2022nisp_photometry}
{Euclid Collaboration}, {Schirmer}, M., {Jahnke}, K., {et~al.}
  2022{\natexlab{c}}, \aap, 662, A92

\bibitem[{{Freudling} {et~al.}(2013){Freudling}, {Romaniello}, {Bramich},
  {Ballester}, {Forchi}, {Garc{\'{\i}}a-Dabl{\'o}}, {Moehler}, \&
  {Neeser}}]{freudling2013esoreflex}
{Freudling}, W., {Romaniello}, M., {Bramich}, D.~M., {et~al.} 2013, \aap, 559,
  A96

\bibitem[{{Gagn{\'e}} {et~al.}(2015){Gagn{\'e}}, {Faherty}, {Cruz},
  {Lafreni{\'e}re}, {Doyon}, {Malo}, {Burgasser}, {Naud}, {Artigau},
  {Bouchard}, {Gizis}, \& {Albert}}]{gagne2015BANYAN_substellar_movinggroup}
{Gagn{\'e}}, J., {Faherty}, J.~K., {Cruz}, K.~L., {et~al.} 2015, \apjs, 219, 33

\bibitem[{{Gaia Collaboration} {et~al.}(2016){Gaia Collaboration}, {Prusti},
  {de Bruijne}, {Brown}, {Vallenari}, {Babusiaux}, {Bailer-Jones}, {Bastian},
  {Biermann}, {Evans}, {Eyer}, {Jansen}, {Jordi}, {Klioner}, {Lammers},
  {Lindegren}, {Luri}, {Mignard}, {Milligan}, {Panem}, {Poinsignon},
  {Pourbaix}, {Randich}, {Sarri}, {Sartoretti}, {Siddiqui}, {Soubiran},
  {Valette}, {van Leeuwen}, {Walton}, {Aerts}, {Arenou}, {Cropper}, {Drimmel},
  {H{\o}g}, {Katz}, {Lattanzi}, {O'Mullane}, {Grebel}, {Holland}, {Huc},
  {Passot}, {Bramante}, {Cacciari}, {Casta{\~n}eda}, {Chaoul}, {Cheek}, {De
  Angeli}, {Fabricius}, {Guerra}, {Hern{\'a}ndez}, {Jean-Antoine-Piccolo},
  {Masana}, {Messineo}, {Mowlavi}, {Nienartowicz}, {Ord{\'o}{\~n}ez-Blanco},
  {Panuzzo}, {Portell}, {Richards}, {Riello}, {Seabroke}, {Tanga},
  {Th{\'e}venin}, {Torra}, {Els}, {Gracia-Abril}, {Comoretto},
  {Garcia-Reinaldos}, {Lock}, {Mercier}, {Altmann}, {Andrae}, {Astraatmadja},
  {Bellas-Velidis}, {Benson}, {Berthier}, {Blomme}, {Busso}, {Carry},
  {Cellino}, {Clementini}, {Cowell}, {Creevey}, {Cuypers}, {Davidson}, {De
  Ridder}, {de Torres}, {Delchambre}, {Dell'Oro}, {Ducourant}, {Fr{\'e}mat},
  {Garc{\'\i}a-Torres}, {Gosset}, {Halbwachs}, {Hambly}, {Harrison}, {Hauser},
  {Hestroffer}, {Hodgkin}, {Huckle}, {Hutton}, {Jasniewicz}, {Jordan},
  {Kontizas}, {Korn}, {Lanzafame}, {Manteiga}, {Moitinho}, {Muinonen},
  {Osinde}, {Pancino}, {Pauwels}, {Petit}, {Recio-Blanco}, {Robin}, {Sarro},
  {Siopis}, {Smith}, {Smith}, {Sozzetti}, {Thuillot}, {van Reeven}, {Viala},
  {Abbas}, {Abreu Aramburu}, {Accart}, {Aguado}, {Allan}, {Allasia},
  {Altavilla}, {{\'A}lvarez}, {Alves}, {Anderson}, {Andrei}, {Anglada Varela},
  {Antiche}, {Antoja}, {Ant{\'o}n}, {Arcay}, {Atzei}, {Ayache}, {Bach},
  {Baker}, {Balaguer-N{\'u}{\~n}ez}, {Barache}, {Barata}, {Barbier}, {Barblan},
  {Baroni}, {Barrado y Navascu{\'e}s}, {Barros}, {Barstow}, {Becciani},
  {Bellazzini}, {Bellei}, {Bello Garc{\'\i}a}, {Belokurov}, {Bendjoya},
  {Berihuete}, {Bianchi}, {Bienaym{\'e}}, {Billebaud}, {Blagorodnova},
  {Blanco-Cuaresma}, {Boch}, {Bombrun}, {Borrachero}, {Bouquillon}, {Bourda},
  {Bouy}, {Bragaglia}, {Breddels}, {Brouillet}, {Br{\"u}semeister},
  {Bucciarelli}, {Budnik}, {Burgess}, {Burgon}, {Burlacu}, {Busonero}, {Buzzi},
  {Caffau}, {Cambras}, {Campbell}, {Cancelliere}, {Cantat-Gaudin}, {Carlucci},
  {Carrasco}, {Castellani}, {Charlot}, {Charnas}, {Charvet}, {Chassat},
  {Chiavassa}, {Clotet}, {Cocozza}, {Collins}, {Collins}, {Costigan}, {Crifo},
  {Cross}, {Crosta}, {Crowley}, {Dafonte}, {Damerdji}, {Dapergolas}, {David},
  {David}, {De Cat}, {de Felice}, {de Laverny}, {De Luise}, {De March}, {de
  Martino}, {de Souza}, {Debosscher}, {del Pozo}, {Delbo}, {Delgado},
  {Delgado}, {di Marco}, {Di Matteo}, {Diakite}, {Distefano}, {Dolding}, {Dos
  Anjos}, {Drazinos}, {Dur{\'a}n}, {Dzigan}, {Ecale}, {Edvardsson}, {Enke},
  {Erdmann}, {Escolar}, {Espina}, {Evans}, {Eynard Bontemps}, {Fabre},
  {Fabrizio}, {Faigler}, {Falc{\~a}o}, {Farr{\`a}s Casas}, {Faye}, {Federici},
  {Fedorets}, {Fern{\'a}ndez-Hern{\'a}ndez}, {Fernique}, {Fienga}, {Figueras},
  {Filippi}, {Findeisen}, {Fonti}, {Fouesneau}, {Fraile}, {Fraser}, {Fuchs},
  {Furnell}, {Gai}, {Galleti}, {Galluccio}, {Garabato}, {Garc{\'\i}a-Sedano},
  {Gar{\'e}}, {Garofalo}, {Garralda}, {Gavras}, {Gerssen}, {Geyer}, {Gilmore},
  {Girona}, {Giuffrida}, {Gomes}, {Gonz{\'a}lez-Marcos},
  {Gonz{\'a}lez-N{\'u}{\~n}ez}, {Gonz{\'a}lez-Vidal}, {Granvik}, {Guerrier},
  {Guillout}, {Guiraud}, {G{\'u}rpide}, {Guti{\'e}rrez-S{\'a}nchez}, {Guy},
  {Haigron}, {Hatzidimitriou}, {Haywood}, {Heiter}, {Helmi}, {Hobbs},
  {Hofmann}, {Holl}, {Holland}, {Hunt}, {Hypki}, {Icardi}, {Irwin}, {Jevardat
  de Fombelle}, {Jofr{\'e}}, {Jonker}, {Jorissen}, {Julbe}, {Karampelas},
  {Kochoska}, {Kohley}, {Kolenberg}, {Kontizas}, {Koposov}, {Kordopatis},
  {Koubsky}, {Kowalczyk}, {Krone-Martins}, {Kudryashova}, {Kull}, {Bachchan},
  {Lacoste-Seris}, {Lanza}, {Lavigne}, {Le Poncin-Lafitte}, {Lebreton},
  {Lebzelter}, {Leccia}, {Leclerc}, {Lecoeur-Taibi}, {Lemaitre}, {Lenhardt},
  {Leroux}, {Liao}, {Licata}, {Lindstr{\o}m}, {Lister}, {Livanou}, {Lobel},
  {L{\"o}ffler}, {L{\'o}pez}, {Lopez-Lozano}, {Lorenz}, {Loureiro},
  {MacDonald}, {Magalh{\~a}es Fernandes}, {Managau}, {Mann}, {Mantelet},
  {Marchal}, {Marchant}, {Marconi}, {Marie}, {Marinoni}, {Marrese},
  {Marschalk{\'o}}, {Marshall}, {Mart{\'\i}n-Fleitas}, {Martino}, {Mary},
  {Matijevi{\v{c}}}, {Mazeh}, {McMillan}, {Messina}, {Mestre}, {Michalik},
  {Millar}, {Miranda}, {Molina}, {Molinaro}, {Molinaro}, {Moln{\'a}r},
  {Moniez}, {Montegriffo}, {Monteiro}, {Mor}, {Mora}, {Morbidelli}, {Morel},
  {Morgenthaler}, {Morley}, {Morris}, {Mulone}, {Muraveva}, {Musella},
  {Narbonne}, {Nelemans}, {Nicastro}, {Noval}, {Ord{\'e}novic},
  {Ordieres-Mer{\'e}}, {Osborne}, {Pagani}, {Pagano}, {Pailler}, {Palacin},
  {Palaversa}, {Parsons}, {Paulsen}, {Pecoraro}, {Pedrosa}, {Pentik{\"a}inen},
  {Pereira}, {Pichon}, {Piersimoni}, {Pineau}, {Plachy}, {Plum}, {Poujoulet},
  {Pr{\v{s}}a}, {Pulone}, {Ragaini}, {Rago}, {Rambaux}, {Ramos-Lerate},
  {Ranalli}, {Rauw}, {Read}, {Regibo}, {Renk}, {Reyl{\'e}}, {Ribeiro},
  {Rimoldini}, {Ripepi}, {Riva}, {Rixon}, {Roelens}, {Romero-G{\'o}mez},
  {Rowell}, {Royer}, {Rudolph}, {Ruiz-Dern}, {Sadowski}, {Sagrist{\`a}
  Sell{\'e}s}, {Sahlmann}, {Salgado}, {Salguero}, {Sarasso}, {Savietto},
  {Schnorhk}, {Schultheis}, {Sciacca}, {Segol}, {Segovia}, {Segransan},
  {Serpell}, {Shih}, {Smareglia}, {Smart}, {Smith}, {Solano}, {Solitro},
  {Sordo}, {Soria Nieto}, {Souchay}, {Spagna}, {Spoto}, {Stampa}, {Steele},
  {Steidelm{\"u}ller}, {Stephenson}, {Stoev}, {Suess}, {S{\"u}veges}, {Surdej},
  {Szabados}, {Szegedi-Elek}, {Tapiador}, {Taris}, {Tauran}, {Taylor},
  {Teixeira}, {Terrett}, {Tingley}, {Trager}, {Turon}, {Ulla}, {Utrilla},
  {Valentini}, {van Elteren}, {Van Hemelryck}, {van Leeuwen}, {Varadi},
  {Vecchiato}, {Veljanoski}, {Via}, {Vicente}, {Vogt}, {Voss}, {Votruba},
  {Voutsinas}, {Walmsley}, {Weiler}, {Weingrill}, {Werner}, {Wevers},
  {Whitehead}, {Wyrzykowski}, {Yoldas}, {{\v{Z}}erjal}, {Zucker}, {Zurbach},
  {Zwitter}, {Alecu}, {Allen}, {Allende Prieto}, {Amorim},
  {Anglada-Escud{\'e}}, {Arsenijevic}, {Azaz}, {Balm}, {Beck}, {Bernstein},
  {Bigot}, {Bijaoui}, {Blasco}, {Bonfigli}, {Bono}, {Boudreault}, {Bressan},
  {Brown}, {Brunet}, {Bunclark}, {Buonanno}, {Butkevich}, {Carret}, {Carrion},
  {Chemin}, {Ch{\'e}reau}, {Corcione}, {Darmigny}, {de Boer}, {de Teodoro}, {de
  Zeeuw}, {Delle Luche}, {Domingues}, {Dubath}, {Fodor}, {Fr{\'e}zouls},
  {Fries}, {Fustes}, {Fyfe}, {Gallardo}, {Gallegos}, {Gardiol}, {Gebran},
  {Gomboc}, {G{\'o}mez}, {Grux}, {Gueguen}, {Heyrovsky}, {Hoar}, {Iannicola},
  {Isasi Parache}, {Janotto}, {Joliet}, {Jonckheere}, {Keil}, {Kim},
  {Klagyivik}, {Klar}, {Knude}, {Kochukhov}, {Kolka}, {Kos}, {Kutka}, {Lainey},
  {LeBouquin}, {Liu}, {Loreggia}, {Makarov}, {Marseille}, {Martayan},
  {Martinez-Rubi}, {Massart}, {Meynadier}, {Mignot}, {Munari}, {Nguyen},
  {Nordlander}, {Ocvirk}, {O'Flaherty}, {Olias Sanz}, {Ortiz}, {Osorio},
  {Oszkiewicz}, {Ouzounis}, {Palmer}, {Park}, {Pasquato}, {Peltzer}, {Peralta},
  {P{\'e}turaud}, {Pieniluoma}, {Pigozzi}, {Poels}, {Prat}, {Prod'homme},
  {Raison}, {Rebordao}, {Risquez}, {Rocca-Volmerange}, {Rosen}, {Ruiz-Fuertes},
  {Russo}, {Sembay}, {Serraller Vizcaino}, {Short}, {Siebert}, {Silva},
  {Sinachopoulos}, {Slezak}, {Soffel}, {Sosnowska}, {Strai{\v{z}}ys}, {ter
  Linden}, {Terrell}, {Theil}, {Tiede}, {Troisi}, {Tsalmantza}, {Tur},
  {Vaccari}, {Vachier}, {Valles}, {Van Hamme}, {Veltz}, {Virtanen}, {Wallut},
  {Wichmann}, {Wilkinson}, {Ziaeepour}, \& {Zschocke}}]{gaia2016}
{Gaia Collaboration}, {Prusti}, T., {de Bruijne}, J.~H.~J., {et~al.} 2016,
  \aap, 595, A1

\bibitem[{{Geballe} {et~al.}(2002){Geballe}, {Knapp}, {Leggett}, {Fan},
  {Golimowski}, {Anderson}, {Brinkmann}, {Csabai}, {Gunn}, {Hawley},
  {Hennessy}, {Henry}, {Hill}, {Hindsley}, {Ivezi{\'c}}, {Lupton}, {McDaniel},
  {Munn}, {Narayanan}, {Peng}, {Pier}, {Rockosi}, {Schneider}, {Smith},
  {Strauss}, {Tsvetanov}, {Uomoto}, {York}, \&
  {Zheng}}]{geballe2002dT_classification}
{Geballe}, T.~R., {Knapp}, G.~R., {Leggett}, S.~K., {et~al.} 2002, \apj, 564,
  466

\bibitem[{{Golimowski} {et~al.}(2004){Golimowski}, {Leggett}, {Marley}, {Fan},
  {Geballe}, {Knapp}, {Vrba}, {Henden}, {Luginbuhl}, {Guetter}, {Munn},
  {Canzian}, {Zheng}, {Tsvetanov}, {Chiu}, {Glazebrook}, {Hoversten},
  {Schneider}, \& {Brinkmann}}]{golimowski2004LM_phot_LT}
{Golimowski}, D.~A., {Leggett}, S.~K., {Marley}, M.~S., {et~al.} 2004, \aj,
  127, 3516

\bibitem[{{Gorlova} {et~al.}(2003){Gorlova}, {Meyer}, {Rieke}, \&
  {Liebert}}]{gorlova2003NIRgravity_BD}
{Gorlova}, N.~I., {Meyer}, M.~R., {Rieke}, G.~H., \& {Liebert}, J. 2003, \apj,
  593, 1074

\bibitem[{{Hainline} {et~al.}(2023){Hainline}, {Helton}, {Johnson}, {Sun},
  {Topping}, {Leisenring}, {Baker}, {Eisenstein}, {Hausen}, {Hviding}, {Lyu},
  {Robertson}, {Tacchella}, {Williams}, {Willmer}, \&
  {Roellig}}]{hainline2023BD_JADES_CEERS}
{Hainline}, K.~N., {Helton}, J.~M., {Johnson}, B.~D., {et~al.} 2023, arXiv
  e-prints, arXiv:2309.03250

\bibitem[{{Holwerda} {et~al.}(2023){Holwerda}, {Pirzkal}, {Burgasser}, \&
  {Hsu}}]{holwerda2023MLTY_Roman}
{Holwerda}, B., {Pirzkal}, N., {Burgasser}, A., \& {Hsu}, C.-C. 2023, arXiv
  e-prints, arXiv:2306.12363

\bibitem[{{Holwerda} {et~al.}(2024){Holwerda}, {Hsu}, {Hathi}, {Bisigello}, {de
  la Vega}, {Haro}, {Bagley}, {Dickinson}, {Finkelstein}, {Kartaltepe},
  {Koekemoer}, {Papovich}, {Pirzkal}, {Cook}, {Robertson}, {Casey}, {Aganze},
  {P{\'e}rez-Gonz{\'a}lez}, {Lucas}, {Jogee}, {Wilkins}, {Burgarella}, \&
  {Kirkpatrick}}]{holwerda2023JWST_CEERS_dwarf}
{Holwerda}, B.~W., {Hsu}, C.-C., {Hathi}, N., {et~al.} 2024, \mnras, 529, 1067

\bibitem[{{Holwerda} {et~al.}(2014){Holwerda}, {Trenti}, {Clarkson}, {Sahu},
  {Bradley}, {Stiavelli}, {Pirzkal}, {De Marchi}, {Andersen}, {Bouwens}, \&
  {Ryan}}]{holwerda2014dwarf_distribution_WFC3}
{Holwerda}, B.~W., {Trenti}, M., {Clarkson}, W., {et~al.} 2014, \apj, 788, 77

\bibitem[{{Hsu} {et~al.}(2021){Hsu}, {Burgasser}, {Theissen}, {Gelino},
  {Birky}, {Diamant}, {Bardalez Gagliuffi}, {Aganze}, {Blake}, \&
  {Faherty}}]{hsu2021dT_RV}
{Hsu}, C.-C., {Burgasser}, A.~J., {Theissen}, C.~A., {et~al.} 2021, \apjs, 257,
  45

\bibitem[{{Kausch} {et~al.}(2015){Kausch}, {Noll}, {Smette}, {Kimeswenger},
  {Barden}, {Szyszka}, {Jones}, {Sana}, {Horst}, \&
  {Kerber}}]{kausch2015molecfit_xshooter}
{Kausch}, W., {Noll}, S., {Smette}, A., {et~al.} 2015, \aap, 576, A78

\bibitem[{{Kirkpatrick} {et~al.}(2011){Kirkpatrick}, {Cushing}, {Gelino},
  {Griffith}, {Skrutskie}, {Marsh}, {Wright}, {Mainzer}, {Eisenhardt},
  {McLean}, {Thompson}, {Bauer}, {Benford}, {Bridge}, {Lake}, {Petty},
  {Stanford}, {Tsai}, {Bailey}, {Beichman}, {Bloom}, {Bochanski}, {Burgasser},
  {Capak}, {Cruz}, {Hinz}, {Kartaltepe}, {Knox}, {Manohar}, {Masters},
  {Morales-Calder{\'o}n}, {Prato}, {Rodigas}, {Salvato}, {Schurr}, {Scoville},
  {Simcoe}, {Stapelfeldt}, {Stern}, {Stock}, \&
  {Vacca}}]{kirkpatrick2011hundredBD_WISE}
{Kirkpatrick}, J.~D., {Cushing}, M.~C., {Gelino}, C.~R., {et~al.} 2011, \apjs,
  197, 19

\bibitem[{{Kirkpatrick} {et~al.}(2021){Kirkpatrick}, {Gelino}, {Faherty},
  {Meisner}, {Caselden}, {Schneider}, {Marocco}, {Cayago}, {Smart},
  {Eisenhardt}, {Kuchner}, {Wright}, {Cushing}, {Allers}, {Bardalez Gagliuffi},
  {Burgasser}, {Gagn{\'e}}, {Logsdon}, {Martin}, {Ingalls}, {Lowrance},
  {Abrahams}, {Aganze}, {Gerasimov}, {Gonzales}, {Hsu}, {Kamraj}, {Kiman},
  {Rees}, {Theissen}, {Ammar}, {Andersen}, {Beaulieu}, {Colin}, {Elachi},
  {Goodman}, {Gramaize}, {Hamlet}, {Hong}, {Jonkeren}, {Khalil}, {Martin},
  {Pendrill}, {Pumphrey}, {Rothermich}, {Sainio}, {Stenner}, {Tanner},
  {Th{\'e}venot}, {Voloshin}, {Walla}, {W{\k{e}}dracki}, \& {Backyard Worlds:
  Planet 9 Collaboration}}]{kirkpatrick2021LTY20pc}
{Kirkpatrick}, J.~D., {Gelino}, C.~R., {Faherty}, J.~K., {et~al.} 2021, \apjs,
  253, 7

\bibitem[{{Kirkpatrick} {et~al.}(1997){Kirkpatrick}, {Henry}, \&
  {Irwin}}]{kirkpatrick1997ultracoolM}
{Kirkpatrick}, J.~D., {Henry}, T.~J., \& {Irwin}, M.~J. 1997, \aj, 113, 1421

\bibitem[{{Kirkpatrick} {et~al.}(1995){Kirkpatrick}, {Henry}, \&
  {Simons}}]{kirkpatrick1995first_solar_neighbour_UCD}
{Kirkpatrick}, J.~D., {Henry}, T.~J., \& {Simons}, D.~A. 1995, \aj, 109, 797

\bibitem[{{Kirkpatrick} {et~al.}(1999){Kirkpatrick}, {Reid}, {Liebert},
  {Cutri}, {Nelson}, {Beichman}, {Dahn}, {Monet}, {Gizis}, \&
  {Skrutskie}}]{kirkpatrick1999L}
{Kirkpatrick}, J.~D., {Reid}, I.~N., {Liebert}, J., {et~al.} 1999, \apj, 519,
  802

\bibitem[{{Kirkpatrick} {et~al.}(2000){Kirkpatrick}, {Reid}, {Liebert},
  {Gizis}, {Burgasser}, {Monet}, {Dahn}, {Nelson}, \&
  {Williams}}]{2000Kirkpatrick}
{Kirkpatrick}, J.~D., {Reid}, I.~N., {Liebert}, J., {et~al.} 2000, \aj, 120,
  447

\bibitem[{{Knapp} {et~al.}(2004){Knapp}, {Leggett}, {Fan}, {Marley}, {Geballe},
  {Golimowski}, {Finkbeiner}, {Gunn}, {Hennawi}, {Ivezi{\'c}}, {Lupton},
  {Schlegel}, {Strauss}, {Tsvetanov}, {Chiu}, {Hoversten}, {Glazebrook},
  {Zheng}, {Hendrickson}, {Williams}, {Uomoto}, {Vrba}, {Henden}, {Luginbuhl},
  {Guetter}, {Munn}, {Canzian}, {Schneider}, \&
  {Brinkmann}}]{knapp2004nir_spectra_LT}
{Knapp}, G.~R., {Leggett}, S.~K., {Fan}, X., {et~al.} 2004, \aj, 127, 3553

\bibitem[{{Langeroodi} \& {Hjorth}(2023)}]{langeroodi2023farthestBD}
{Langeroodi}, D. \& {Hjorth}, J. 2023, \apjl, 957, L27

\bibitem[{{Laureijs} {et~al.}(2011){Laureijs}, {Amiaux}, {Arduini},
  {Augu{\`e}res}, {Brinchmann}, {Cole}, {Cropper}, {Dabin}, {Duvet}, {Ealet},
  {Garilli}, {Gondoin}, {Guzzo}, {Hoar}, {Hoekstra}, {Holmes}, {Kitching},
  {Maciaszek}, {Mellier}, {Pasian}, {Percival}, {Rhodes}, {Saavedra Criado},
  {Sauvage}, {Scaramella}, {Valenziano}, {Warren}, {Bender}, {Castander},
  {Cimatti}, {Le F{\`e}vre}, {Kurki-Suonio}, {Levi}, {Lilje}, {Meylan},
  {Nichol}, {Pedersen}, {Popa}, {Rebolo Lopez}, {Rix}, {Rottgering},
  {Zeilinger}, {Grupp}, {Hudelot}, {Massey}, {Meneghetti}, {Miller}, {Paltani},
  {Paulin-Henriksson}, {Pires}, {Saxton}, {Schrabback}, {Seidel}, {Walsh},
  {Aghanim}, {Amendola}, {Bartlett}, {Baccigalupi}, {Beaulieu}, {Benabed},
  {Cuby}, {Elbaz}, {Fosalba}, {Gavazzi}, {Helmi}, {Hook}, {Irwin}, {Kneib},
  {Kunz}, {Mannucci}, {Moscardini}, {Tao}, {Teyssier}, {Weller}, {Zamorani},
  {Zapatero Osorio}, {Boulade}, {Foumond}, {Di Giorgio}, {Guttridge}, {James},
  {Kemp}, {Martignac}, {Spencer}, {Walton}, {Bl{\"u}mchen}, {Bonoli},
  {Bortoletto}, {Cerna}, {Corcione}, {Fabron}, {Jahnke}, {Ligori}, {Madrid},
  {Martin}, {Morgante}, {Pamplona}, {Prieto}, {Riva}, {Toledo}, {Trifoglio},
  {Zerbi}, {Abdalla}, {Douspis}, {Grenet}, {Borgani}, {Bouwens}, {Courbin},
  {Delouis}, {Dubath}, {Fontana}, {Frailis}, {Grazian}, {Koppenh{\"o}fer},
  {Mansutti}, {Melchior}, {Mignoli}, {Mohr}, {Neissner}, {Noddle}, {Poncet},
  {Scodeggio}, {Serrano}, {Shane}, {Starck}, {Surace}, {Taylor},
  {Verdoes-Kleijn}, {Vuerli}, {Williams}, {Zacchei}, {Altieri}, {Escudero
  Sanz}, {Kohley}, {Oosterbroek}, {Astier}, {Bacon}, {Bardelli}, {Baugh},
  {Bellagamba}, {Benoist}, {Bianchi}, {Biviano}, {Branchini}, {Carbone},
  {Cardone}, {Clements}, {Colombi}, {Conselice}, {Cresci}, {Deacon}, {Dunlop},
  {Fedeli}, {Fontanot}, {Franzetti}, {Giocoli}, {Garcia-Bellido}, {Gow},
  {Heavens}, {Hewett}, {Heymans}, {Holland}, {Huang}, {Ilbert}, {Joachimi},
  {Jennins}, {Kerins}, {Kiessling}, {Kirk}, {Kotak}, {Krause}, {Lahav}, {van
  Leeuwen}, {Lesgourgues}, {Lombardi}, {Magliocchetti}, {Maguire}, {Majerotto},
  {Maoli}, {Marulli}, {Maurogordato}, {McCracken}, {McLure}, {Melchiorri},
  {Merson}, {Moresco}, {Nonino}, {Norberg}, {Peacock}, {Pello}, {Penny},
  {Pettorino}, {Di Porto}, {Pozzetti}, {Quercellini}, {Radovich}, {Rassat},
  {Roche}, {Ronayette}, {Rossetti}, {Sartoris}, {Schneider}, {Semboloni},
  {Serjeant}, {Simpson}, {Skordis}, {Smadja}, {Smartt}, {Spano}, {Spiro},
  {Sullivan}, {Tilquin}, {Trotta}, {Verde}, {Wang}, {Williger}, {Zhao},
  {Zoubian}, \& {Zucca}}]{laureijs2011euclid}
{Laureijs}, R., {Amiaux}, J., {Arduini}, S., {et~al.} 2011, arXiv e-prints,
  arXiv:1110.3193

\bibitem[{{Leggett} {et~al.}(2003){Leggett}, {Golimowski}, {Fan}, {Geballe}, \&
  {Knapp}}]{leggett2003NIRcolorLT}
{Leggett}, S.~K., {Golimowski}, D.~A., {Fan}, X., {Geballe}, T.~R., \& {Knapp},
  G.~R. 2003, in Cambridge Workshop on Cool Stars, Stellar Systems, and the
  Sun, Vol.~12, The Future of Cool-Star Astrophysics: 12th Cambridge Workshop
  on Cool Stars, Stellar Systems, and the Sun, ed. A.~{Brown}, G.~M. {Harper},
  \& T.~R. {Ayres}, 120--127

\bibitem[{{Li} {et~al.}(2016){Li}, {Smith}, {Zhong}, {Hou}, {Carlin},
  {Newberg}, {Liu}, {Chen}, {Li}, {Shao}, {Small}, \&
  {Tian}}]{li2016redgiantsIRphot}
{Li}, J., {Smith}, M.~C., {Zhong}, J., {et~al.} 2016, \apj, 823, 59

\bibitem[{{Lodieu} {et~al.}(2008){Lodieu}, {Hambly}, {Jameson}, \&
  {Hodgkin}}]{lodieu2008NIR_upSCo}
{Lodieu}, N., {Hambly}, N.~C., {Jameson}, R.~F., \& {Hodgkin}, S.~T. 2008,
  \mnras, 383, 1385

\bibitem[{{Lodieu} {et~al.}(2018){Lodieu}, {Zapatero Osorio}, {B{\'e}jar}, \&
  {Pe{\~n}a Ram{\'\i}rez}}]{lodieu2018dL_upperSco}
{Lodieu}, N., {Zapatero Osorio}, M.~R., {B{\'e}jar}, V.~J.~S., \& {Pe{\~n}a
  Ram{\'\i}rez}, K. 2018, \mnras, 473, 2020

\bibitem[{{Lucas} {et~al.}(2001){Lucas}, {Roche}, {Allard}, \&
  {Hauschildt}}]{lucas2001Ori_young}
{Lucas}, P.~W., {Roche}, P.~F., {Allard}, F., \& {Hauschildt}, P.~H. 2001,
  \mnras, 326, 695

\bibitem[{{Luhman}(2014)}]{luhman2014W0855}
{Luhman}, K.~L. 2014, \apjl, 786, L18

\bibitem[{{Mace} {et~al.}(2013){Mace}, {Kirkpatrick}, {Cushing}, {Gelino},
  {Griffith}, {Skrutskie}, {Marsh}, {Wright}, {Eisenhardt}, {McLean},
  {Thompson}, {Mix}, {Bailey}, {Beichman}, {Bloom}, {Burgasser}, {Fortney},
  {Hinz}, {Knox}, {Lowrance}, {Marley}, {Morley}, {Rodigas}, {Saumon},
  {Sheppard}, \& {Stock}}]{mace2013WISE_dT}
{Mace}, G.~N., {Kirkpatrick}, J.~D., {Cushing}, M.~C., {et~al.} 2013, \apjs,
  205, 6

\bibitem[{{Manjavacas} {et~al.}(2014){Manjavacas}, {Bonnefoy}, {Schlieder},
  {Allard}, {Rojo}, {Goldman}, {Chauvin}, {Homeier}, {Lodieu}, \&
  {Henning}}]{manjavacas2014atm_young_ML}
{Manjavacas}, E., {Bonnefoy}, M., {Schlieder}, J.~E., {et~al.} 2014, \aap, 564,
  A55

\bibitem[{{Manjavacas} {et~al.}(2024){Manjavacas}, {Tremblin}, {Birkmann},
  {Valenti}, {Alves de Oliveira}, {Beck}, {Giardino}, {Luetzgendorf},
  {Rauscher}, \& {Sirianni}}]{manjavacas2024JWST_youngBD}
{Manjavacas}, E., {Tremblin}, P., {Birkmann}, S., {et~al.} 2024, arXiv
  e-prints, arXiv:2402.04230

\bibitem[{{Mart{\'\i}n} {et~al.}(1997){Mart{\'\i}n}, {Basri}, {Delfosse}, \&
  {Forveille}}]{1997Martin}
{Mart{\'\i}n}, E.~L., {Basri}, G., {Delfosse}, X., \& {Forveille}, T. 1997,
  \aap, 327, L29

\bibitem[{{Mart{\'\i}n} {et~al.}(1998){Mart{\'\i}n}, {Basri},
  {Zapatero-Osorio}, {Rebolo}, \& {L{\'o}pez}}]{1998Martin}
{Mart{\'\i}n}, E.~L., {Basri}, G., {Zapatero-Osorio}, M.~R., {Rebolo}, R., \&
  {L{\'o}pez}, R.~J.~G. 1998, \apjl, 507, L41

\bibitem[{{Mart{\'\i}n} {et~al.}(1999){Mart{\'\i}n}, {Delfosse}, {Basri},
  {Goldman}, {Forveille}, \& {Zapatero Osorio}}]{martin1999Lclassification}
{Mart{\'\i}n}, E.~L., {Delfosse}, X., {Basri}, G., {et~al.} 1999, \aj, 118,
  2466

\bibitem[{{Mart{\'\i}n} {et~al.}(2021){Mart{\'\i}n}, {Zhang}, {Esparza},
  {Gracia}, {Rasilla}, {Masseron}, \& {Burgasser}}]{martin2021ch4nh3}
{Mart{\'\i}n}, E.~L., {Zhang}, J.~Y., {Esparza}, P., {et~al.} 2021, \aap, 655,
  L3

\bibitem[{{McGovern} {et~al.}(2004){McGovern}, {Kirkpatrick}, {McLean},
  {Burgasser}, {Prato}, \& {Lowrance}}]{mcgovern2004lowgravity_BD}
{McGovern}, M.~R., {Kirkpatrick}, J.~D., {McLean}, I.~S., {et~al.} 2004, \apj,
  600, 1020

\bibitem[{{McLean} {et~al.}(1998){McLean}, {Becklin}, {Bendiksen}, {Brims},
  {Canfield}, {Figer}, {Graham}, {Hare}, {Lacayanga}, {Larkin}, {Larson},
  {Levenson}, {Magnone}, {Teplitz}, \& {Wong}}]{mclean1998nirspec}
{McLean}, I.~S., {Becklin}, E.~E., {Bendiksen}, O., {et~al.} 1998, in Society
  of Photo-Optical Instrumentation Engineers (SPIE) Conference Series, Vol.
  3354, Infrared Astronomical Instrumentation, ed. A.~M. {Fowler}, 566--578

\bibitem[{{McLean} {et~al.}(2000){McLean}, {Graham}, {Becklin}, {Figer},
  {Larkin}, {Levenson}, \& {Teplitz}}]{mclean2000nirspec}
{McLean}, I.~S., {Graham}, J.~R., {Becklin}, E.~E., {et~al.} 2000, in Society
  of Photo-Optical Instrumentation Engineers (SPIE) Conference Series, Vol.
  4008, Optical and IR Telescope Instrumentation and Detectors, ed. M.~{Iye} \&
  A.~F. {Moorwood}, 1048--1055

\bibitem[{{McLean} {et~al.}(2003){McLean}, {McGovern}, {Burgasser},
  {Kirkpatrick}, {Prato}, \& {Kim}}]{mclean2003BDSS}
{McLean}, I.~S., {McGovern}, M.~R., {Burgasser}, A.~J., {et~al.} 2003, \apj,
  596, 561

\bibitem[{{McMahon} {et~al.}(2013){McMahon}, {Banerji}, {Gonzalez}, {Koposov},
  {Bejar}, {Lodieu}, {Rebolo}, \& {VHS Collaboration}}]{mcmahon2013vhs}
{McMahon}, R.~G., {Banerji}, M., {Gonzalez}, E., {et~al.} 2013, The Messenger,
  154, 35

\bibitem[{{Mu{\v{z}}i{\'c}} {et~al.}(2015){Mu{\v{z}}i{\'c}}, {Scholz}, {Geers},
  \& {Jayawardhana}}]{muzic2015young_substellar_lupus3}
{Mu{\v{z}}i{\'c}}, K., {Scholz}, A., {Geers}, V.~C., \& {Jayawardhana}, R.
  2015, \apj, 810, 159

\bibitem[{{Nonino} {et~al.}(2023){Nonino}, {Glazebrook}, {Burgasser},
  {Polenta}, {Morishita}, {Lepinzan}, {Castellano}, {Fontana}, {Merlin},
  {Bonchi}, {Paris}, {Treu}, {Vulcani}, {Wang}, {Santini}, {Vanzella},
  {Nanayakkara}, {Mercurio}, {Rosati}, {Grillo}, \&
  {Bradac}}]{nonino2023GLASS_JWST}
{Nonino}, M., {Glazebrook}, K., {Burgasser}, A.~J., {et~al.} 2023, \apjl, 942,
  L29

\bibitem[{{Pirzkal} {et~al.}(2009){Pirzkal}, {Burgasser}, {Malhotra},
  {Holwerda}, {Sahu}, {Rhoads}, {Xu}, {Bochanski}, {Walsh}, {Windhorst},
  {Hathi}, \& {Cohen}}]{pirzkal2009PEARS}
{Pirzkal}, N., {Burgasser}, A.~J., {Malhotra}, S., {et~al.} 2009, \apj, 695,
  1591

\bibitem[{{Pirzkal} {et~al.}(2005){Pirzkal}, {Sahu}, {Burgasser}, {Moustakas},
  {Xu}, {Malhotra}, {Rhoads}, {Koekemoer}, {Nelan}, {Windhorst}, {Panagia},
  {Gronwall}, {Pasquali}, \& {Walsh}}]{pirzkal2005HubbleDeepField_star}
{Pirzkal}, N., {Sahu}, K.~C., {Burgasser}, A., {et~al.} 2005, \apj, 622, 319

\bibitem[{{Rayner} {et~al.}(2009){Rayner}, {Cushing}, \&
  {Vacca}}]{rayner2009IRTF_coolstars}
{Rayner}, J.~T., {Cushing}, M.~C., \& {Vacca}, W.~D. 2009, \apjs, 185, 289

\bibitem[{{Reyl{\'e}}(2018)}]{reyle2018GaiaDR2UCD}
{Reyl{\'e}}, C. 2018, \aap, 619, L8

\bibitem[{{Rhodes} {et~al.}(2017){Rhodes}, {Nichol}, {Aubourg}, {Bean},
  {Boutigny}, {Bremer}, {Capak}, {Cardone}, {Carry}, {Conselice}, {Connolly},
  {Cuillandre}, {Hatch}, {Helou}, {Hemmati}, {Hildebrandt}, {Hlo{\v{z}}ek},
  {Jones}, {Kahn}, {Kiessling}, {Kitching}, {Lupton}, {Mandelbaum}, {Markovic},
  {Marshall}, {Massey}, {Maughan}, {Melchior}, {Mellier}, {Newman},
  {Robertson}, {Sauvage}, {Schrabback}, {Smith}, {Strauss}, {Taylor}, \& {Von
  Der Linden}}]{rhodes2017lsst_euclid}
{Rhodes}, J., {Nichol}, R.~C., {Aubourg}, {\'E}., {et~al.} 2017, \apjs, 233, 21

\bibitem[{{Ryan} {et~al.}(2005){Ryan}, {Hathi}, {Cohen}, \&
  {Windhorst}}]{ryan2005HST_LTdistribution}
{Ryan}, R.~E., J., {Hathi}, N.~P., {Cohen}, S.~H., \& {Windhorst}, R.~A. 2005,
  \apjl, 631, L159

\bibitem[{{Ryan} {et~al.}(2011){Ryan}, {Thorman}, {Yan}, {Fan}, {Yan},
  {Mechtley}, {Hathi}, {Cohen}, {Windhorst}, {McCarthy}, \&
  {Wittman}}]{Ryan2011HST_UCD_highLat}
{Ryan}, R.~E., {Thorman}, P.~A., {Yan}, H., {et~al.} 2011, \apj, 739, 83

\bibitem[{{Sarro} {et~al.}(2023){Sarro}, {Berihuete}, {Smart}, {Reyl{\'e}},
  {Barrado}, {Garc{\'\i}a-Torres}, {Cooper}, {Jones}, {Marocco}, {Creevey},
  {Sordo}, {Bailer-Jones}, {Montegriffo}, {Carballo}, {Andrae}, {Fouesneau},
  {Lanzafame}, {Pailler}, {Th{\'e}venin}, {Lobel}, {Delchambre}, {Korn},
  {Recio-Blanco}, {Schultheis}, {De Angeli}, {Brouillet}, {Casamiquela},
  {Contursi}, {de Laverny}, {Garc{\'\i}a-Lario}, {Kordopatis}, {Lebreton},
  {Livanou}, {Lorca}, {Palicio}, {Slezak-Oreshina}, {Soubiran}, {Ulla}, \&
  {Zhao}}]{sarro2023gaiaDR3_UCD}
{Sarro}, L.~M., {Berihuete}, A., {Smart}, R.~L., {et~al.} 2023, \aap, 669, A139

\bibitem[{{Schmidt} {et~al.}(2015){Schmidt}, {Hawley}, {West}, {Bochanski},
  {Davenport}, {Ge}, \& {Schneider}}]{schmidt2015MLcolor}
{Schmidt}, S.~J., {Hawley}, S.~L., {West}, A.~A., {et~al.} 2015, \aj, 149, 158

\bibitem[{{Scholz} {et~al.}(2012){Scholz}, {Muzic}, {Geers}, {Bonavita},
  {Jayawardhana}, \& {Tamura}}]{scholz2012young_substellar_NGC1333}
{Scholz}, A., {Muzic}, K., {Geers}, V., {et~al.} 2012, \apj, 744, 6

\bibitem[{{Skrutskie} {et~al.}(2006){Skrutskie}, {Cutri}, {Stiening},
  {Weinberg}, {Schneider}, {Carpenter}, {Beichman}, {Capps}, {Chester},
  {Elias}, {Huchra}, {Liebert}, {Lonsdale}, {Monet}, {Price}, {Seitzer},
  {Jarrett}, {Kirkpatrick}, {Gizis}, {Howard}, {Evans}, {Fowler}, {Fullmer},
  {Hurt}, {Light}, {Kopan}, {Marsh}, {McCallon}, {Tam}, {Van Dyk}, \&
  {Wheelock}}]{skrutskie2006_2MASS}
{Skrutskie}, M.~F., {Cutri}, R.~M., {Stiening}, R., {et~al.} 2006, \aj, 131,
  1163

\bibitem[{{Skrzypek} {et~al.}(2016){Skrzypek}, {Warren}, \&
  {Faherty}}]{skrzypeck2016phometric_classification_LTcolor}
{Skrzypek}, N., {Warren}, S.~J., \& {Faherty}, J.~K. 2016, \aap, 589, A49

\bibitem[{{Smette} {et~al.}(2015){Smette}, {Sana}, {Noll}, {Horst}, {Kausch},
  {Kimeswenger}, {Barden}, {Szyszka}, {Jones}, {Gallenne}, {Vinther},
  {Ballester}, \& {Taylor}}]{smette2015molecfit}
{Smette}, A., {Sana}, H., {Noll}, S., {et~al.} 2015, \aap, 576, A77

\bibitem[{{Solano} {et~al.}(2021){Solano}, {G{\'a}lvez-Ortiz}, {Mart{\'\i}n},
  {G{\'o}mez Mu{\~n}oz}, {Rodrigo}, {Burgasser}, {Lodieu}, {B{\'e}jar},
  {Hu{\'e}lamo}, {Morales-Calder{\'o}n}, \& {Bouy}}]{solano2021virtualUCDs}
{Solano}, E., {G{\'a}lvez-Ortiz}, M.~C., {Mart{\'\i}n}, E.~L., {et~al.} 2021,
  \mnras, 501, 281

\bibitem[{{Stanway} {et~al.}(2008){Stanway}, {Bremer}, {Lehnert}, \&
  {Eldridge}}]{stanway2008GOODS_dM}
{Stanway}, E.~R., {Bremer}, M.~N., {Lehnert}, M.~D., \& {Eldridge}, J.~J. 2008,
  \mnras, 384, 348

\bibitem[{{Stevenson} {et~al.}(2023){Stevenson}, {Haswell}, {Barnes}, \&
  {Barstow}}]{stevenson2023BD_desert_binary_Gaia}
{Stevenson}, A.~T., {Haswell}, C.~A., {Barnes}, J.~R., \& {Barstow}, J.~K.
  2023, \mnras, 526, 5155

\bibitem[{{Taylor}(2005)}]{taylor2005topcat}
{Taylor}, M.~B. 2005, in Astronomical Society of the Pacific Conference Series,
  Vol. 347, Astronomical Data Analysis Software and Systems XIV, ed.
  P.~{Shopbell}, M.~{Britton}, \& R.~{Ebert}, 29

\bibitem[{{Tody}(1986)}]{tody1986iraf}
{Tody}, D. 1986, in Society of Photo-Optical Instrumentation Engineers (SPIE)
  Conference Series, Vol. 627, Instrumentation in astronomy VI, ed. D.~L.
  {Crawford}, 733

\bibitem[{{van Vledder} {et~al.}(2016){van Vledder}, {van der Vlugt},
  {Holwerda}, {Kenworthy}, {Bouwens}, \& {Trenti}}]{vanVledder2016BDdM_BoRG}
{van Vledder}, I., {van der Vlugt}, D., {Holwerda}, B.~W., {et~al.} 2016,
  \mnras, 458, 425

\bibitem[{{Vernet} {et~al.}(2011){Vernet}, {Dekker}, {D'Odorico}, {Kaper},
  {Kjaergaard}, {Hammer}, {Randich}, {Zerbi}, {Groot}, {Hjorth}, {Guinouard},
  {Navarro}, {Adolfse}, {Albers}, {Amans}, {Andersen}, {Andersen}, {Binetruy},
  {Bristow}, {Castillo}, {Chemla}, {Christensen}, {Conconi}, {Conzelmann},
  {Dam}, {de Caprio}, {de Ugarte Postigo}, {Delabre}, {di Marcantonio},
  {Downing}, {Elswijk}, {Finger}, {Fischer}, {Flores}, {Fran{\c{c}}ois},
  {Goldoni}, {Guglielmi}, {Haigron}, {Hanenburg}, {Hendriks}, {Horrobin},
  {Horville}, {Jessen}, {Kerber}, {Kern}, {Kiekebusch}, {Kleszcz}, {Klougart},
  {Kragt}, {Larsen}, {Lizon}, {Lucuix}, {Mainieri}, {Manuputy}, {Martayan},
  {Mason}, {Mazzoleni}, {Michaelsen}, {Modigliani}, {Moehler}, {M{\o}ller},
  {Norup S{\o}rensen}, {N{\o}rregaard}, {P{\'e}roux}, {Patat}, {Pena}, {Pragt},
  {Reinero}, {Rigal}, {Riva}, {Roelfsema}, {Royer}, {Sacco}, {Santin},
  {Schoenmaker}, {Spano}, {Sweers}, {Ter Horst}, {Tintori}, {Tromp}, {van
  Dael}, {van der Vliet}, {Venema}, {Vidali}, {Vinther}, {Vola}, {Winters},
  {Wistisen}, {Wulterkens}, \& {Zacchei}}]{vernet2011xshooter}
{Vernet}, J., {Dekker}, H., {D'Odorico}, S., {et~al.} 2011, \aap, 536, A105

\bibitem[{{Wang} {et~al.}(2023){Wang}, {Goto}, {Ho}, {Lin}, {Wu}, {Ling},
  {Hashimoto}, {Kim}, \& {Hsiao}}]{wang_poya2023dT_NIRCamEarly}
{Wang}, P.-Y., {Goto}, T., {Ho}, S. C.~C., {et~al.} 2023, \mnras, 523, 4534

\bibitem[{{Wright} {et~al.}(2010){Wright}, {Eisenhardt}, {Mainzer}, {Ressler},
  {Cutri}, {Jarrett}, {Kirkpatrick}, {Padgett}, {McMillan}, {Skrutskie},
  {Stanford}, {Cohen}, {Walker}, {Mather}, {Leisawitz}, {Gautier}, {McLean},
  {Benford}, {Lonsdale}, {Blain}, {Mendez}, {Irace}, {Duval}, {Liu}, {Royer},
  {Heinrichsen}, {Howard}, {Shannon}, {Kendall}, {Walsh}, {Larsen}, {Cardon},
  {Schick}, {Schwalm}, {Abid}, {Fabinsky}, {Naes}, \& {Tsai}}]{wright2010WISE}
{Wright}, E.~L., {Eisenhardt}, P. R.~M., {Mainzer}, A.~K., {et~al.} 2010, \aj,
  140, 1868

\bibitem[{{Zapatero Osorio} {et~al.}(2000){Zapatero Osorio}, {B{\'e}jar},
  {Mart{\'\i}n}, {Rebolo}, {Barrado y Navascu{\'e}s}, {Bailer-Jones}, \&
  {Mundt}}]{zapatero2000PMO_sOri}
{Zapatero Osorio}, M.~R., {B{\'e}jar}, V.~J.~S., {Mart{\'\i}n}, E.~L., {et~al.}
  2000, Science, 290, 103

\end{thebibliography}

\onecolumn

\begingroup
\setlength{\tabcolsep}{3pt} 

\fontsize{8}{10}\selectfont

\begin{appendix}
\section{Lists of ultracool dwarf candidates in Euclid Deep Field North}
\LTcapwidth=\textwidth

\begin{longtable}{ccccccccccc}
    \caption{\label{EDFNcatalogM} 81 late-M-type UCD candidates as photometric standards in EDF North and their photometry measurements. }\\  
  \hline\hline
$\alpha$ (hms.ss) &
$\delta$ (dms.s)&
  $i$&
  $z$&
  $Y$&
  $J$&
  $H$&
    $K$&
  $W1$&
  $W2$&
  \\
\hline
\endfirsthead
\caption{continued.}\\
\hline\hline
$\alpha$ (hhmmss.ss) &
$\delta$ (ddmmss.s)&
  $i$&
  $z$&
  $Y$&
  $J$&
  $H$&
    $K$&
  $W1$&
  $W2$&
 \\
\hline
\endhead
\hline
\endfoot

\fontsize{8}{10}\selectfont

17:35:39.86 & +65:35:41.7 & 18.79$\pm$0.01 & 17.84$\pm$0.01 & 17.35$\pm$0.01 & 15.99$\pm$0.08 & 15.3$\pm$0.09 & 15.09$\pm$0.16 & 14.69$\pm$0.01 & 14.52$\pm$0.02\\
  17:36:03.27 & +65:35:11.8 & 18.53$\pm$0.0 & 17.61$\pm$0.01 & 17.14$\pm$0.01 & 15.81$\pm$0.07 & 15.24$\pm$0.09 & 14.71$\pm$0.12 & 14.64$\pm$0.01 & 14.4$\pm$0.04\\
  17:36:24.77 & +65:57:35.1 & 18.64$\pm$0.01 & 17.55$\pm$0.01 & 17.02$\pm$0.01 & 15.68$\pm$0.08 & 15.05$\pm$0.1 & 14.51$\pm$0.11 & 14.31$\pm$0.01 & 13.9$\pm$0.01\\
  17:37:25.84 & +66:00:00.3 & 19.47$\pm$0.01 & 18.51$\pm$0.01 & 18.02$\pm$0.02 & 16.47$\pm$0.14 & 15.73$\pm$0.18 & 15.36 & 15.23$\pm$0.04 & 14.94$\pm$0.04\\
  17:38:31.21 & +65:57:25.7 & 18.19$\pm$0.01 & 17.07$\pm$0.0 & 16.42$\pm$0.0 & 14.83$\pm$0.04 & 14.25$\pm$0.05 & 13.92$\pm$0.07 & 13.61$\pm$0.01 & 13.39$\pm$0.01\\
  17:38:36.60 & +66:53:18.8 & 19.15$\pm$0.01 & 18.26$\pm$0.01 & 17.77$\pm$0.02 & 16.38$\pm$0.13 & 15.84$\pm$0.19 & 15.4$\pm$0.21 & 14.95$\pm$0.03 & 14.68$\pm$0.06\\
  17:41:18.44 & +65:51:41.9 & 19.01$\pm$0.01 & 17.9$\pm$0.01 & 17.32$\pm$0.01 & 15.93$\pm$0.08 & 15.45$\pm$0.09 & 15.19$\pm$0.17 & 14.74$\pm$0.03 & 14.51$\pm$0.03\\
  17:42:40.51 & +66:14:24.9 & 17.86$\pm$0.0 & 16.79$\pm$0.0 & 16.23$\pm$0.01 & 14.76$\pm$0.04 & 14.11$\pm$0.05 & 13.79$\pm$0.05 & 13.47$\pm$0.01 & 13.3$\pm$0.01\\
  17:43:02.14 & +66:42:00.9 & 17.66$\pm$0.0 & 16.46$\pm$0.0 & 15.79$\pm$0.01 & 14.25$\pm$0.03 & 13.6$\pm$0.04 & 13.2$\pm$0.04 & 12.95$\pm$0.01 & 12.76$\pm$0.01\\
  17:45:02.13 & +67:10:45.6 & 19.74$\pm$0.01 & 18.36$\pm$0.0 & 17.56$\pm$0.01 & 16.01$\pm$0.09 & 15.24$\pm$0.11 & 14.77$\pm$0.12 & 14.12$\pm$0.02 & 13.84$\pm$0.03\\
  17:46:07.58 & +67:43:13.1 & 19.24$\pm$0.01 & 18.3$\pm$0.01 & 17.79$\pm$0.01 & 16.51$\pm$0.12 & 15.94$\pm$0.19 & 15.35$\pm$0.22 & 15.27$\pm$0.04 & 15.13$\pm$0.07\\
  17:46:33.48 & +67:31:15.2 & 19.28$\pm$0.01 & 18.51$\pm$0.01 & 18.14$\pm$0.03 & 16.58$\pm$0.12 & 16.05$\pm$0.21 & 15.41$\pm$0.2 & 15.24$\pm$0.03 & 15.07$\pm$0.06\\
  17:47:13.27 & +65:48:49.8 & 19.28$\pm$0.02 & 18.52$\pm$0.04 & 18.22$\pm$0.05 & 16.29$\pm$0.11 & 15.71$\pm$0.13 & 14.89 & 14.93$\pm$0.02 & 14.72$\pm$0.05\\
  17:47:28.29 & +64:42:49.5 & 18.77$\pm$0.01 & 17.91$\pm$0.04 & 17.47$\pm$0.02 & 16.09$\pm$0.09 & 15.3$\pm$0.09 & 14.9$\pm$0.13 & 14.83$\pm$0.03 & 14.59$\pm$0.03\\
  17:48:14.55 & +64:31:37.9 & 19.83$\pm$0.01 & 18.82$\pm$0.02 & 18.37$\pm$0.03 & 17.01$\pm$0.21 & 16.15$\pm$0.17 & 15.77$\pm$0.28 & 15.93$\pm$0.07 & 15.61$\pm$0.08\\
  17:48:55.26 & +67:08:18.0 & 19.98$\pm$0.01 & 18.87$\pm$0.01 & 18.19$\pm$0.01 & 16.54$\pm$0.14 & 15.98$\pm$0.18 & 15.49$\pm$0.22 & 14.79$\pm$0.02 & 14.52$\pm$0.03\\
  17:51:02.52 & +66:26:00.3 & 18.88$\pm$0.01 & 18.01$\pm$0.01 & 17.55$\pm$0.01 & 16.13$\pm$0.09 & 15.56$\pm$0.14 & 15.28$\pm$0.18 & 14.65$\pm$0.02 & 14.43$\pm$0.03\\
  17:51:14.75 & +64:57:18.9 & 18.62$\pm$0.01 & 17.71$\pm$0.01 & 17.26$\pm$0.01 & 15.9$\pm$0.08 & 15.41$\pm$0.1 & 14.89$\pm$0.13 & 14.63$\pm$0.02 & 14.49$\pm$0.04\\
  17:51:19.39 & +64:05:47.2 & 19.33$\pm$0.01 & 18.15$\pm$0.02 & 17.54$\pm$0.01 & 16.05$\pm$0.09 & 15.31$\pm$0.09 & 14.89$\pm$0.14 & 14.61$\pm$0.01 & 14.37$\pm$0.03\\
  17:51:38.41 & +65:40:14.9 & 19.54$\pm$0.02 & 18.7$\pm$0.01 & 18.23$\pm$0.02 & 16.72$\pm$0.15 & 16.22$\pm$0.21 & 15.56$\pm$0.24 & 15.55$\pm$0.05 & 15.21$\pm$0.07\\
  17:51:42.95 & +68:20:11.9 & 19.34$\pm$0.02 & 18.32$\pm$0.02 & 17.81$\pm$0.01 & 16.54$\pm$0.12 & 15.65$\pm$0.13 & 15.19$\pm$0.18 & 15.27$\pm$0.03 & 15.14$\pm$0.05\\
  17:51:50.30 & +64:54:32.0 & 18.54$\pm$0.01 & 17.58$\pm$0.0 & 17.07$\pm$0.01 & 15.7$\pm$0.06 & 15.08$\pm$0.08 & 14.66$\pm$0.12 & 13.96$\pm$0.01 & 13.81$\pm$0.04\\
  17:53:23.33 & +66:53:29.8 & 19.42$\pm$0.01 & 18.47$\pm$0.01 & 17.97$\pm$0.02 & 16.68$\pm$0.15 & 15.91$\pm$0.17 & 15.47$\pm$0.17 & 14.78$\pm$0.03 & 14.54$\pm$0.03\\
  17:53:58.60 & +65:26:58.2 & 17.58$\pm$0.0 & 16.66$\pm$0.0 & 16.2$\pm$0.01 & 14.84$\pm$0.04 & 14.18$\pm$0.05 & 13.83$\pm$0.06 & 13.55$\pm$0.01 & 13.38$\pm$0.01\\
  $^{(1)}$17:54:40.48 & +64:08:09.2 & 16.31$\pm$0.0 & 15.2$\pm$0.0 & 14.64$\pm$0.01 & 13.12$\pm$0.02 & 12.59$\pm$0.02 & 12.27$\pm$0.03 & 12.01$\pm$0.02 & 11.78$\pm$0.02\\
  17:55:15.62 & +66:54:22.8 & 18.67$\pm$0.01 & 17.41$\pm$0.01 & 16.75$\pm$0.01 & 15.24$\pm$0.05 & 14.57$\pm$0.05 & 14.22$\pm$0.09 & 13.94$\pm$0.01 & 13.8$\pm$0.02\\
  17:55:15.85 & +64:10:51.3 & 17.44$\pm$0.0 & 16.5$\pm$0.01 & 16.02$\pm$0.0 & 14.58$\pm$0.03 & 14.01$\pm$0.04 & 13.65$\pm$0.05 & 13.11$\pm$0.01 & 12.97$\pm$0.01\\
  17:56:08.33 & +67:10:36.7 & 19.31$\pm$0.01 & 18.2$\pm$0.01 & 17.66$\pm$0.01 & 16.25$\pm$0.11 & 15.58$\pm$0.12 & 15.2$\pm$0.19 & 15.09$\pm$0.02 & 14.86$\pm$0.04\\
  17:56:26.28 & +63:52:07.1 & 16.92$\pm$0.0 & 16.04$\pm$0.0 & 15.59$\pm$0.0 & 14.23$\pm$0.03 & 13.61$\pm$0.03 & 13.4$\pm$0.04 & 12.87$\pm$0.01 & 12.7$\pm$0.01\\
  17:56:49.34 & +68:25:57.2 & 18.59$\pm$0.01 & 17.71$\pm$0.01 & 17.22$\pm$0.01 & 15.85$\pm$0.08 & 15.35$\pm$0.11 & 14.8$\pm$0.13 & 14.7$\pm$0.02 & 14.45$\pm$0.04\\
  \textbf{17:57:30.73} & \textbf{+67:41:40.1} & 16.86$\pm$0.0 & 15.74$\pm$0.01 & 15.15$\pm$0.01 & 13.7$\pm$0.03 & 13.15$\pm$0.02 & 12.74$\pm$0.03 & 12.53$\pm$0.01 & 12.32$\pm$0.01\\
  17:57:31.04 & +64:39:19.1 & 19.16$\pm$0.0 & 18.48$\pm$0.01 & 18.2$\pm$0.01 & 16.47$\pm$0.12 & 15.89$\pm$0.16 & 14.66 & 14.95$\pm$0.03 & 14.79$\pm$0.03\\
  17:57:35.66 & +67:18:52.0 & 19.77$\pm$0.02 & 18.59$\pm$0.02 & 17.92$\pm$0.02 & 16.51$\pm$0.14 & 15.82$\pm$0.15 & 15.22$\pm$0.2 & 15.1$\pm$0.03 & 14.9$\pm$0.05\\
  17:57:55.04 & +64:18:05.7 & 19.95$\pm$0.01 & 18.59$\pm$0.01 & 17.78$\pm$0.02 & 16.23$\pm$0.09 & 15.13$\pm$0.07 & 14.61$\pm$0.1 & 14.46$\pm$0.01 & 14.32$\pm$0.05\\
  17:58:35.99 & +65:43:06.2 & 18.44$\pm$0.0 & 17.46$\pm$0.01 & 16.95$\pm$0.0 & 15.63$\pm$0.06 & 14.92$\pm$0.07 & 14.7$\pm$0.11 & 14.34$\pm$0.01 & 14.12$\pm$0.02\\
  17:59:13.72 & +65:24:09.2 & 20.28$\pm$0.02 & 19.04$\pm$0.01 & 18.28$\pm$0.01 & 16.82$\pm$0.15 & 15.95$\pm$0.17 & 15.47$\pm$0.21 & 15.07$\pm$0.05 & 14.9$\pm$0.08\\
  18:00:56.18 & +65:52:23.1 & 19.1$\pm$0.01 & 17.74$\pm$0.01 & 16.93$\pm$0.01 & 15.22$\pm$0.04 & 14.53$\pm$0.04 & 14.17$\pm$0.06 & 13.8$\pm$0.01 & 13.59$\pm$0.01\\
  18:01:09.10 & +63:53:39.4 & 19.42$\pm$0.01 & 18.49$\pm$0.01 & 17.87$\pm$0.06 & 16.53$\pm$0.11 & 15.95$\pm$0.14 & 15.23$\pm$0.16 & 15.44$\pm$0.05 & 15.16$\pm$0.08\\
  18:01:41.24 & +64:00:09.7 & 19.48$\pm$0.01 & 18.57$\pm$0.01 & 18.11$\pm$0.02 & 16.6$\pm$0.11 & 16.03$\pm$0.16 & 15.57$\pm$0.22 & 15.19$\pm$0.04 & 14.89$\pm$0.05\\
  18:01:55.15 & +67:37:03.2 & 19.24$\pm$0.01 & 18.32$\pm$0.01 & 17.84$\pm$0.01 & 16.49$\pm$0.13 & 15.72$\pm$0.14 & 15.5 & 15.07$\pm$0.03 & 14.87$\pm$0.04\\
  18:02:13.83 & +65:00:21.3 & 18.88$\pm$0.0 & 18.03$\pm$0.01 & 17.56$\pm$0.01 & 16.14$\pm$0.07 & 15.62$\pm$0.11 & 15.04$\pm$0.12 & 14.9$\pm$0.03 & 14.69$\pm$0.05\\
  18:02:30.40 & +67:00:13.8 & 19.04$\pm$0.01 & 18.16$\pm$0.01 & 17.68$\pm$0.01 & 16.28$\pm$0.12 & 15.78$\pm$0.17 & 15.38$\pm$0.23 & 14.88$\pm$0.01 & 14.67$\pm$0.05\\
  18:02:40.56 & +65:46:26.9 & 19.97$\pm$0.01 & 18.96$\pm$0.01 & 18.4$\pm$0.02 & 17.07$\pm$0.18 & 16.08$\pm$0.15 & 15.72$\pm$0.23 & 15.29$\pm$0.03 & 15.01$\pm$0.04\\
  $^{(1)}$18:03:14.25 & +65:32:08.1 & 18.77$\pm$0.01 & 17.65$\pm$0.0 & 17.05$\pm$0.01 & 15.54$\pm$0.05 & 14.85$\pm$0.06 & 14.47$\pm$0.08 & 14.37$\pm$0.03 & 14.17$\pm$0.03\\
  18:03:29.52 & +65:02:28.2 & 19.85$\pm$0.01 & 18.83$\pm$0.02 & 18.26$\pm$0.01 & 16.8$\pm$0.13 & 16.23$\pm$0.18 & 15.84 & 15.34$\pm$0.02 & 15.19$\pm$0.09\\
  18:03:51.03 & +66:48:06.4 & 19.26$\pm$0.01 & 18.28$\pm$0.01 & 17.76$\pm$0.01 & 16.35$\pm$0.11 & 15.73$\pm$0.13 & 14.84 & 15.16$\pm$0.02 & 14.93$\pm$0.04\\
  18:04:38.48 & +64:40:30.2 & 20.44$\pm$0.03 & 19.28$\pm$0.02 & 18.63$\pm$0.03 & 17.23$\pm$0.19 & 16.11$\pm$0.16 & 15.76$\pm$0.25 & 15.82$\pm$0.05 & 15.66$\pm$0.08\\
  18:04:40.83 & +65:03:30.8 & 19.14$\pm$0.01 & 18.32$\pm$0.01 & 17.88$\pm$0.01 & 16.41$\pm$0.11 & 15.94$\pm$0.13 & 15.4 & 15.56$\pm$0.03 & 15.03$\pm$0.04\\
  18:04:41.87 & +65:27:06.9 & 20.22$\pm$0.02 & 19.35$\pm$0.03 & 18.92$\pm$0.04 & 17.33$\pm$0.22 & 16.55$\pm$0.23 & 15.58$\pm$0.21 & 16.01$\pm$0.05 & 15.71$\pm$0.1\\
  18:05:03.33 & +63:56:26.6 & 19.8$\pm$0.01 & 18.78$\pm$0.01 & 18.25$\pm$0.01 & 16.79$\pm$0.14 & 16.1$\pm$0.17 & 15.68 & 15.53$\pm$0.02 & 15.27$\pm$0.07\\
  18:05:32.10 & +68:20:01.8 & 19.63$\pm$0.01 & 18.55$\pm$0.01 & 17.95$\pm$0.02 & 16.27$\pm$0.11 & 15.67$\pm$0.12 & 15.2 & 15.0$\pm$0.03 & 14.72$\pm$0.05\\
  18:05:41.60 & +63:38:17.8 & 20.24$\pm$0.02 & 19.09$\pm$0.02 & 18.47$\pm$0.02 & 16.77$\pm$0.14 & 16.22$\pm$0.19 & 15.36$\pm$0.18 & 15.21$\pm$0.03 & 15.02$\pm$0.05\\
  18:05:44.81 & +66:22:59.8 & 19.92$\pm$0.01 & 18.77$\pm$0.02 & 18.12$\pm$0.02 & 16.78$\pm$0.15 & 15.82$\pm$0.14 & 15.03 & 14.97$\pm$0.01 & 14.73$\pm$0.04\\
  18:06:01.72 & +67:53:11.4 & 19.47$\pm$0.01 & 18.6$\pm$0.01 & 18.11$\pm$0.02 & 16.67$\pm$0.15 & 16.04$\pm$0.18 & 15.34 & 15.53$\pm$0.03 & 15.27$\pm$0.07\\
  18:06:15.25 & +65:23:25.6 & 19.39$\pm$0.01 & 18.44$\pm$0.01 & 17.88$\pm$0.01 & 16.44$\pm$0.11 & 15.96$\pm$0.15 & 15.4 & 15.03$\pm$0.02 & 14.82$\pm$0.05\\
  18:06:59.30 & +66:10:51.6 & 18.81$\pm$0.01 & 17.88$\pm$0.01 & 17.47$\pm$0.01 & 16.13$\pm$0.09 & 15.48$\pm$0.11 & 15.27$\pm$0.19 & 14.47$\pm$0.04 & 14.28$\pm$0.04\\
  18:07:34.58 & +64:02:22.5 & 19.73$\pm$0.01 & 18.76$\pm$0.02 & 18.28$\pm$0.02 & 16.96$\pm$0.15 & 15.9 & 15.57$\pm$0.19 & 15.36$\pm$0.03 & 15.12$\pm$0.06\\
  18:08:08.20 & +67:01:09.7 & 19.09$\pm$0.0 & 18.18$\pm$0.01 & 17.73$\pm$0.01 & 16.27$\pm$0.1 & 15.65$\pm$0.12 & 15.36$\pm$0.22 & 14.48$\pm$0.02 & 14.2$\pm$0.03\\
  18:08:45.44 & +64:06:22.1 & 19.82$\pm$0.01 & 18.59$\pm$0.02 & 17.93$\pm$0.02 & 16.43$\pm$0.11 & 15.85$\pm$0.13 & 15.4$\pm$0.19 & 14.99$\pm$0.02 & 14.84$\pm$0.04\\
  18:08:51.52 & +68:00:04.3 & 17.95$\pm$0.0 & 17.02$\pm$0.01 & 16.46$\pm$0.01 & 15.09$\pm$0.04 & 14.51$\pm$0.05 & 14.1$\pm$0.07 & 13.73$\pm$0.01 & 13.52$\pm$0.01\\
  18:09:34.21 & +66:21:13.5 & 19.18$\pm$0.01 & 18.15$\pm$0.01 & 17.6$\pm$0.01 & 16.16$\pm$0.09 & 15.55$\pm$0.11 & 15.29$\pm$0.21 & 14.96$\pm$0.02 & 14.77$\pm$0.05\\
  18:09:34.94 & +66:21:41.2 & 19.03$\pm$0.02 & 17.66$\pm$0.01 & 16.89$\pm$0.01 & 15.4$\pm$0.05 & 14.86$\pm$0.06 & 14.39$\pm$0.09 & 14.05$\pm$0.01 & 13.79$\pm$0.01\\
  18:09:40.07 & +64:48:33.6 & 19.32$\pm$0.01 & 18.31$\pm$0.02 & 17.78$\pm$0.01 & 16.42$\pm$0.11 & 15.63$\pm$0.13 & 15.23 & 15.18$\pm$0.04 & 14.91$\pm$0.06\\
  18:09:41.00 & +63:51:41.2 & 18.17$\pm$0.01 & 17.15$\pm$0.01 & 16.6$\pm$0.01 & 15.16$\pm$0.04 & 14.59$\pm$0.05 & 14.26$\pm$0.07 & 13.91$\pm$0.01 & 13.76$\pm$0.04\\
  18:10:22.13 & +64:16:37.5 & 19.36$\pm$0.01 & 18.48$\pm$0.01 & 18.13$\pm$0.02 & 16.65$\pm$0.12 & 16.02$\pm$0.17 & 15.32 & 15.26$\pm$0.02 & 15.05$\pm$0.07\\
  18:10:22.24 & +64:04:37.2 & 19.06$\pm$0.01 & 18.3$\pm$0.02 & 17.85$\pm$0.01 & 16.35$\pm$0.11 & 15.88$\pm$0.15 & 14.94 & 15.03$\pm$0.03 & 14.83$\pm$0.05\\
  18:10:30.13 & +65:13:38.8 & 19.64$\pm$0.01 & 18.58$\pm$0.01 & 18.07$\pm$0.01 & 16.87$\pm$0.15 & 16.16$\pm$0.2 & 15.55 & 15.15$\pm$0.03 & 14.91$\pm$0.06\\
  18:10:48.82 & +65:32:39.9 & 19.61$\pm$0.01 & 18.79$\pm$0.02 & 18.36$\pm$0.03 & 16.68$\pm$0.14 & 16.1 & 15.69 & 14.95$\pm$0.02 & 14.75$\pm$0.07\\
  18:12:31.83 & +66:41:50.8 & 16.9$\pm$0.0 & 15.97$\pm$0.0 & 15.51$\pm$0.0 & 14.19$\pm$0.03 & 13.57$\pm$0.03 & 13.3$\pm$0.04 & 13.04$\pm$0.01 & 12.82$\pm$0.01\\
  18:13:43.62 & +65:48:10.5 & 19.37$\pm$0.01 & 18.49$\pm$0.01 & 17.95$\pm$0.02 & 16.45$\pm$0.1 & 15.95$\pm$0.14 & 15.39$\pm$0.17 & 15.28$\pm$0.02 & 15.06$\pm$0.06\\
  18:15:03.88 & +64:24:10.7 & 19.85$\pm$0.01 & 18.81$\pm$0.02 & 18.25$\pm$0.03 & 16.88$\pm$0.14 & 16.06$\pm$0.16 & 14.99 & 15.43$\pm$0.03 & 15.24$\pm$0.07\\
  18:15:10.85 & +64:34:27.8 & 18.74$\pm$0.01 & 17.64$\pm$0.01 & 17.05$\pm$0.01 & 15.77$\pm$0.06 & 15.02$\pm$0.07 & 14.73$\pm$0.1 & 14.55$\pm$0.02 & 14.36$\pm$0.04\\
  18:15:21.44 & +64:47:21.3 & 19.22$\pm$0.01 & 17.99$\pm$0.01 & 17.29$\pm$0.01 & 15.73$\pm$0.07 & 15.06$\pm$0.07 & 14.49$\pm$0.08 & 14.35$\pm$0.01 & 14.15$\pm$0.03\\
  18:15:30.11 & +67:45:28.7 & 17.97$\pm$0.01 & 17.04$\pm$0.01 & 16.56$\pm$0.01 & 15.16$\pm$0.04 & 14.58$\pm$0.05 & 14.33$\pm$0.09 & 13.71$\pm$0.01 & 13.5$\pm$0.02\\
  18:15:30.92 & +67:08:13.7 & 19.45$\pm$0.01 & 18.42$\pm$0.01 & 17.88$\pm$0.02 & 16.4$\pm$0.11 & 15.62$\pm$0.12 & 15.23 & 15.18$\pm$0.03 & 15.03$\pm$0.07\\
  18:16:08.19 & +67:50:26.8 & 18.71$\pm$0.01 & 17.83$\pm$0.01 & 17.39$\pm$0.01 & 16.02$\pm$0.08 & 15.48$\pm$0.11 & 15.11$\pm$0.18 & 14.62$\pm$0.01 & 14.48$\pm$0.05\\
  18:17:15.82 & +66:27:00.1 & 19.29$\pm$0.01 & 18.26$\pm$0.0 & 17.7$\pm$0.02 & 16.25$\pm$0.1 & 15.4$\pm$0.1 & 15.12$\pm$0.17 & 14.93$\pm$0.02 & 14.76$\pm$0.05\\
  18:18:18.38 & +65:04:30.0 & 19.43$\pm$0.01 & 18.56$\pm$0.01 & 18.09$\pm$0.01 & 16.67$\pm$0.13 & 16.15$\pm$0.17 & 15.07 & 15.43$\pm$0.06 & 15.17$\pm$0.12\\
  18:20:49.32 & +65:47:44.5 & 18.98$\pm$0.01 & 17.96$\pm$0.0 & 17.41$\pm$0.01 & 16.08$\pm$0.09 & 15.43$\pm$0.1 & 15.06$\pm$0.13 & 14.51$\pm$0.02 & 14.31$\pm$0.02\\
  18:21:42.11 & +65:37:24.4 & 19.71$\pm$0.01 & 18.58$\pm$0.01 & 18.05$\pm$0.02 & 16.62$\pm$0.14 & 16.14$\pm$0.22 & 15.13 & 15.38$\pm$0.04 & 15.22$\pm$0.09\\
  18:21:54.00 & +66:49:00.2 & 19.55$\pm$0.01 & 18.67$\pm$0.01 & 18.15$\pm$0.01 & 16.8$\pm$0.15 & 15.98$\pm$0.18 & 15.72$\pm$0.24 & 15.57$\pm$0.06 & 15.28$\pm$0.09\\
  
\end{longtable}
  \vspace{1ex}
     {\raggedright 
         \textbf{Note:} Objects are listed in the order of RA value. The coordinates are from the 2MASS. The object with bold coordinates is the one selected for the spectroscopy observation. \par}
         
    \tablebib{(1). Sample only from \citet{reyle2018GaiaDR2UCD}, with \textit{Gaia} coordinates, $i$, $z$, $Y$ photometry from the PS1 and $J$, $H$, $K$ photometry from the 2MASS. } 


\begin{table}[htbp]
\fontsize{8}{10}\selectfont
\centering
\caption{\label{EDFNcatalogL}Eight L-type UCD candidates as photometric standards in the 
EDF North and their photometry measurements.}
\begin{tabular}{cccccccccccc}
  \hline\hline
  $\alpha$ (hms.ss) & 
  $\delta$ (dms.s) &
  $i$&
  $z$&
  $Y$&
  $J$&
  $H$&
  $K$&
  $W1$&
  $W2$&
  \\
\hline
  17:44:09.36 & +64:06:39.4 & 20.86$\pm$0.04 & 19.44$\pm$0.02 & 18.58$\pm$0.0 & 16.79$\pm$0.16 & 15.78$\pm$0.14 & 15.57$\pm$0.24 & 14.81$\pm$0.02 & 14.56$\pm$0.04\\
  \textbf{17:54:10.29} & \textbf{+67:12:11.7} & 20.35$\pm$0.03 & 18.91$\pm$0.02 & 18.03$\pm$0.03 & 16.13$\pm$0.1 & 15.48$\pm$0.12 & 15.09$\pm$0.18 & 14.77$\pm$0.01 & 14.61$\pm$0.05\\
  17:59:47.45 & +63:31:24.5 & 20.53$\pm$0.04 & 19.06$\pm$0.01 & 18.2$\pm$0.02 & 16.51$\pm$0.12 & 15.91$\pm$0.15 & 15.48$\pm$0.19 & 15.22$\pm$0.12 & 14.85$\pm$0.07\\
  18:09:04.41 & +64:08:44.4 & 21.03$\pm$0.03 & 19.59$\pm$0.04 & 18.69$\pm$0.02 & 16.73$\pm$0.14 & 16.14$\pm$0.17 & 15.44$\pm$0.19 & 15.03$\pm$0.02 & 14.77$\pm$0.06\\
  18:14:40.32 & +64:08:50.9 & 20.17$\pm$0.01 & 18.79$\pm$0.01 & 17.92$\pm$0.01 & 16.19$\pm$0.08 & 15.68$\pm$0.11 & 14.94$\pm$0.11 & 14.76$\pm$0.02 & 14.48$\pm$0.05\\
  18:18:32.17 & +64:51:17.8 & 21.02$\pm$0.02 & 19.56$\pm$0.02 & 18.7$\pm$0.03 & 16.91$\pm$0.17 & 16.02$\pm$0.14 & 15.03 & 15.62$\pm$0.15 & 15.18$\pm$0.11\\
  18:19:10.46 & +67:16:21.5 & 20.79$\pm$0.02 & 19.4$\pm$0.03 & 18.45$\pm$0.03 & 16.76$\pm$0.16 & 16.05$\pm$0.22 & 15.25$\pm$0.17 & 15.17$\pm$0.04 & 14.92$\pm$0.06\\
  18:19:49.43 & +66:00:54.8 & 20.24$\pm$0.02 & 18.65$\pm$0.01 & 17.62$\pm$0.01 & 15.44$\pm$0.06 & 14.6$\pm$0.06 & 14.13$\pm$0.06 & 13.46$\pm$0.01 & 13.18$\pm$0.02\\
  \hline
\end{tabular}

\vspace{1ex}
     {\raggedright 
         \textbf{Note:} Objects are listed in the order of RA value. The coordinates are from the 2MASS. The object with bold coordinates is the one selected for the spectroscopy observation. \par}
\end{table}

\FloatBarrier

\clearpage

\section{Lists of ultracool dwarf candidates in Euclid Deep Field Fornax}

\begin{table}[htbp]
\fontsize{8}{10}\selectfont
\centering
\caption{\label{EDFFcatalogM}49 M-type UCD candidates as photometric standards in EDF Fornax and their photometry measurements.}
\begin{tabular}{cccccccccccc}
  \hline\hline
  $\alpha$ (hms.ss) & 
  $\delta$ (dms.s) &
  SpT.&
  $i$&
  $z$&
  $Y$&
  $J$&
  $H$&
  $K$&
  $W1$&
  $W2$
  \\
\hline
  $^{(1)}$03:27:11.85 & $-$28:28:23.5 & 9.0 & 19.05$\pm$0.0 & 17.55$\pm$0.0 & 17.09$\pm$0.0 & 15.34$\pm$0.05 & 14.62$\pm$0.05 & 14.19$\pm$0.06 & 13.91$\pm$0.03 & 13.65$\pm$0.03\\
  $^{(1)}$03:27:46.17 & $-$29:06:59.5 & 8.5 & 17.95$\pm$0.0 & 16.54$\pm$0.0 & 16.09$\pm$0.0 & 14.53$\pm$0.04 & 13.8$\pm$0.04 & 13.41$\pm$0.04 & 13.14$\pm$0.02 & 12.87$\pm$0.03\\
  03:29:34.75 & $-$29:11:52.1 & 8.0 & 21.15$\pm$0.01 & 19.94$\pm$0.01 & 19.62$\pm$0.04 & 18.02$\pm$0.03 & 17.41$\pm$0.03 & 16.93$\pm$0.05 & - & -\\
  03:30:13.20 & $-$26:30:18.8 & 9.0 & 21.37$\pm$0.02 & 20.05$\pm$0.01 & 19.68$\pm$0.04 & 17.95$\pm$0.03 & 17.48$\pm$0.04 & 17.04$\pm$0.06 & 16.69$\pm$0.06 & 16.74$\pm$0.22\\
  03:30:27.26 & $-$27:01:51.5 & 8.0 & 21.86$\pm$0.03 & 20.53$\pm$0.01 & 20.21$\pm$0.07 & 18.57$\pm$0.06 & 18.27$\pm$0.09 & 17.73$\pm$0.12 & 18.07$\pm$0.18 & 17.21\\
  03:30:44.91 & $-$26:24:35.0 & 9.0 & 22.98$\pm$0.07 & 21.53$\pm$0.03 & 21.29$\pm$0.2 & 19.51$\pm$0.13 & 19.5$\pm$0.27 & - & - & -\\
  03:31:13.55 & $-$27:27:12.1 & 9.0 & 22.62$\pm$0.04 & 21.3$\pm$0.02 & 20.95$\pm$0.12 & 19.24$\pm$0.1 & 18.75$\pm$0.14 & 18.37$\pm$0.22 & - & -\\
  03:31:33.08 & $-$29:23:41.9 & 9.0 & 22.19$\pm$0.03 & 20.82$\pm$0.02 & 20.4$\pm$0.07 & 18.69$\pm$0.05 & 18.04$\pm$0.05 & 17.56$\pm$0.08 & - & -\\
  03:31:49.70 & $-$26:55:13.0 & 8.0 & 20.09$\pm$0.01 & 18.81$\pm$0.0 & 18.5$\pm$0.02 & 16.84$\pm$0.01 & 16.28$\pm$0.02 & 15.9$\pm$0.02 & 15.59$\pm$0.03 & 15.55$\pm$0.08\\
  03:31:50.21 & $-$29:30:22.1 & 9.0 & 22.86$\pm$0.07 & 21.44$\pm$0.04 & 21.1$\pm$0.16 & 19.46$\pm$0.1 & 18.86$\pm$0.12 & 18.3$\pm$0.16 & 18.15$\pm$0.21 & 17.48\\
  03:31:52.86 & $-$29:20:38.0 & 8.0 & 21.82$\pm$0.03 & 20.61$\pm$0.02 & 20.45$\pm$0.09 & 18.66$\pm$0.05 & 18.09$\pm$0.06 & 17.89$\pm$0.11 & 17.54$\pm$0.12 & 17.51\\
  03:32:12.47 & $-$29:46:39.6 & 8.0 & 21.77$\pm$0.02 & 20.45$\pm$0.02 & 20.19$\pm$0.06 & 18.45$\pm$0.04 & 18.02$\pm$0.06 & 17.84$\pm$0.11 & 17.35$\pm$0.1 & 16.87$\pm$0.24\\
  03:32:14.37 & $-$29:40:20.9 & 9.0 & 22.65$\pm$0.07 & 21.19$\pm$0.04 & 20.83$\pm$0.17 & 19.17$\pm$0.08 & 18.8$\pm$0.11 & 17.93$\pm$0.12 & - & -\\
  03:32:19.94 & $-$28:52:44.3 & 8.0 & 20.31$\pm$0.01 & 19.07$\pm$0.0 & 18.75$\pm$0.02 & 17.06$\pm$0.02 & 16.54$\pm$0.02 & 16.1$\pm$0.03 & 15.96$\pm$0.04 & 15.73$\pm$0.09\\
  03:32:24.16 & $-$27:42:11.4 & 8.0 & 22.26$\pm$0.04 & 21.01$\pm$0.02 & 20.76$\pm$0.11 & 19.06$\pm$0.09 & 18.68$\pm$0.13 & 18.0$\pm$0.16 & 18.15$\pm$0.2 & 17.56\\
  03:32:26.71 & $-$28:02:47.8 & 9.0 & 22.04$\pm$0.03 & 20.59$\pm$0.01 & 20.21$\pm$0.06 & 18.46$\pm$0.08 & 18.04$\pm$0.09 & 17.39$\pm$0.09 & - & -\\
  03:32:27.28 & $-$27:06:59.8 & 9.0 & 21.39$\pm$0.01 & 20.06$\pm$0.01 & 19.72$\pm$0.04 & 18.04$\pm$0.03 & 17.43$\pm$0.04 & 17.01$\pm$0.06 & 16.71$\pm$0.06 & 16.06$\pm$0.12\\
  03:32:32.52 & $-$28:17:11.7 & 9.0 & 22.57$\pm$0.05 & 21.31$\pm$0.03 & 20.98$\pm$0.13 & 19.26$\pm$0.15 & 18.86$\pm$0.19 & 18.42$\pm$0.24 & 17.72$\pm$0.13 & 17.04\\
  03:32:33.19 & $-$26:42:05.7 & 9.0 & 20.84$\pm$0.01 & 19.52$\pm$0.0 & 19.19$\pm$0.03 & 17.46$\pm$0.02 & 16.87$\pm$0.02 & 16.48$\pm$0.04 & 16.09$\pm$0.05 & 15.88$\pm$0.11\\
  \textbf{03:32:34.37} & \textbf{$-$27:33:33.9} & 9.0 & 20.54$\pm$0.01 & 19.25$\pm$0.0 & 18.88$\pm$0.02 & 17.2$\pm$0.02 & 16.66$\pm$0.02 & 16.24$\pm$0.03 & 15.84$\pm$0.04 & 15.58$\pm$0.08\\
  03:32:34.40 & $-$29:34:01.0 & 8.0 & 21.86$\pm$0.03 & 20.61$\pm$0.02 & 20.41$\pm$0.09 & 18.7$\pm$0.05 & 18.09$\pm$0.06 & 17.62$\pm$0.09 & - & -\\
  03:32:58.14 & $-$27:58:29.9 & 9.0 & 22.42$\pm$0.04 & 21.11$\pm$0.02 & 20.85$\pm$0.12 & 18.97$\pm$0.12 & 18.7$\pm$0.16 & 18.25$\pm$0.21 & 17.77$\pm$0.14 & 17.0$\pm$0.26\\
  03:32:59.19 & $-$29:26:03.7 & 8.0 & 22.09$\pm$0.03 & 20.86$\pm$0.02 & 20.64$\pm$0.11 & 18.99$\pm$0.06 & 18.52$\pm$0.09 & 17.97$\pm$0.12 & 17.85$\pm$0.15 & 17.53\\
  03:33:02.41 & $-$27:09:27.9 & 9.0 & 23.12$\pm$0.07 & 21.77$\pm$0.04 & 21.26$\pm$0.16 & 19.66$\pm$0.14 & 19.0$\pm$0.17 & 18.68$\pm$0.28 & - & -\\
  03:33:18.66 & $-$26:39:31.8 & 9.0 & 21.44$\pm$0.02 & 20.1$\pm$0.01 & 19.88$\pm$0.05 & 18.1$\pm$0.04 & 17.54$\pm$0.04 & 17.07$\pm$0.06 & - & -\\
  03:33:32.76 & $-$29:12:10.5 & 9.0 & 22.34$\pm$0.04 & 20.91$\pm$0.02 & 20.64$\pm$0.11 & 18.86$\pm$0.06 & 18.37$\pm$0.08 & 18.06$\pm$0.13 & 17.83$\pm$0.15 & 17.5\\
  03:33:41.46 & $-$29:36:08.7 & 9.0 & 21.15$\pm$0.01 & 19.67$\pm$0.01 & 19.33$\pm$0.04 & 17.57$\pm$0.02 & 16.98$\pm$0.02 & 16.53$\pm$0.03 & 16.34$\pm$0.05 & 16.04$\pm$0.11\\
  03:33:41.71 & $-$27:24:31.6 & 8.0 & 21.0$\pm$0.01 & 19.78$\pm$0.01 & 19.39$\pm$0.04 & 17.74$\pm$0.03 & 17.17$\pm$0.03 & 16.88$\pm$0.05 & 16.44$\pm$0.05 & 16.61$\pm$0.21\\
  03:33:53.27 & $-$27:49:15.1 & 8.0 & 21.85$\pm$0.03 & 20.56$\pm$0.01 & 20.28$\pm$0.07 & 18.6$\pm$0.08 & 18.29$\pm$0.11 & 17.7$\pm$0.13 & 17.41$\pm$0.11 & 17.41\\
  03:34:06.32 & $-$28:25:05.0 & 8.0 & 21.6$\pm$0.02 & 20.32$\pm$0.01 & 20.0$\pm$0.05 & 18.39$\pm$0.07 & 17.68$\pm$0.06 & 17.31$\pm$0.09 & 17.33$\pm$0.1 & 17.23$\pm$0.35\\
  03:34:09.89 & $-$26:47:29.5 & 9.0 & 22.89$\pm$0.06 & 21.35$\pm$0.02 & 20.96$\pm$0.12 & 19.1$\pm$0.09 & 18.69$\pm$0.13 & 18.48$\pm$0.23 & 18.14$\pm$0.21 & 17.59\\
  03:34:21.38 & $-$26:59:08.5 & 9.0 & 21.88$\pm$0.02 & 20.43$\pm$0.01 & 20.2$\pm$0.07 & 18.34$\pm$0.05 & 17.82$\pm$0.06 & 17.46$\pm$0.09 & 17.05$\pm$0.08 & 17.48\\
  03:34:34.36 & $-$29:43:56.3 & 8.0 & 21.21$\pm$0.01 & 19.93$\pm$0.01 & 19.68$\pm$0.05 & 18.03$\pm$0.03 & 17.59$\pm$0.04 & 17.25$\pm$0.07 & 17.14$\pm$0.09 & 16.74$\pm$0.21\\
  03:34:41.17 & $-$27:46:08.6 & 9.0 & 21.86$\pm$0.03 & 20.51$\pm$0.01 & 20.16$\pm$0.07 & 18.39$\pm$0.07 & 17.86$\pm$0.08 & 17.35$\pm$0.1 & 17.13$\pm$0.09 & 16.73$\pm$0.21\\
  03:34:50.80 & $-$26:32:31.3 & 9.0 & 21.85$\pm$0.03 & 20.56$\pm$0.02 & 20.2$\pm$0.08 & 18.55$\pm$0.06 & 18.03$\pm$0.07 & 17.57$\pm$0.11 & 17.52$\pm$0.14 & 16.61$\pm$0.2\\
  03:34:52.21 & $-$27:06:22.1 & 8.0 & 21.28$\pm$0.02 & 20.03$\pm$0.01 & 19.7$\pm$0.04 & 18.06$\pm$0.04 & 17.54$\pm$0.05 & 17.15$\pm$0.07 & 17.02$\pm$0.08 & 16.4$\pm$0.17\\
  03:34:52.42 & $-$28:04:28.6 & 8.0 & 22.52$\pm$0.05 & 21.25$\pm$0.03 & 21.07$\pm$0.14 & 19.22$\pm$0.15 & 18.82$\pm$0.18 & 18.26$\pm$0.22 & - & -\\
  03:34:56.52 & $-$29:20:19.1 & 8.0 & 21.73$\pm$0.02 & 20.48$\pm$0.01 & 20.24$\pm$0.07 & 18.59$\pm$0.05 & 18.12$\pm$0.06 & 17.67$\pm$0.11 & 17.47$\pm$0.12 & 17.33\\
  03:35:07.60 & $-$28:48:25.9 & 8.0 & 21.5$\pm$0.02 & 20.22$\pm$0.01 & 20.0$\pm$0.07 & 18.28$\pm$0.05 & - & - & 17.1$\pm$0.09 & 16.95$\pm$0.27\\
  03:35:17.37 & $-$28:43:41.4 & 9.0 & 21.82$\pm$0.01 & 20.56$\pm$0.01 & 20.34$\pm$0.07 & 18.62$\pm$0.07 & - & - & 16.97$\pm$0.08 & 16.37$\pm$0.16\\
  03:35:21.13 & $-$27:43:39.4 & 9.0 & 22.31$\pm$0.04 & 20.93$\pm$0.02 & 20.62$\pm$0.09 & 18.89$\pm$0.11 & - & - & 17.62$\pm$0.13 & 17.11\\
  03:35:40.31 & $-$29:17:08.7 & 9.0 & 21.65$\pm$0.01 & 20.26$\pm$0.01 & 19.8$\pm$0.05 & 18.14$\pm$0.03 & 17.62$\pm$0.04 & 17.12$\pm$0.07 & - & -\\
  03:35:48.51 & $-$29:21:14.7 & 9.0 & 21.96$\pm$0.02 & 20.63$\pm$0.01 & 20.28$\pm$0.07 & 18.6$\pm$0.05 & 18.02$\pm$0.06 & 17.63$\pm$0.1 & 17.31$\pm$0.1 & 16.96$\pm$0.27\\
  $^{(1)}$03:36:19.11 & $-$29:24:49.6 & 7.0 & 18.44$\pm$0.0 & 17.26$\pm$0.0 & 16.93$\pm$0.0 & 15.38$\pm$0.06 & 14.71$\pm$0.07 & 14.39$\pm$0.08 & 14.19$\pm$0.03 & 13.95$\pm$0.03\\
  03:36:24.43 & $-$27:24:31.0 & 9.0 & 21.56$\pm$0.02 & 20.2$\pm$0.01 & 19.89$\pm$0.05 & 18.27$\pm$0.06 & 17.55$\pm$0.07 & 17.29$\pm$0.1 & 16.98$\pm$0.08 & 16.81$\pm$0.25\\
  03:36:41.95 & $-$26:44:25.6 & 8.0 & 22.35$\pm$0.03 & 21.08$\pm$0.02 & 20.86$\pm$0.11 & 19.11$\pm$0.15 & 18.77$\pm$0.21 & 18.21$\pm$0.22 & - & -\\
  03:37:58.85 & $-$26:59:26.4 & 8.0 & 21.5$\pm$0.02 & 20.29$\pm$0.01 & 19.97$\pm$0.05 & 18.29$\pm$0.07 & 17.83$\pm$0.09 & 17.33$\pm$0.1 & 17.02$\pm$0.09 & 17.14\\
  $^{(1)}$03:38:25.46 & $-$27:26:31.3 & 7.0 & 18.19$\pm$0.0 & 17.08$\pm$0.0 & 16.76$\pm$0.0 & 15.25$\pm$0.05 & 14.63$\pm$0.06 & 14.38$\pm$0.08 & 14.09$\pm$0.03 & 13.9$\pm$0.03\\
  03:39:11.72 & $-$28:21:14.5 & 9.0 & 20.98$\pm$0.01 & 19.62$\pm$0.01 & 19.24$\pm$0.03 & 17.52$\pm$0.03 & - & - & 16.4$\pm$0.05 & 16.06$\pm$0.14\\
  \hline
\end{tabular}

\vspace{1ex}
     {\raggedright 
         \textbf{Note:} Objects are listed in the order of RA value. The coordinates are from the DES. SpT indicates the spectral type beginning with M0.0 as 0.0. The object with bold coordinates is the one selected for the spectroscopy observation. \par}

        \tablebib{(1). Sample and SpT. only from \citet{reyle2018GaiaDR2UCD}, with \textit{Gaia} coordinates, $i$, $z$, $Y$ photometry from the DES and $J$, $H$, $K$ photometry from the 2MASS. } 
\end{table}

\FloatBarrier

\begin{table}
\fontsize{8}{10}\selectfont
\centering
\caption{\label{EDFFcatalogL}29 L-type UCD candidates as photometric standards in the 
EDF Fornax and their photometry measurements.}
\begin{tabular}{cccccccccccc}
  \hline\hline
  $\alpha$ (hms.ss) & 
  $\delta$ (dms.s) &
  SpT.&
  $i$&
  $z$&
  $Y$&
  $J$&
  $H$&
  $K$&
  $W1$&
  $W2$
  \\
\hline
  03:30:26.86 & $-$28:56:38.1 & 10.0 & 21.95$\pm$0.03 & 20.45$\pm$0.02 & 20.12$\pm$0.07 & 18.36$\pm$0.07 & 17.68$\pm$0.07 & 17.09$\pm$0.07 & 16.96$\pm$0.08 & 16.91$\pm$0.25\\
  03:30:29.06 & $-$27:35:49.7 & 10.0 & 22.92$\pm$0.08 & 21.44$\pm$0.04 & 20.92$\pm$0.17 & 19.3$\pm$0.11 & 18.62$\pm$0.12 & 18.31$\pm$0.2 & 17.93$\pm$0.16 & 17.09$\pm$0.29\\
  03:30:42.26 & $-$28:40:11.8 & 10.0 & 22.93$\pm$0.07 & 21.42$\pm$0.03 & 21.14$\pm$0.17 & 19.18$\pm$0.13 & 18.5$\pm$0.14 & 18.27$\pm$0.22 & - & -\\
  03:31:13.43 & $-$28:58:15.7 & 11.0 & 23.63$\pm$0.12 & 21.99$\pm$0.05 & 21.48$\pm$0.17 & 19.32$\pm$0.16 & 19.01$\pm$0.22 & - & 18.18$\pm$0.21 & 17.73\\
  03:31:16.21 & $-$28:28:17.2 & 10.0 & 23.01$\pm$0.06 & 21.61$\pm$0.03 & 21.14$\pm$0.2 & 19.32$\pm$0.15 & 18.61$\pm$0.15 & 18.2$\pm$0.2 & - & -\\
  03:31:17.74 & $-$26:21:49.0 & 11.0 & 22.45$\pm$0.05 & 20.9$\pm$0.02 & 20.54$\pm$0.1 & 18.58$\pm$0.06 & 17.85$\pm$0.06 & 17.45$\pm$0.09 & 16.99$\pm$0.08 & 17.05$\pm$0.29\\
  03:31:59.07 & $-$27:39:25.7 & 10.0 & 22.48$\pm$0.06 & 21.04$\pm$0.03 & 20.53$\pm$0.14 & 18.79$\pm$0.07 & 18.18$\pm$0.08 & 17.68$\pm$0.11 & 17.52$\pm$0.12 & 16.98$\pm$0.26\\
\hline
\end{tabular}

\vspace{1ex}
     {\raggedright 
         \textbf{Note:} Objects are listed in the order of RA value. The coordinates are from the DES. SpT indicates the spectral type beginning with L0.0 as 10.0. The object with bold coordinates is the one selected for the spectroscopy observation. \par}
\end{table}

\clearpage
\section{Lists of ultracool dwarf candidates in Euclid Deep Field South}
\LTcapwidth=\textwidth

\begin{longtable}{ccccccccccccc}
    \caption{\label{EDFScatalogM} 231 late-M-type UCD candidates as photometric standards in EDF South and their photometry measurements. }\\  
  \hline\hline
$\alpha$ (hhmmss.ss) &
$\delta$ (ddmmss.s)&
SpT.&
  $i$&
  $z$&
  $Y$&
  $J$&
  $H$&
  $K_s$&
  $W1$&
  $W2$&
  \\
\hline
\endfirsthead
\caption{continued.}\\
\hline\hline
$\alpha$ (hms.ss) &
$\delta$ (dms.s)&
SpT.&
  $i$&
  $z$&
  $Y$&
  $J$&
  $H$&
  $K_s$&
  $W1$&
  $W2$&
 \\
\hline
\endhead
\hline
\endfoot

\fontsize{8}{10}\selectfont

  03:45:19.13 & $-$48:46:43.8 & 8.0 & 21.15$\pm$0.01 & 19.87$\pm$0.01 & 19.6$\pm$0.03 & 17.88$\pm$0.04 & 17.41$\pm$0.06 & 17.01$\pm$0.06 & 16.91$\pm$0.06 & 16.72$\pm$0.19\\
  03:45:19.40 & $-$49:58:55.7 & 9.0 & 22.09$\pm$0.03 & 20.85$\pm$0.01 & 20.64$\pm$0.07 & 18.82$\pm$0.07 & 18.24$\pm$0.09 & 17.74$\pm$0.14 & 17.56$\pm$0.09 & 18.0\\
  03:45:39.80 & $-$48:50:05.0 & 8.0 & 21.84$\pm$0.02 & 20.52$\pm$0.01 & 20.25$\pm$0.06 & 18.49$\pm$0.06 & 18.06$\pm$0.1 & 17.52$\pm$0.1 & 17.39$\pm$0.09 & 17.1$\pm$0.25\\
  03:46:06.58 & $-$48:51:15.0 & 9.0 & 21.79$\pm$0.03 & 20.38$\pm$0.01 & 19.98$\pm$0.05 & 18.28$\pm$0.05 & 17.73$\pm$0.08 & 17.23$\pm$0.08 & 17.05$\pm$0.07 & 17.21$\pm$0.26\\
  03:46:23.36 & $-$50:02:43.4 & 9.0 & 22.95$\pm$0.08 & 21.46$\pm$0.03 & 21.17$\pm$0.15 & 19.34$\pm$0.11 & 19.11$\pm$0.19 & - & - & -\\
  03:47:07.41 & $-$49:57:30.2 & 8.0 & 22.81$\pm$0.07 & 21.56$\pm$0.03 & 21.3$\pm$0.11 & 19.61$\pm$0.15 & 19.37$\pm$0.24 & - & - & -\\
  03:47:19.27 & $-$50:39:19.5 & 9.0 & 23.15$\pm$0.06 & 21.87$\pm$0.04 & 21.62$\pm$0.2 & 19.77$\pm$0.17 & 18.96$\pm$0.16 & 18.65$\pm$0.28 & - & -\\
  03:47:56.08 & $-$49:07:41.8 & 9.0 & 21.69$\pm$0.03 & 20.31$\pm$0.01 & 19.89$\pm$0.05 & 18.25$\pm$0.05 & 17.64$\pm$0.07 & 17.15$\pm$0.07 & 17.18$\pm$0.08 & 17.05$\pm$0.24\\
\end{longtable}
  \vspace{1ex}
     {\raggedright 
         \textbf{Note:} Objects are listed in the order of RA value. The coordinates are from the DES. The object with bold coordinates is the one selected for the spectroscopy observation. \par}
    
    \tablebib{(1). Sample and SpT. only from \citet{reyle2018GaiaDR2UCD}, with \textit{Gaia} coordinates, $i$, $z$, $Y$ photometry from the DES and $J$, $H$, $K$ photometry from the 2MASS. } 

\newpage         
\begin{longtable}{ccccccccccccc}
    \caption{\label{EDFScatalogL} 115 L-type UCD candidates as photometric standards in EDF South and their photometry measurements. }\\  
  \hline\hline
$\alpha$ (hhmmss.ss) &
$\delta$ (ddmmss.s)&
SpT.&
  $i$&
  $z$&
  $Y$&
  $J$&
  $H$&
  $K_s$&
  $W1$&
  $W2$&
  \\
\hline
\endfirsthead
\caption{continued.}\\
\hline\hline
$\alpha$ (hms.ss) &
$\delta$ (dms.s)&
SpT.&
  $i$&
  $z$&
  $Y$&
  $J$&
  $H$&
  $K_s$&
  $W1$&
  $W2$&
 \\
\hline
\endhead
\hline
\endfoot

\fontsize{8}{10}\selectfont

  03:45:35.98 & $-$49:07:26.6 & 10.0 & 23.0$\pm$0.1 & 21.39$\pm$0.04 & 21.13$\pm$0.19 & 19.31$\pm$0.13 & 18.61$\pm$0.17 & 18.28$\pm$0.19 & - & -\\
  03:45:39.39 & $-$49:34:32.7 & 12.0 & 22.32$\pm$0.06 & 20.58$\pm$0.02 & 20.1$\pm$0.07 & 17.99$\pm$0.04 & 17.38$\pm$0.06 & 16.97$\pm$0.06 & 16.48$\pm$0.05 & 16.14$\pm$0.1\\
  03:46:38.59 & $-$48:37:16.4 & 10.0 & 21.52$\pm$0.03 & 19.89$\pm$0.01 & 19.46$\pm$0.04 & 17.63$\pm$0.03 & 17.08$\pm$0.04 & 16.68$\pm$0.05 & 16.44$\pm$0.05 & 16.1$\pm$0.11\\
  03:49:04.31 & $-$49:35:41.9 & 11.0 & 22.74$\pm$0.07 & 21.28$\pm$0.03 & 20.87$\pm$0.08 & 18.86$\pm$0.09 & 18.28$\pm$0.13 & 17.49$\pm$0.1 & 17.3$\pm$0.08 & 17.56$\pm$0.36\\
  03:49:20.90 & $-$49:53:58.3 & 10.0 & 20.72$\pm$0.01 & 19.26$\pm$0.01 & 18.79$\pm$0.02 & 16.97$\pm$0.01 & 16.34$\pm$0.02 & 15.92$\pm$0.03 & 15.58$\pm$0.03 & 15.54$\pm$0.07\\
  03:49:41.70 & $-$50:30:46.0 & 13.0 & 22.27$\pm$0.05 & 20.46$\pm$0.02 & 19.96$\pm$0.06 & 17.82$\pm$0.03 & 17.07$\pm$0.03 & 16.61$\pm$0.05 & 16.0$\pm$0.04 & 15.72$\pm$0.07\\
  \textbf{03:50:25.77} & \textbf{$-$49:32:20.7} & 16.0 & 22.62$\pm$0.05 & 20.72$\pm$0.02 & 20.17$\pm$0.05 & 17.96$\pm$0.04 & 16.98$\pm$0.04 & 16.28$\pm$0.03 & 15.55$\pm$0.03 & 15.24$\pm$0.06\\
  03:50:50.36 & $-$50:21:39.0 & 10.0 & 21.65$\pm$0.02 & 20.16$\pm$0.01 & 19.71$\pm$0.04 & 17.96$\pm$0.03 & 17.35$\pm$0.04 & 16.95$\pm$0.07 & 16.59$\pm$0.05 & 16.41$\pm$0.12\\
\end{longtable}
  \vspace{1ex}
     {\raggedright 
         \textbf{Note:} Objects are listed in the order of RA value. The coordinates are from the DES. The object with bold coordinates is the one selected for the spectroscopy observation. \par}

\vspace{2cm}
\begin{table}[htbp]
\fontsize{8}{10}\selectfont
\centering
\caption{\label{EDFScatalogT}Two T-type UCD candidates as photometric standards in the 
EDF South and their photometry measurements.}
\begin{tabular}{cccccccccccc}
  \hline\hline
  $\alpha$ (hms.ss) & 
  $\delta$ (dms.s) &
  SpT.&
  $i$&
  $z$&
  $Y$&
  $J$&
  $H$&
  $K_s$&
  $W1$&
  $W2$&
  \\
\hline
  \textbf{03:52:31.83} & \textbf{$-$49:10:59.4} & 27.0 & 23.78$\pm$0.21 & 21.53$\pm$0.04 & 20.28$\pm$0.08 & 17.82$\pm$0.04 & 18.23$\pm$0.12 & 18.19$\pm$0.17 & 17.88$\pm$0.14 & 15.3$\pm$0.06\\
  \textbf{03:59:09.81} & \textbf{$-$47:40:57.1} & 28.0 & - & 21.82$\pm$0.03 & 20.59$\pm$0.06 & 18.06$\pm$0.03 & 18.3$\pm$0.06 & 18.53$\pm$0.19 & 16.87$\pm$0.06 & 15.19$\pm$0.05\\
  \hline
\end{tabular}

\vspace{1ex}
     {\raggedright 
         \textbf{Note:} Objects are listed in the order of RA value. The coordinates are from the DES. SpT indicates the spectral type beginning with T0.0 as 20.0. The object with bold coordinates is the one selected for the spectroscopy observation. \par}
\end{table}

\end{appendix}

\endgroup
\end{document}